\documentclass[onecolumn,amsmath,amssymb,12pt,superscriptaddress,nofootinbib,floatfix]{revtex4}
\pdfoutput=1

\usepackage[latin1]{inputenc}
\usepackage[english]{babel}
\usepackage{amssymb}
\usepackage{amsmath}
\usepackage{amsthm}
\usepackage[]{graphicx}
\usepackage[]{subfigure}
\usepackage{tensor}
\usepackage{color}
\usepackage{cancel}
\usepackage{setspace}
\usepackage{fancyhdr}
\usepackage[bookmarks,linktocpage, colorlinks=true, plainpages = false, citecolor = blue,  linkcolor=blue, urlcolor = blue, filecolor = blue]{hyperref}

\def\1p{{(1p)}}

\def\p0{\phi_0}

\def\be{\begin{equation}}
\def\ee{\end{equation}}
\def\beq{\begin{eqnarray}}
\def\eeq{\end{eqnarray}}

\begin{document}

\title{Homogeneous Transitions during Inflation: a Description in Quantum Cosmology}
\author{Sebastian F. Bramberger}
\email{sebastian.bramberger@aei.mpg.de}
\affiliation{Max Planck Institute for Gravitational Physics \\ (Albert Einstein Institute), 14476 Potsdam-Golm, Germany}
\author{Alice Di Tucci}
\email{alice.di-tucci@aei.mpg.de}
\affiliation{Max Planck Institute for Gravitational Physics \\ (Albert Einstein Institute), 14476 Potsdam-Golm, Germany}
\author{Jean-Luc Lehners}
\email{jlehners@aei.mpg.de}
\affiliation{Max Planck Institute for Gravitational Physics \\ (Albert Einstein Institute), 14476 Potsdam-Golm, Germany}

\begin{abstract}
\vspace{1cm}
\noindent The usual description of inflationary fluctuations uses the framework of quantum field theory (QFT) in curved spacetime, in which quantum fluctuations are superimposed on a classical background spacetime. Even for large fluctuations, such as those envisioned during a regime of eternal inflation, this framework is frequently used. In the present work we go one step beyond this description by quantising both the scalar field and the scale factor of the universe. Employing the Lorentzian path integral formulation of semi-classical gravity we restrict to a simplified minisuperspace setting by considering homogeneous transitions. This approach allows us to determine the dominant geometry and inflaton evolution contributing to such amplitudes. We find that for precisely specified initial scale factor and inflaton values (and uncertain momenta), two distinct saddle point geometries contribute to the amplitude, leading to interference effects. However, when the momenta of both scale factor and inflaton are specified with sufficient certainty, only a single saddle point is relevant and QFT in curved spacetime is applicable. In particular we find that for inflaton transitions up the potential, meaningful results are only obtained when the initial uncertainty in the inflaton value is large enough, allowing the dominant evolution to be a complexified slow-roll solution \emph{down} from a comparatively unlikely position higher up in the potential.
\end{abstract}

\maketitle
\newpage
	\tableofcontents
	
\section{Introduction}

A beautiful idea of modern cosmology is that the origin of the largest structures in the universe may lie in primordial quantum fluctuations \cite{Sakharov:1966aja,Mukhanov:1981xt}. Inflation and ekpyrosis provide concrete mechanisms that can amplify quantum fluctuations into essentially classical density perturbations, which can then act as seeds for the formation of structure via gravitational collapse \cite{Starobinsky:1979ty,Mukhanov:1981xt,Guth:1982ec,Hawking:1982cz,Bardeen:1983qw,Finelli:2001sr,Finelli:2002we,Lehners:2007ac,Qiu:2013eoa,Li:2013hga,Ijjas:2014fja,Kiefer:1998qe,Battarra:2013cha}. The amplification itself is calculated within the framework of quantum field theory (QFT) in curved spacetime. In this formalism, one fixes a classical background spacetime (and a classical background matter configuration) and then quantises small fluctuations around this background \cite{Birrell:1982ix}. This approach is reminiscent of the Born-Oppenheimer approximation, where light electronic excitations are quantised around a heavy atomic nucleus which to a first approximation is treated classically. This analogy suggests that for many applications this approximation scheme should be valid and yield precise results. However, there are also good reasons to try to go beyond this first approximation: conceptually, it makes little sense to think of the background as classical and the fluctuations as quantum. All of nature should be described by the same theory, and thus the background should be thought of as being just as much part of the quantum wavefunction as the fluctuations. Beyond this conceptual consideration, it is important to gain an understanding of the quantisation of the entire system in order to assess under what circumstances the approximation of QFT in curved spacetime breaks down, and to see what might replace it in such a regime. In the context of inflation, which we will focus on in this paper, the calculation of quantum fluctuations is used not only for small fluctuations, but also for large fluctuations deep in the tails of the distribution. This is especially relevant for eternal inflation, where it is assumed that the quantum fluctuations of the inflaton can be larger than its changes due to classical evolution \cite{Steinhardt:1982kg,Vilenkin:1983xq}. Although such large fluctuations are rare, they may play an important role in the cosmological context as they can alter the global structure of spacetime: in a region where the inflaton jumps up the potential, the expansion rate of the universe will be larger than before, and this will cause that region to grow significantly more than the classical evolution would have suggested. It is notoriously hard to make predictions for observables under these circumstances (see e.g. \cite{Aguirre:2007gy,Johnson:2011aa} and references therein), and this provides further motivation for trying to understand such large quantum fluctuations in more detail.

In this work we will undertake a first step in the direction of understanding inflationary fluctuations in semi-classical gravity, where the background is quantised alongside the fluctuations. We achieve this by working with the path integral formulation of gravity and, more specifically, with the Lorentzian path integral \cite{Teitelboim:1981ua,Teitelboim:1983fk}. Moreover, we will make use of an exactly solvable minisuperspace model in which gravity is coupled to a scalar field with a specific inflationary potential \cite{Garay:1990re}. The fact that we are working in minisuperspace, and that we consequently only consider homogeneous fluctuations of the fields, is a restriction that we hope to improve on in future work. However, on super-Hubble scales such an approximation should be rather accurate by simple virtue of causality (cf. also the stochastic picture of super-Hubble fluctuations \cite{Starobinsky:1986fx}). 

Our goal then is to describe homogeneous inflationary transitions, both small and large, in a fully quantum manner. The framework that we employ allows us to see how the fields evolve ``during'' a quantum transition, and we will see how the transition amplitude depends not only on the change in the scalar field, but also (though to a lesser extent) on the change in the scale factor. The key feature of our calculation is the use of Robin boundary conditions. This allows us to follow the semiclassical evolution of a universe which has a large enough initial size and is initially inflating. In order for these requirements to be compatible with Heisenberg's uncertainty principle, the initial size and velocity are specified only with some uncertainty. This is implemented by the Robin condition which is in fact equivalent to an initial coherent state. A general feature that we observe is that the transition amplitude is governed by contributions from two saddle points when the uncertainty in the initial value of the scalar field is small, but with large uncertainty in the inflaton velocity. In this case a description in terms of QFT in curved spacetime in fact breaks down, as two separate backgrounds contribute significantly. However, as soon as the uncertainty in the field value is increased to the expected level ($H/(2\pi)$) while the uncertainty in the field momentum is correspondingly reduced, we generically see that a so-called Stokes phenomenon happens: this is a topological change in the (steepest descent) flow lines, beyond which only a single saddle point remains relevant to the path integral, and where consequently the approximation in terms of QFT in curved spacetime is vindicated. However, in the flattest region of the potential even this is not quite enough, and some additional uncertainty in the size of the universe is required in order to obtain consistent results.

The plan of our paper is as follows: in section \ref{review} we will first review the standard intuition that for very flat potentials, inflationary fluctuations may be thought of as transitions of the scalar field in a fixed background geometry. We will then present our formalism, and the specific minisuperspace model we will focus on, in section \ref{model}. In order to test this formalism, we will apply it in section \ref{test} to boundary conditions that correspond to a scalar field classically rolling down an inflationary potential. This example turns out to be non-trivial already, in that it demonstrates the need for, and the use of, an appropriate initial state. Equipped with these realisations we can then explore transitions during which the scalar field evolves up the potential, in section \ref{jump}. A further constraint on the validity of our calculations is analysed in section \ref{zeros}. We conclude with a discussion of our results in section \ref{discussion}.

\section{Some Aspects of QFT in Curved Spacetime} \label{review}

We start our discussion with a brief review of a few salient features of the theory of cosmological perturbations. Readers familiar with this material may skip to the next section. We will consider theories of gravity minimally coupled to a scalar field $\phi$ with a potential $V(\phi).$ Thus the action is given by
\begin{align}
S = \int d^4x \sqrt{-g} \left[ \frac{R}{2} - \frac{1}{2}(\partial\phi)^2 - V(\phi)\right]\,,
\end{align}
where we have set $8\pi G =1.$ In the cosmological context we are interested in Friedmann-Lema\^{i}tre-Robertson-Walker (FLRW) solutions and perturbations around them. In this section we will focus on spatially flat backgrounds, $\mathrm{d}\bar{s}^2=-\mathrm{d}t^2+a^2(t)\delta_{ij}\mathrm{d}x^i \mathrm{d}x^j,$ where $a(t)$ denotes the background scale factor and $H=\dot{a}/a$ characterises the expansion rate. An inflationary phase then corresponds to a phase of accelerated expansion, $\ddot{a}>0,$ which can also be formulated as the requirement that $\epsilon < 1,$ where we have introduced the slow-roll parameter $\epsilon \equiv - \dot{H}/H^2 = \dot\phi^2/(2H^2).$ The condition for inflation can be met when the potential is sufficiently flat. For a very flat potential, we have the approximate relation $\epsilon \approx V_{,\phi}^2/(2V^2)$, which is valid when $\epsilon \ll 1.$

Now we can consider perturbations of this background spacetime. Retaining only scalar perturbations, we can write the metric as
\begin{eqnarray}
\mathrm{d}s^2=-(1+2A)\mathrm{d}t^2+2a(t)B_{,i}\mathrm{d}x^i \mathrm{d}t+a^2(t)[(1+2\psi)\delta_{ij}+2\partial_i\partial_j E]\mathrm{d}x^i \mathrm{d}x^j \label{metric1}\,,
\end{eqnarray}
where $A,B,\psi,E$ are the perturbations. One additional scalar perturbation arises from the perturbation of the scalar field, $\delta \phi.$ A small local change in the coordinates can be written as $x^\mu \to  x ^{\prime \mu}= x ^{\mu}+\xi ^{\mu}$,
where the vector $\xi^\mu$ can be decomposed as $\xi^{\mu}=(\xi^0,\xi^i)$ with $\xi^i=\xi_T^i+\partial^i\xi.$ Here $\xi$ is a scalar and 
$\partial_i \xi_T^i=0$ is a divergence free 3-vector. Thus $\xi^0$ and $\xi$ are the two scalar transformation parameters. The associated gauge transformations of the metric perturbations are given by
\begin{eqnarray}
A &\to& A+\dot \xi_0\\
B &\to& B+\frac{1}{a}(-\xi_0-\dot \xi+2H \xi) \label{GaugeB}\\
\psi &\to& \psi+H\xi_0\\
E &\to& E- \frac{1}{a^2}\xi,
\end{eqnarray}
while the scalar field perturbation transforms as $\delta\phi \to \delta\phi - \dot\phi \xi^0.$

We will perform our calculation in flat gauge where the spatial metric $h_{ij} = a(t)^2 \delta_{ij}$ is kept fixed as the spatial section of a flat FLRW universe ($\xi^0$ can be chosen to eliminate $\psi$ and $\xi$ to eliminate $E$). At linear order the constraints, which can be thought of as the $00$ and $0i$ Einstein equations, are given by (see e.g. \cite{Koehn:2015vvy})
\begin{align}
A=& \frac{\dot\phi}{2 H}\,\delta\phi  = \sqrt{\frac{\epsilon}{2}} \delta \phi \label{eq:alpha}\\
\partial^i \partial_i B =& -\frac{1}{2 H}(V_{,\phi}+\frac{\dot\phi}{H}V)\delta\phi-\frac{\dot\phi}{2 H} \dot{\delta\phi} = -\epsilon \frac{\mathrm{d}}{\mathrm{d}t}\left( \frac{\delta\phi}{\sqrt{2\epsilon}}\right),
\end{align} 
where in the constraint for $B$ we have already used \eqref{eq:alpha} to replace $A.$  The constraints show that when the slow-roll parameter is very small, $\epsilon \ll 1,$ the metric perturbations are negligible compared to the scalar field fluctuations $\delta\phi$ since they are suppressed by factors of $\sqrt{\epsilon}.$ This is the basis for the standard intuition that in slow-roll inflation one may think of the background spacetime as being constant, with only the scalar field fluctuating.

This picture is reinforced by the fact that at cubic order in interactions, up to a numerical factor of order one the leading contribution in the Lagrangian is a term of the form $\sqrt{\epsilon}(\dot{\delta\phi})^2\delta\phi,$ which is also small in the slow-roll limit. Hence, in the presence of a very flat potential, the system is perturbative. In other words, to a first approximation the system is described by free scalar field fluctuations in a fixed geometry. 

In flat gauge the comoving curvature perturbation is given by $\mathcal{R} = \psi - \frac{H}{\dot \phi}\delta  \phi = - \frac{H}{\dot \phi}\delta  \phi \approx -\frac{1}{\sqrt{2\epsilon}} \delta\phi.$ A classic calculation shows that inflation amplifies quantum fluctuations and induces a variance of the curvature perturbation which on super-Hubble scales and in the slow-roll limit is given by \cite{Mukhanov:1981xt,Guth:1982ec,Hawking:1982cz,Bardeen:1983qw}
\begin{align}
\Delta_{\mathcal{R}}^2 = \frac{H^2}{8\pi^2 \epsilon}\,.
\end{align} 
The relation between the curvature perturbation and the scalar field perturbation then implies that the variance of the scalar field is given by 
\begin{equation}
\Delta\phi_{qu} \equiv \langle (\delta \phi)^2 \rangle^{1/2} = \frac{H}{2\pi}\,.
\end{equation}
This is the typical quantum induced change in the scalar field value during one Hubble time. By comparison, the classical rolling of the scalar field during the same time interval induces a change
\begin{equation}
\Delta\phi_{cl} \equiv \frac{|\dot\phi|}{H} 
\end{equation}
Note that the quantum change dominates over the classical rolling when 
\begin{equation}
\Delta\phi_{qu} > \Delta\phi_{cl} \quad\leftrightarrow \quad\frac{H^2}{2\pi |\dot\phi|} \approx \frac{H}{\sqrt{8\pi \epsilon}}>1 \quad\leftrightarrow\quad \Delta_{\mathcal{R}}^2 > 1\,,
\end{equation}
i.e. precisely when the variance of the curvature perturbation is larger than one, and when perturbation theory becomes questionable. In this regime inflation  is thought to be eternal, and this leads to severe paradoxes in its interpretation \cite{Ijjas:2014nta}. One motivation for the present study is to verify the intuitions from QFT in curved spacetime: does quantum cosmology, where the scale factor of the universe is also quantised, support the view that the scalar field fluctuations evolve in a fixed background spacetime. Does this picture become better or worse as the potential becomes flatter? Is there a qualitative difference between the eternal and non-eternal regimes?

\section{Exactly Soluble Scalar Field Minisuperspace Models} \label{model}

For gravity minimally coupled to a scalar field with a potential, the Feynman propagator in minisuperspace is given by
\begin{align}
G[a_1,\phi_1;a_0,\phi_0] = \int_{0^+}^{\infty} dN \int_{a_0}^{a_1}  \int_{\phi_0}^{\phi_1} Da D\phi e^{iS(a,\phi,N)/\hbar}\,.
\end{align}
This propagator describes the amplitude to go from an initial 3-surface with scale factor $a_0$ and scalar field $\phi_0$ to a final 3-surface specified by $a_1$ and $\phi_1$. The action here is given by the Einstein-Hilbert functional with a minimally coupled scalar field and the Gibbons-Hawking-York boundary term. Note that the last term is crucial to make the variational principle compatible with the mentioned Dirichlet boundary conditions. The full action reads
\begin{align}
\label{action-phys-coord}
S = 6 \pi^2 \int dt_p N \left( -\frac{a \dot{a}^2}{N^2} + a + \frac{a^3}{3} \left(\frac{1}{2}\frac{\dot{\phi}^2}{N^2} - V \right) \right) 
\end{align}
where we used the usual metric of a closed FLRW universe with lapse $N$
\begin{align}
ds^2 = -N^2 dt_p^2 + a(t_p)^2d\Omega_3^2\,.
\end{align}
We take the range of integration of the lapse function to be over strictly positive and real values only. (A detailed discussion of the attractive properties of the Lorentzian path integral was provided in \cite{Feldbrugge:2017kzv,Feldbrugge:2017mbc}.) While the path integral is a very intuitive tool in computing amplitudes for the evolution of the universe, it is not used very much because in most situations it is difficult or impossible to compute it explicitly. In particular it is impossible to solve the above analytically for generic potentials of the scalar field $V(\phi)$. For certain specific forms of $V(\phi)$, however, exact solutions may be obtained. One class has been studied in \cite{Garay:1990re} and we shall review their approach here. Our goal is to transform the action \eqref{action-phys-coord} into a form that is quadratic in its variables such that we can solve the resulting path integral exactly. To do this, first consider a rescaling of the time coordinate,
\begin{align}
ds^2 = -\frac{N^2}{a(t)^2}dt^2 +a(t)^2 d\Omega_3^2,
\end{align}
followed by a redefinition of the fields \cite{Garay:1990re},
\begin{align}
x(t) \equiv a^2(t) \cosh \left( \sqrt{\frac{2}{3}} \phi(t) \right)\,, \label{redef1}\\
y(t) \equiv a^2(t) \sinh \left( \sqrt{\frac{2}{3}}  \phi(t) \right)\,. \label{redef2}
\end{align}
The inverse transformations are given by 
\begin{align}
a(t)=\left(x^2(t)-y^2(t) \right)^{1/4}\,, \qquad \phi(t) = \sqrt{\frac{3}{2}} \tanh^{-1} \left( \frac{y(t)}{x(t)}\right)\label{invtrfm}\,.
\end{align}
Then, for a potential of the form
\begin{align}
V(\phi) = \alpha \cosh \sqrt{\frac{2}{3}}  \phi \,, \label{Pot}
\end{align}
the action reduces to the remarkably compact form \cite{Garay:1990re}
\begin{align}
S = V_3 \int_0^1 dt N\left[ \frac{3}{4 N^2}  \left(y'(t)^2 - x'(t)^2 \right) +3 - \alpha x(t)  \right]\,, \label{eq:action}
\end{align}
where a prime refers to derivation with respect to the coordinate time $t,$ and we are choosing the range of the time coordinate between the initial and final hypersurface to be $0 \leq t \leq 1.$ Here we wrote the coordinate volume of the three-dimensional spatial slice as $V_3$ -- for the standard three-sphere we have $V_3=2\pi^2$ but here, for notational simplicity, we will use re-scaled coordinates such that $V_3=1$ (since we will be interested in situations where the scale factor is large, our calculations also apply with good accuracy to FLRW metrics with flat spatial slices, as long as the spatial volume is regulated to a finite value). The resulting equations of motion are 
\begin{align}
x''(t) = \frac{2\alpha}{3}N^2\,,   \indent
y''(t) = 0\,.
\end{align}
Imposing Dirichlet boundary conditions $x(0) = x_0$, $x(1) = x_1$, and $y(0) = y_0$, $y(1) = y_1$ (where these boundary values are related to the original boundary conditions $a_{0,1}, \phi_{0,1}$ via the definitions \eqref{redef1} and \eqref{redef2}), the resulting solutions are given by
\begin{align}
\bar{x}(t) &= \frac{\alpha}{3} N^2 t^2 + (x_1 - x_0 - \frac{\alpha}{3} N^2)t + x_0\,, \label{sol1}\\
\bar{y}(t) &= (y_1 - y_0)t + y_0\,. \label{sol2}
\end{align}
A general path that is summed over in the path integral can now be written as $x(t) = \bar{x}(t) + X(t)$ and similarly for $y(t).$ The path integral over $x$ can then be performed by shifting variables to $X,$ where the integral over $X$ is a simple Gaussian that can be evaluated exactly. After solving the $x$ and $y$ integrals in this manner we are left with an ordinary one-dimensional integral over the lapse only,
\begin{align}
G[x_1,y_1;x_0,y_0] = \int_{0^+}^{\infty} dN  P(N) e^{iS_0(x_0,x_1,y_0,y_1,N)/\hbar} \label{lapseintegral}
\end{align}
where $P(N)$ is a non-exponential prefactor (scaling as $1/N$), and the action $S_0$ is obtained by substitution of the solutions \eqref{sol1} and \eqref{sol2}, yielding
\begin{align}
S_0 = \frac{\alpha^2}{36} N^3  + N \left( 3 - \frac{1}{2}\alpha (x_0 + x_1) \right) + \frac{3}{4N}  \left( (y_1 -y_0)^2 - (x_1 - x_0)^2 \right)\,. 
\label{action-integr}
\end{align}
In order to evaluate the above integral, which is a conditionally convergent integral, we will make use of Picard-Lefschetz theory, which may be seen as a systematic way of evaluating a saddle point approximation of oscillatory integrals. We will briefly review the salient features in what follows. 

The main idea of Picard-Lefschetz theory is to complexify the integral of interest and then deform the original contour of integration (here the contour for the lapse integral) in such a way as to render the resulting integral manifestly convergent. It may be useful to consider a simple example, say $\tilde{I}=\int_{\mathbb{R}} dx e^{ix^2}.$ Along the defining contour, namely the real line, this is a highly oscillating integral. But now we can deform the contour by defining $x=e^{i\pi/4}y,$ such that $\tilde{I}=e^{i\pi/4}\int dy e^{-y^2}.$ Along the new contour, the integral has stopped oscillating, and in fact the magnitude of the integrand decreases as rapidly as possible. The integral is now manifestly convergent, and one may check that the arcs at infinity linking the original contour to the new one yield zero contribution. Note that along the steepest descent path, there is an overall constant phase factor (here $e^{i\pi/4}$) -- this is a general feature of such paths.

\begin{figure}[h] 
		\includegraphics[width=0.85\textwidth]{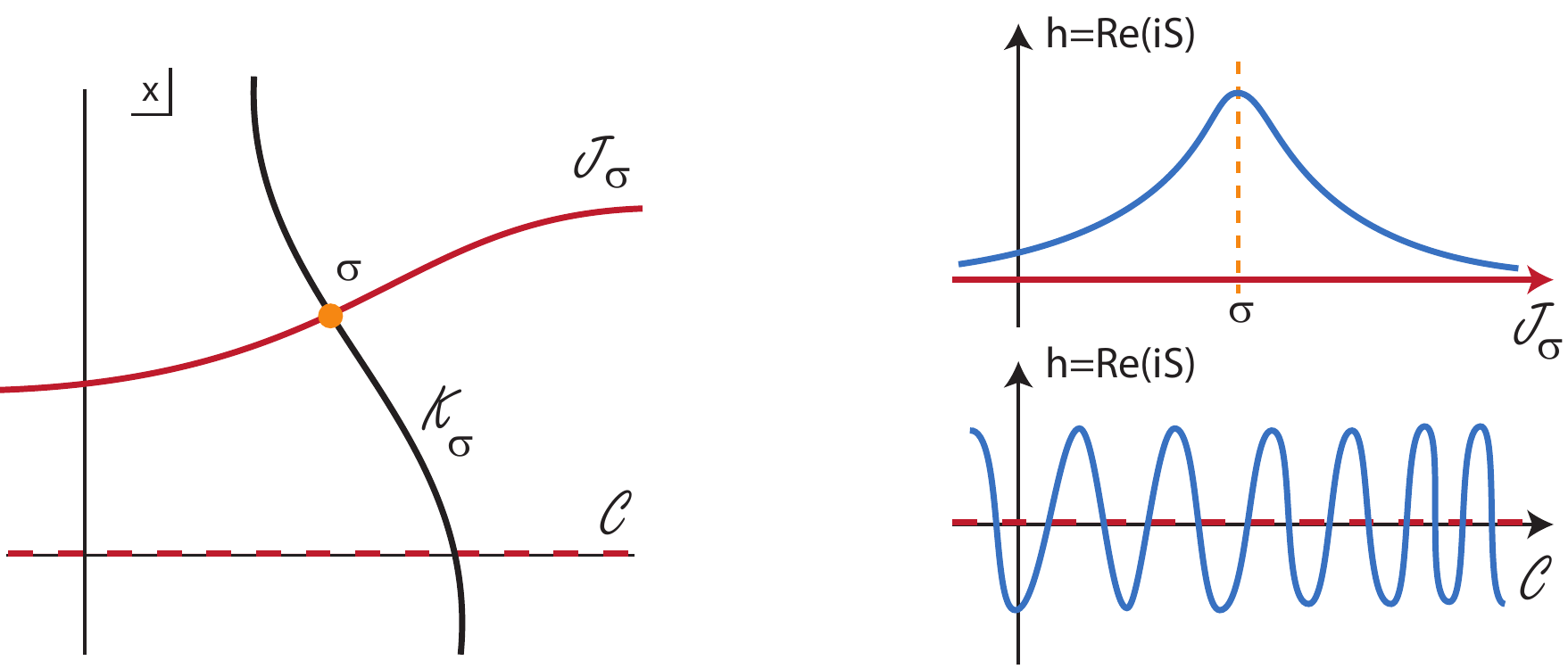}
	\caption{Picard-Lefschetz theory instructs us how to deform a contour of integration such that an oscillating integral along a contour $\mathcal{C}$ gets replaced by a steepest descent contour (or in general a sum thereof) along a Lefschetz thimble $\mathcal{J}_\sigma$ associated with a saddle point $\sigma$. Only those saddle points contribute for which the flow of steepest ascent $\mathcal{K}_\sigma$ intersects $\mathcal{C}.$}
	\protect
	\label{fig:thimble}
\end{figure}

More formally, we can write the exponent $iS[x]/\hbar$ and its argument, taken to be $x$ here, in terms of their real and imaginary parts, $iS/\hbar=h+i H$ and $x=u^1+iu^2$ --  see Fig.~\ref{fig:thimble} for an illustration of the concepts. Downward flow of the magnitude of the integrand is then defined by 
\begin{equation}
\frac{\mathrm{d}u^i}{\mathrm{d}\lambda} = -g^{ij}\frac{\partial h}{\partial u^j}\,,
\label{eq:dw}
\end{equation}
with $\lambda$ denoting a parameter (along the flow) and $g_{ij}$ denoting a metric on the complexified plane of the original variable $x$ (here we can take this metric to be the trivial one, $ds^2 = d|u|^2$). The real part of the exponent $h$  is also called the Morse function. It decreases along the flow, since  $\frac{\mathrm{d}h}{\mathrm{d} \lambda} = \sum_i\frac{\partial h}{\partial u^i}\frac{\mathrm{d}u^i}{\mathrm{d}\lambda} = -\sum_i\left(\frac{\partial h}{\partial u^i}\right)^2<0.$ The downward flow Eq. (\ref{eq:dw}) can be rewritten as 
\begin{equation}
\frac{\mathrm{d}u}{\mathrm{d}\lambda} = - \frac{\partial {\bar{\cal I}}}{\partial \bar{u}}, \quad \frac{\mathrm{d}\bar{u}}{\mathrm{d}\lambda} = - \frac{\partial {{\cal I}}}{\partial {u}}\,,
\end{equation} 
and this form of the equations is useful in that it straightforwardly implies that the phase of the integrand,  $H = \text{Im}[iS/\hbar],$ is conserved along a flow,
\begin{equation}
\label{eq:imh}
\frac{\mathrm{d} H}{\mathrm{d}\lambda} = \frac{1}{2i}\frac{\mathrm{d}({\cal I} - \bar{\cal I})}{\mathrm{d}\lambda} = \frac{1}{2i}\left( \frac{\partial {\cal I}}{\partial u}\frac{\mathrm{d}u}{\mathrm{d}\lambda} - \frac{\partial \bar{\cal I}}{\partial \bar{u}}\frac{\mathrm{d}\bar{u}}{\mathrm{d}\lambda}\right) = 0\,.
\end{equation}
Thus, along a flow the integrand does not oscillate, rather its amplitude decreases as fast as possible. Such a downwards flow emanating from a saddle point $\sigma$ is denoted by $\mathcal{J}_\sigma$ and is often called a ``Lefschetz thimble''. 

In much the same way one can define an upwards flow
\begin{equation}
\frac{\mathrm{d}u^i}{\mathrm{d}\lambda} = +g^{ij}\frac{\partial h}{\partial u^j}\,,
\label{eq:uw}
\end{equation}
with $H$ likewise being constant along such flows. Upwards flows are denoted by ${\cal K}_\sigma,$ and they intersect the thimbles at the saddle points. Thus we can write
\begin{equation}
{\rm Int}({\cal J}_\sigma, {\cal K}_{\sigma'})=\delta_{\sigma \sigma'}. \label{eq:intersection}
\end{equation}
Our goal then is to express the original integration contour $\mathcal{C}$ as a sum over Lefschetz thimbles, 
\begin{equation}
\label{eq:contourexp}
{\cal C} = \sum_\sigma n_\sigma {\cal J}_\sigma\,.
\end{equation}
Multiplying this equation on both sides by ${\cal K}_{\sigma}$ we obtain that $n_\sigma= {\rm Int}(\mathcal{C}, {\cal K}_{\sigma}).$ Thus a saddle point, and its associated thimble, are relevant if and only if one can reach the original integration contour via an upwards flow from the saddle point in question. Intuitively, this makes sense: we are replacing an oscillating integral, with many cancellations, by one which does not contain cancellations, and thus the amplitude along the non-oscillating path must be lower. Putting everything together, we can then re-express the conditionally convergent integral by a sum over convergent integrals, 
\begin{align}
\label{eq:contour}
\int_{\cal C} \mathrm{d} x \, e^{iS[x]/\hbar} & = \sum_\sigma n_\sigma \int_{{\cal J}_\sigma} \mathrm{d} x \, e^{iS[x]/\hbar} \\
& = \sum_\sigma n_\sigma \, e^{i \, H(x_\sigma)}\int_{{\cal J}_\sigma} e^h \mathrm{d}x \\
& \approx \sum_\sigma n_\sigma \, e^{i S(x_\sigma)/\hbar}\,.
\end{align}
The last line expresses the fact that the integral along each thimble may easily be approximated via the saddle point approximation, the leading term being the value at the saddle point itself.  If required, one can then evaluate sub-leading terms by expanding in $\hbar,$ but in the present work this will not be necessary. This concludes our mini-review of Picard-Lefschetz theory -- for a detailed discussion see \cite{Witten:2010cx}, and for applications in a similar context than the present one see \cite{Feldbrugge:2017kzv,Feldbrugge:2017fcc,FLT4,DiTucci:2018fdg}.

The first step in evaluating the propagator \eqref{lapseintegral} then is to identify the saddle points of the integrand. Since we will be interested in the leading semi-classical approximation, we can neglect the prefactor $P(N)$ from this point onwards, as it will not affect the saddle points of the integrand at leading order in $\hbar.$ The saddle points obey the condition
\begin{align}
\frac{\partial S_0}{\partial N} = \frac{\alpha^2}{12} N^2 + \left( 3 - \frac{1}{2}\alpha (x_0 + x_1) \right) - \frac{3}{4N^2}  \left( (y_1 -y_0)^2 - (x_1 - x_0)^2 \right) = 0\,,
\end{align}
which has four solutions
\begin{align}
N_{c_1,c_2} = c_1  \sqrt{\frac{3}{\alpha^2 }} \sqrt{-6 + \alpha(x_0 + x_1) - c_2 \sqrt{I} }\,, \label{saddles}
\end{align}
where 
\begin{align}
I = \alpha^2 \left((y_1 -y_0)^2 - (x_1 - x_0)^2 \right) + \left(6 -\alpha (x_0 + x_1) \right)^2
\end{align}
and $c_1,c_2 \in \{-1,1\}$. As we will see below, for the cases of interest to us, these saddle points will either be all real, or two real and two pure imaginary. The subsequent analysis depends on the boundary conditions that are chosen. 

We will be interested in inflationary evolution, in two distinct cases: first, to set up our calculation and to check its validity, we will investigate the description of purely rolling down the potential. Afterwards, we will consider the case where the universe inflates, and then we will demand that the scalar field jump up the potential. 

\begin{figure}[h]
	\centering
	\includegraphics[width=0.4\textwidth]{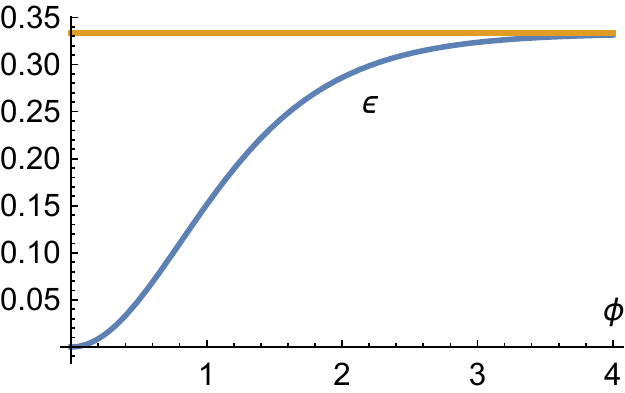} \hspace{1.5cm}
	\includegraphics[width=0.4\textwidth]{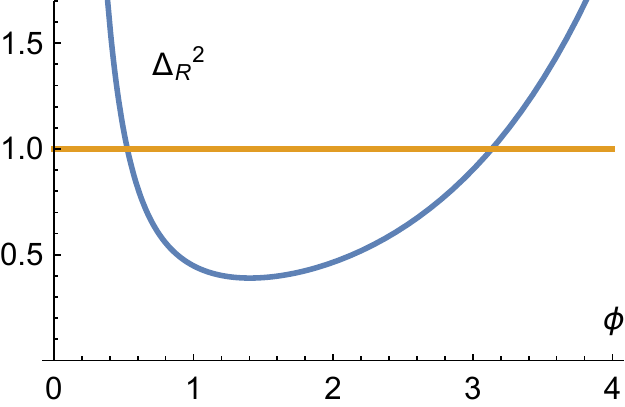}
		\caption{These plots show the flatness $V_{,\phi}^2/(2V^2)$ (left panel) and the variance of the curvature perturbation (right panel) for our potential \eqref{Pot} with $\alpha=1/10.$ Slow-roll is achieved only for small values of $\phi,$ and for small $\phi$ we are also in the conjectured regime of eternal inflation. There is a second regime of large variance at larger values of $\phi \gtrapprox 3,$ but here the potential quickly exceeds the Planck energy density, so that we will ignore this region in the present work. The yellow line on the left indicates the asymptotic value of $\epsilon$ for large $\phi$. On the right the yellow line separates the regimes where eternal inflation is expected from those where it is not.} \label{fig:eternal-infl-regime}
\end{figure}

Before continuing, we should add a note about the potential we are using, namely $V(\phi) = \alpha \cosh \left(\sqrt{\frac{2}{3}}\phi \right).$ In Fig. \ref{fig:eternal-infl-regime} we have plotted the flatness of the potential (more specifically, we have plotted $V_{,\phi}^2/(2V^2)$ which in the slow-roll limit coincides with $\epsilon$) as well as the variance of the curvature perturbation, for $\alpha=1/10$. Here we can see that inflationary solutions can be achieved throughout, but slow roll is only applicable for very small $\phi \lessapprox 0.2.$ Meanwhile the variance becomes large both for small field values $\phi \lessapprox 0.5$ and for very large values $\phi \gtrapprox 3,$ although these specific numbers will change for other choices of $\alpha.$


\section{Inflation - Rolling Down the Potential} \label{test}

\begin{figure}
\begin{center}
\includegraphics[width=0.5\textwidth]{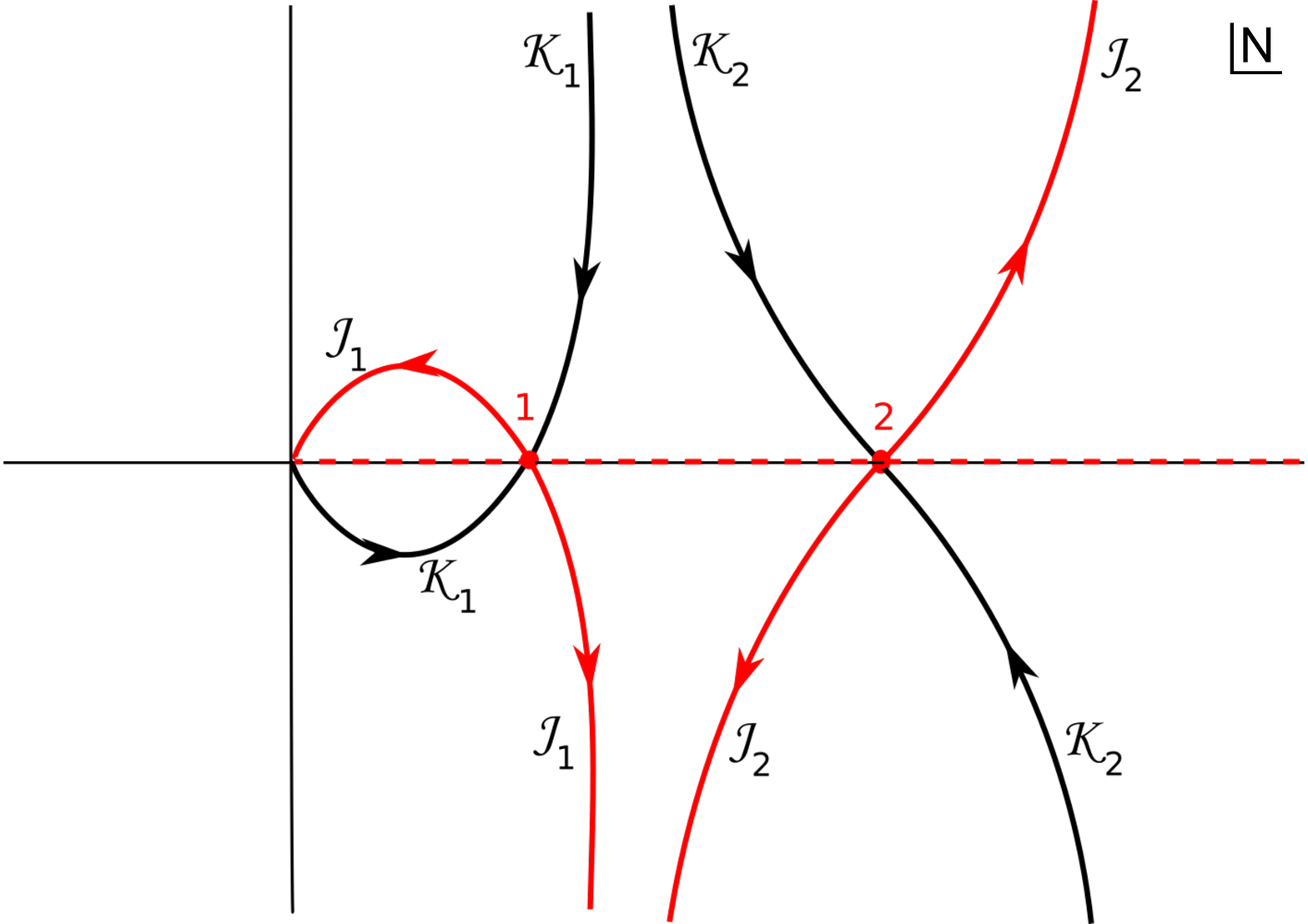}
\caption{The figure shows a typical example of the saddle points and the flow lines in the complex $N$ plane. The $\mathcal{J}_s$/$\mathcal{K}_s$ lines are the steepest descent/ascent paths associated with the saddle point $s,$ where arrows indicate downwards flow. The integral along the positive real $N$ line (dashed line) is equivalent to the integral along the path $\mathcal{J}_1 + \mathcal{J}_2$ (full red line). Both saddle points are relevant to the path integral.}
\label{fig:nfl-pl}
\end{center}
\end{figure}

Now that we have set up our model, we can evaluate transition amplitudes with various boundary conditions. In fact, in the present paper we will only look at homogeneous configurations. This is because on the one hand, this restriction brings about a significant technical simplification, and on the other hand it is suggested as a reasonable approximation (in a suitably sized patch of the universe) by the calculations of stochastic inflation, as discussed in the introduction. In order to test our formalism, we will start with a situation in which the universe is expanding while the scalar field is rolling down the potential, i.e. we start with a situation in which we expect there to exist a classical inflationary solution. Thus at first we will pick Dirichlet boundary conditions with
\begin{align}
a_1 > a_0 \,,\qquad  \qquad \phi_1 < \phi_0 \,,
\end{align}
where we will stick to the $\phi \geq 0$ side of the potential, and we will assume that the scale factors are larger than the de Sitter radius implied by the potential, $a_{0,1} > \sqrt{3/V(\phi_{0,1})}$. For boundary conditions such as these, the action \eqref{action-integr} admits four real saddle points, two at positive values of the lapse function, and two at negative values, as given by Eq. \eqref{saddles}. The two saddle points at positive $N$ are trivially relevant to our path integral, since they lie on the original integration contour -- see Fig. \ref{fig:nfl-pl} for an illustration. The figure also shows the associated paths of steepest descent, and the original integration contour along $\mathbb{R}^+$ can indeed be deformed into the sum of these two steepest descent contours.  Superficially, it may be surprising that there are {\it two} relevant saddle points because we expect only the inflationary solution, but upon analysing the saddle point geometries it becomes clear what is happening.

\begin{figure}[h]
	\centering
	\includegraphics[width=0.45\textwidth]{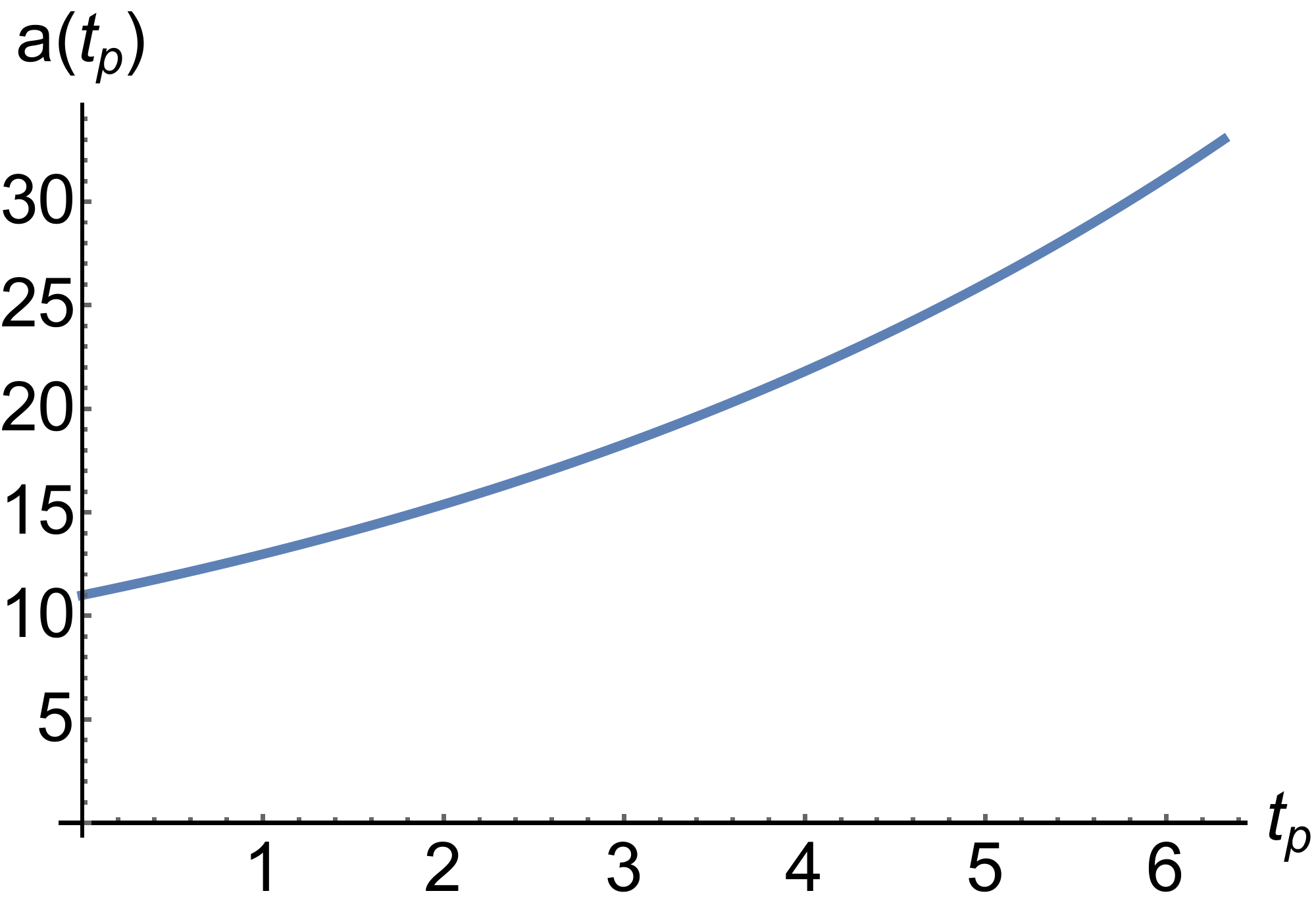}
	\includegraphics[width=0.45\textwidth]{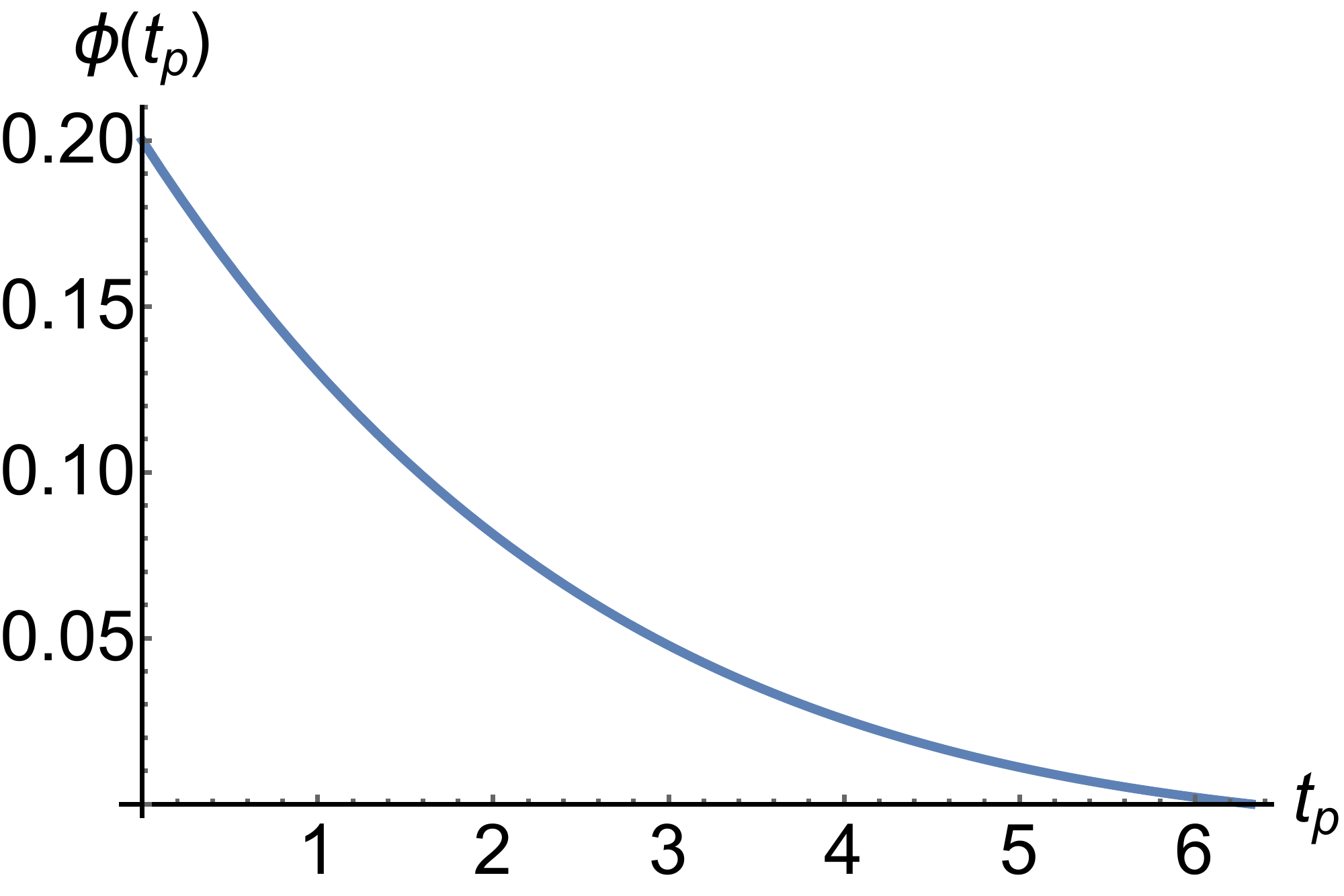}\\
	\caption{A typical example of the geometry at the saddle point $N_1$. In particular, here we have $\phi_0 = 2/10$, $\phi_1 = 0$, $a_0 = 11$, and $a_1 = 33$ corresponding to 1 e-fold of inflation and, as expected, we find inflationary behaviour of the scale factor and scalar field.} \label{fig:infl-sp1-geom}
\end{figure}

\begin{figure}[h]
	\centering
	\includegraphics[width=0.45\textwidth]{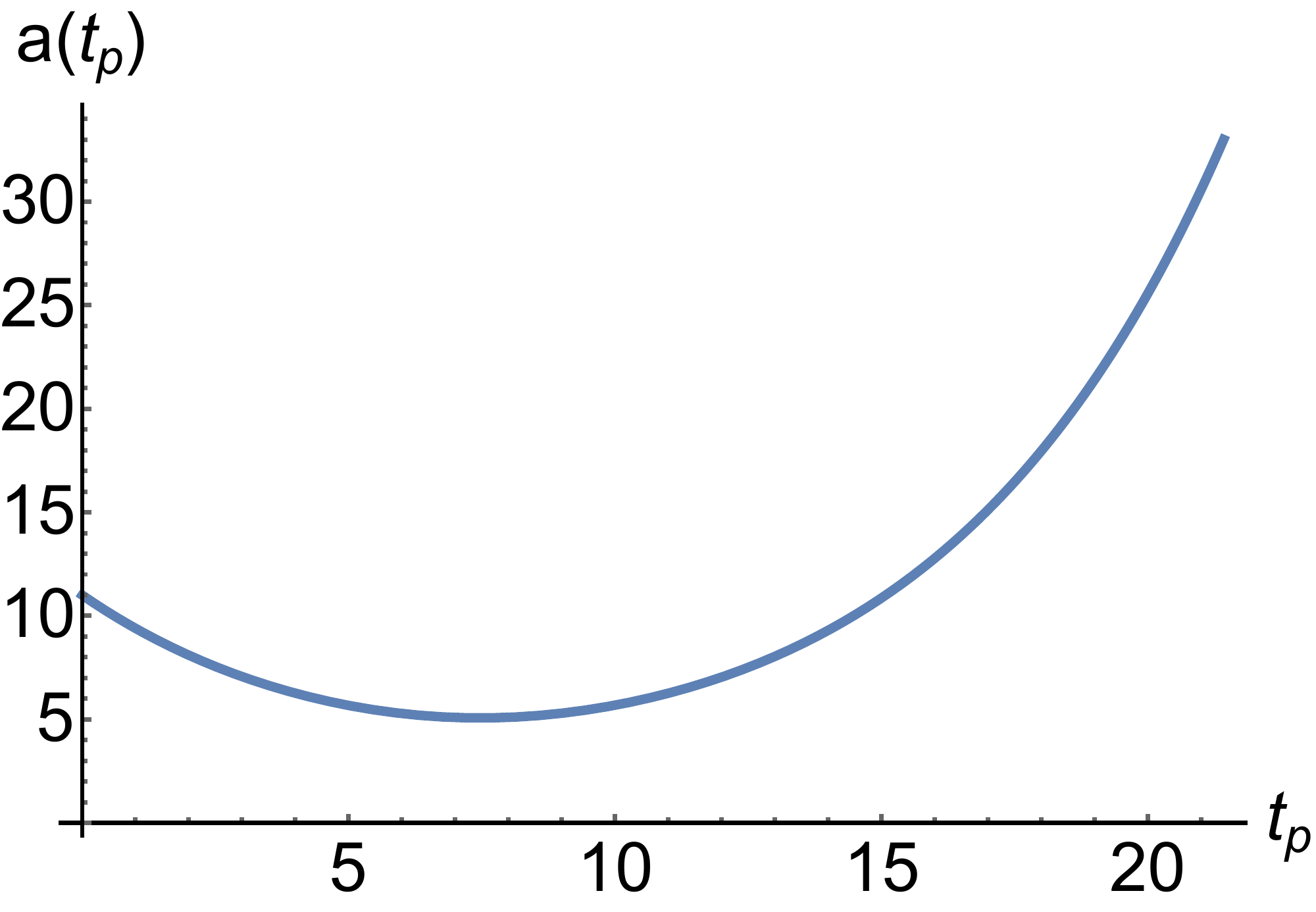}
	\includegraphics[width=0.45\textwidth]{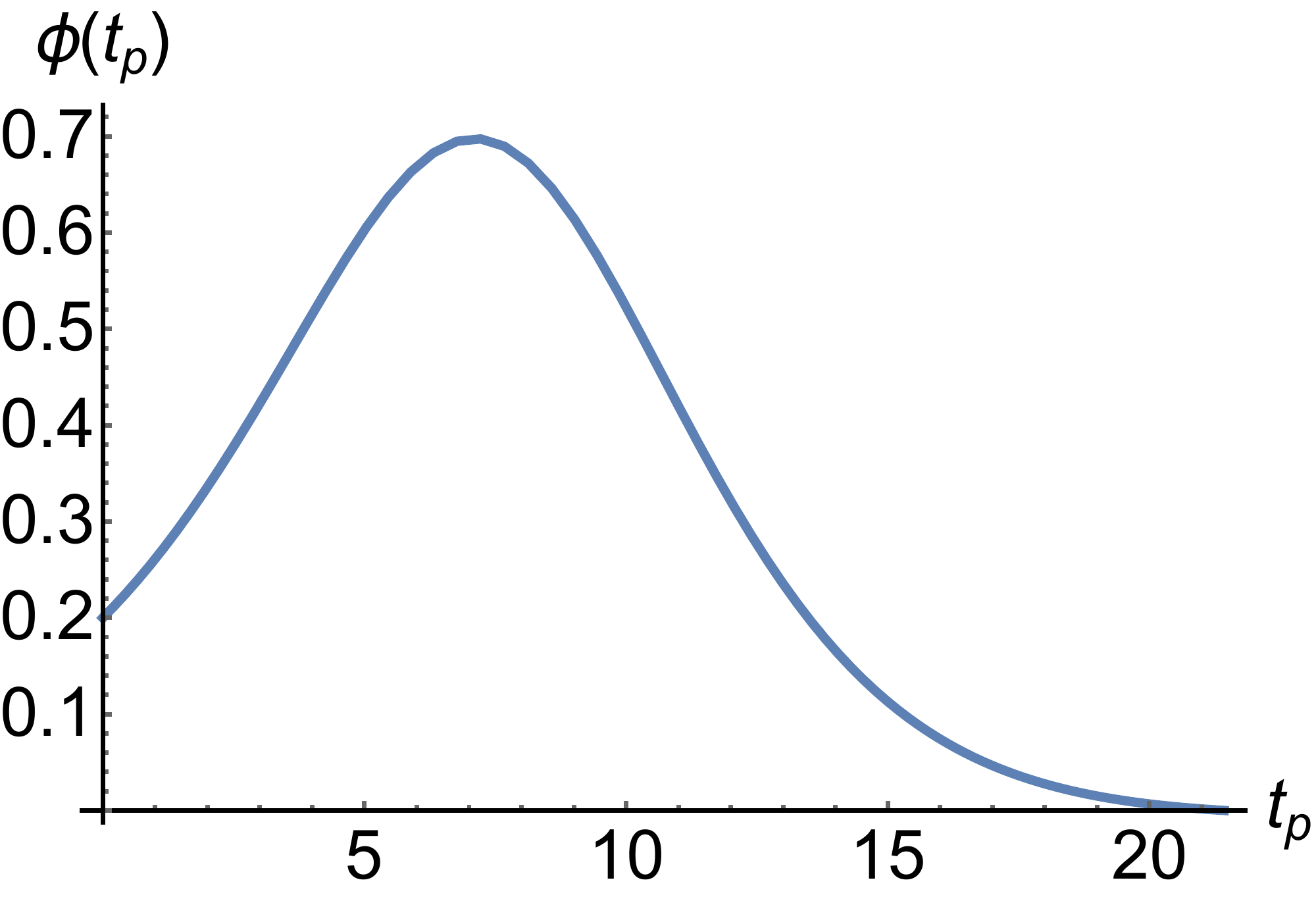}\\
	\caption{A typical example of the geometry of the saddle point $N_2$. In particular, here we have $\phi_0 = 2/10$, $\phi_1 = 1/10$, $a_0 = 11$, and $a_1 = 33$ corresponding to bouncing behaviour of the scale factor and scalar field.} \label{fig:infl-sp2-geom}
\end{figure}

The first solution, for smaller $N$, corresponds to an inflationary universe (an example of which is given in Fig. \ref{fig:infl-sp1-geom}). The second solution, the one for larger $N$, corresponds to a bouncing universe (see Fig. \ref{fig:infl-sp2-geom}). Note that due to the blue-shifting that occurs during contraction, the scalar field can initially roll up the potential, and then roll down again during the expanding phase. From these geometrical properties it also becomes clear why there are two solutions: the path integral simply finds all solutions corresponding to the given boundary conditions. It does not know about the prior evolution of the universe and hence picks out solutions consistent both with initial expansion and contraction. Note that a classical bouncing solution exists because we took the spatial sections of the metric to be closed, and hence the solution can be thought of as being a deformation of the de Sitter hyperboloid with the waist sitting in between the initial and final hypersurfaces. We should emphasise that in this situation, where two (real) saddle points contribute, an approximation in terms fo QFT in curved spacetime does \emph{not} hold, since we are in the presence of \emph{two} relevant background spacetimes (cf. the analogous discussion regarding pure de Sitter space in \cite{DiTucci:2019xcr}).

If we would like to single out the purely expanding inflationary solution we have to impose that the universe was already expanding with the scalar field rolling down the potential before we consider the transition computed through the path integral. In other words we need to include information not only about the initial values of the fields but also about their initial velocities. So far we have calculated the propagator with Dirichlet boundary conditions
\begin{align}
G[x_1,y_1;x_0,y_0] = \int_{0^+}^{\infty} dN  e^{iS(x_0,x_1,y_0,y_1,N)/\hbar}\,.
\end{align}
In this description we have complete certainty of the initial and final values of $x$ and $y$ or correspondingly of $a$ and $\phi$. On the other hand, the uncertainty principle implies that we have no knowledge of the initial and final velocities. 

We would now like to spread the uncertainty between positions and momenta imposing initial conditions where neither the value of the fields nor their conjugate momenta are specified but rather a linear combination of the two: 
 \begin{align}
c_1 x(0) + c_2 P_x(0) = c_3\,, \\
c_4 y(0) + c_5 P_y(0) = c_6\,.
\label{eq:robin-bcs-general}
 \end{align}
These are initial conditions of Robin type which require boundary terms in the action different from the Gibbons-Hawking-York one. To this effect, we will augment the action by additional boundary terms \cite{DiTucci:2019xcr,DiTucci:2019dji}, 
\begin{align}
S_{R} = S + p_x x_0 + p_y y_0 + \frac{i \hbar}{4 \sigma_x^2} (x_0 - x_i)^2 + \frac{i \hbar}{4 \sigma_y^2} (y_0 - y_i)^2\,,
\end{align}  
where $p_x ,p_y , \sigma_x$ and $\sigma_y$ are constants. The variation of the action now reads
\begin{align}
\delta S_{R} &= \int_0^1 N \left[ \frac{3}{2 N^2} \left( x''(t)\delta x - y''(t) \delta y \right) - \alpha \delta x \right] dt  \notag \\ &  - \frac{3}{2N}x'(t)  \delta x \Bigl |_0^1 +  \frac{3}{2N}y'(t)   \delta y \Bigl |_0^1+  \left(p_x + \frac{i \hbar}{2 \sigma_x^2} \left( x_0 - x_i\right) \right)  \delta x_0 +  \left(p_y + \frac{i \hbar}{2 \sigma_y^2} \left( y_0 - y_i\right) \right)  \delta y_0\,.
\end{align} 
Substituting the definitions of the momenta $P_x = - \frac{3}{2N}x'(t)$ and $P_y =  \frac{3}{2N}y'(t) $, the variational principle is satisfied if
\begin{align}
x_0 - \frac{2 \sigma_x^2}{i \hbar} P_x(0) = x_i - \frac{2 \sigma_x^2}{i \hbar} p_x \,,  \\
y_0 - \frac{2 \sigma_y^2}{i \hbar} P_y(0) = y_i - \frac{2 \sigma_y^2}{i \hbar} p_y 
 \label{eq:robin-bcs-special}
\end{align} 
at the initial boundary and if $x(1) = x_1, y(1) = y_1$ at the final boundary. Hence, comparing to the conditions (\ref{eq:robin-bcs-general}),
the action $S_R$ defines a mixed boundary value problem with a Dirichlet condition at $t=1$ and a Robin one at $t=0$. The Robin condition interpolates between Dirichlet (where the positions are known exactly) and Neumann (where the momenta are known exactly) as the parameters $\sigma_x$ and $\sigma_y$ are changed. 
For $\sigma_x, \sigma_y \to 0$ the boundary condition reduces to Dirichlet while for $\sigma_x, \sigma_y \to \infty$ it reduces to Neumann. 

In the following we will evaluate the path integral 
 \begin{equation}
 \int dN \int \,  \delta x  \, \int \, \delta y\, e^{i S_R/ \hbar} \label{robiN} 
   \end{equation}
with the mixed boundary conditions defined by $S_R$ for various values of $\sigma_x$ and $\sigma_y$ and explore the consequences in terms of the structure of the flow lines. Notice that the propagator (\ref{robiN}) can be interpreted as a convolution with an initial state 
\begin{align}
G[x_1,y_1;\psi_0] = \int \int G[x_1,y_1;x_0,y_0] \psi_0(x_0,y_0) dx_0 dy_0
\label{eq:convolved-propagator}
\end{align}
where $G[x_1,y_1;x_0,y_0]$ is the propagator evaluated with Dirichlet boundary conditions and the initial wave function reads
\begin{align}
\psi_0(x_0,y_0) = e^{\frac{i}{\hbar} \left(p_x x_0 + p_y y_0 \right) - \frac{(x_0 - x_i)^2}{4\sigma_x^2}- \frac{(y_0 - y_i)^2}{4\sigma_y^2}}\,.
\label{eq:wavefunction}
\end{align}
The functional form of this initial state is that of a coherent, Gaussian state, which allows us to express our knowledge of the initial uncertainty in the field values and their momenta\footnote{A detailed discussion of the use of initial and final (off-shell) states will be published in upcoming work by Angelika Fertig, Job Feldbrugge, Laura Sberna and Neil Turok \cite{FFST}.}. By construction the initial positions are peaked around the values $x_i$,$y_i$, with a Gaussian spread around them. In the limit where $\sigma_x = \sigma_y = 0$ the initial positions simply become $x_i$ and $y_i$ by construction. We are then back to the position representation which we were (implicitly) using up to now. Performing the Gaussian integrals over $x_0$ and $y_0$ gives us the saddle point solutions
\begin{align}
\bar{x}_0 &= \frac{\hbar N x_i - \alpha i N^2 \sigma_x^2 +2 i N p_x \sigma_x^2 + 3 x_1 i \sigma_x^2}{\hbar N+3 i\sigma_x^2 } \label{eq:x0-sp-approx} \,,\\
\bar{y}_0 &= \frac{\hbar N y_i +2 i N p_y \sigma_y^2 - 3 y_1 i \sigma_y^2}{\hbar N-3 i\sigma_y^2 }\,.
\end{align}
For small spreads $\sigma,$ we have $\bar{x}_0 \approx x_i, \bar{y}_0 \approx y_i,$ while for very large $\sigma$ we obtain 
\begin{align}
\bar{x}_0 \approx x_1 - \frac{\alpha}{3}N^2+ \frac{2N}{3} p_x\,, \qquad \bar{y}_0 \approx y_1 - \frac{2N}{3}p_y \qquad (\sigma_{x,y} \gg 1)\,. \label{largesigma}
\end{align}
Thus at large spreads $x_i, y_i$ disappear from the formula, which is an indication that the position is less well known. In fact at large $\sigma$ the momentum is determined with increasing precision.  To show this in more detail we focus on one of the momenta and variables ($p_x$ and $x$ respectively) but the result holds for both. Hamilton's equations give 
\begin{align}
P_x(t) = - \frac{3}{2 N}x'(t)= - N \alpha t - \frac{3}{2N} \left(x_1 - x_0 - \frac{1}{3} N^2 \alpha \right)
\end{align}
where the last line was obtained by plugging in the solution of the equations of motion for $x$. Thus, at the saddle points the initial momentum simply reduces to
\begin{align}
P_x(0) = - \frac{3}{2N} \left(x_1 - x_0 - \frac{1}{3} N^2 \alpha \right)\,,
\end{align}
which agrees with Eq. \eqref{largesigma}. We may also find the sub-leading terms by making use of Eq. \eqref{eq:x0-sp-approx}, plugging it into the general expression for the momentum and expanding for large $\sigma_x,$ to obtain
\begin{align}
P_x(0) = p_x + \frac{i \hbar}{\sigma_x^2}\frac{1}{6} \left( 3x_1 - 3x_i + 2Np_x - \alpha N^2 \right) + O\left(\frac{1}{\sigma_x^4} \right)\,,
\end{align}
which confirms that in the large $\sigma_x$ limit we reach the pure momentum representation.

\begin{figure}
\includegraphics[width=1\textwidth]{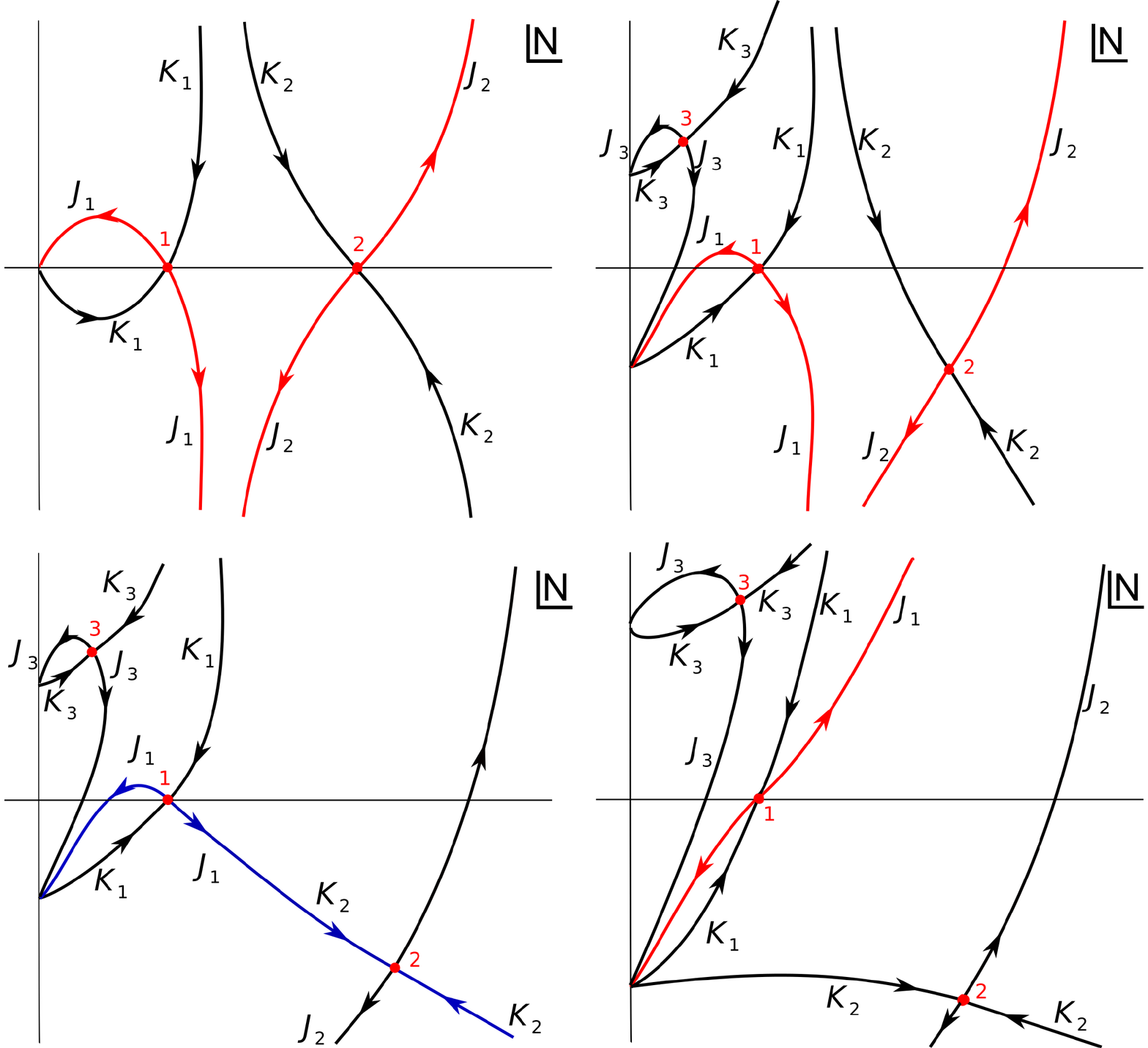}
\caption{The structure of the flow lines is shown as a function of the uncertainty $\sigma_\phi$ for inflationary boundary conditions, with $\sigma_{x,y}$ determined via Eqs. \eqref{sigxphi}, \eqref{sigyphi}. In order to draw these graphs we have used the boundary conditions $a_0=100, \phi_0=1/10, a_1 = 200, \phi_1=1/100$ and $\alpha = 1/10$, with the corresponding momenta being given by $(\dot{a}(t_p),\dot{\phi}(t_p)) = (1.7953, -0.0820864)$. The numerically determined flow lines would have been difficult to put on a legible graph due to the large distances between saddle points, hence we have re-drawn these graphs to show the qualitative behaviour of the flow lines. Only the saddle points with $Re(N_i) >0$ are considered, being the only ones relevant for the flow analysis. Top left panel: For $\sigma_\phi=0$  both the expanding and the bouncing solutions are relevant to the path integral (corresponding to the saddle points $N_1$ and $N_2$). Top right panel: For non-zero $\sigma$ a new saddle point appears, the saddle point $N_2$ moves off the real line while $N_1$ maintains its original position (here $\sigma_\phi=0.0100$). For small enough $\sigma$ the original integration contour is deformed to the Lefschetz thimble $\mathcal{J}_1 + \mathcal{J}_2$. Bottom left panel: For a critical value of $\sigma = \sigma_c$ a Stokes phenomenon happens (here $\sigma_c \approx 0.0154$). The steepest descent path associated to $N_1$ ($\mathcal{J}_1$) coincides now with the steepest ascent through $N_2$ ($\mathcal{J}_2$). This is the Stokes line, the blue line in the figure. Bottom right panel: For $\sigma> \sigma_c$ the bouncing solution ($N_2$) no longer contributes to the path integral and only the inflating one ($N_1$) survives (here $\sigma_\phi=0.0200$).   }
 \label{fig:infl-pl-conv}
\end{figure}

Let us now return to our inflationary example. We choose initial momenta $p_x$ and $p_y$ such that $a$ is expanding and $\phi$ is rolling down the potential. The values of $p_x$ and $p_y$ are fixed such that they correspond to the classical inflationary solution that links our initial and final boundary conditions.   After performing the integrals over $x_0$ and $y_0$, we are again left with an integral over the lapse function $N,$
\begin{align}
G[x_1,y_1;\psi_0] \approx \int_{0^+}^{\infty} dN  e^{i\tilde{S}(x_i,x_1,y_i,y_1,p_x,p_y,\sigma_x,\sigma_y,N)/\hbar}
\end{align}
This new action results from having replaced $x_0$ and $y_0$ with their saddle point values $\bar{x}_0, \bar{y}_0$ in Eq. \eqref{action-integr} and including the contributions from the initial state. More explicitly, we have 
\begin{align}
\frac{i}{\hbar}\tilde{S}& =\frac{i}{\hbar} S_0 + \frac{i}{\hbar}  (p_x x_0 + p_y y_0) - \frac{1}{4 \sigma_x^2} (x_0 - x_i)^2 - \frac{1}{4 \sigma_y^2} (y_0 - y_i)^2 \nonumber \\ &= 
\frac{i}{\hbar}\left[\frac{\alpha^2}{36} N^3  + N \left( 3 -  \frac{\hbar \alpha N (x_i+x_1) - \alpha^2 i N^2 \sigma_x^2 +2 i \alpha N p_x \sigma_x^2 + 6 \alpha x_1 i \sigma_x^2}{2(\hbar N+3 i\sigma_x^2) } \right) \right. \nonumber \\ & \left. \qquad+ \frac{3N}{4}  \left(\frac{\hbar  (y_i-y_1) +2 i  p_y \sigma_y^2}{\hbar N-3 i\sigma_y^2 }\right)^2 - \frac{3N}{4} \left(\frac{\hbar  (x_i-x_0) - \alpha i N \sigma_x^2 +2 i  p_x \sigma_x^2}{\hbar N+3 i\sigma_x^2 }\right)^2 \right] \nonumber \\
& + \frac{i}{\hbar}(p_x x_i + p_y y_i)  + \frac{ \alpha  N^2 \sigma_x^2 p_x -2  N p_x^2 \sigma_x^2 - 3 x_1 p_x \sigma_x^2}{\hbar(\hbar N+3 i\sigma_x^2) } +  \frac{-2  N p_y^2 \sigma_y^2 + 3 y_1  \sigma_y^2 p_y}{\hbar(\hbar N-3 i\sigma_y^2) } \nonumber \\
& + \left(\frac{- \alpha  N^2 +2  N p_x  + 3 (x_1-x_i)  }{4(\hbar N+3 i\sigma_x^2) }\right)^2 + \left( \frac{2  N p_y - 3 (y_1-y_i) }{4(\hbar N-3 i\sigma_y^2) }\right)^2 \label{actionfull}
\end{align}
The replacements of $x_0$ and $y_0$ have as a consequence that the dependence of the action on the lapse function $N$ has become more complicated. But once again we can solve this integral using Picard-Lefschetz theory. Let us start from small values of $\sigma_x$ and $ \sigma_y$ and investigate what happens as the spreads $\sigma_{x,y}$ are increased, see Fig. \ref{fig:infl-pl-conv}. At zero spread, we are in the pure position representation, with two relevant saddle points (upper left panel in the figure). But as soon as the spreads are turned on, the situation changes:  we now have six complex saddle points (three with positive real part, and three with negative real part) out of which two are relevant to the Lorentzian path integral, see the upper right panel in Fig. \ref{fig:infl-pl-conv}.  As we increase our certainty about the values of the initial momenta, the saddle points and flow lines change their location in the complex plane. Eventually a drastic transition occurs where the topology of the flow lines changes. This, so-called Stokes phenomenon, happens when a flow line connects two saddle points, for example in this case when
\begin{align}
Im(\tilde{S}(N_1)) = Im(\tilde{S}(N_2))
\end{align}
for two distinct saddle points $N_1$ and $N_2$. After this transition only one saddle point ($N_1$) remains relevant to the path integral, while the second one ($N_2$) has become irrelevant. The saddle point $N_1$, the only relevant critical point after the Stokes phenomenon, does not move at all as a function of $\sigma_{x}$ and $\sigma_{y}$. Furthermore the behaviour of the scale factor and scalar field at this location is inflationary as desired (see Fig. \ref{fig:infl-sp1-geom}), while the bouncing solution (Fig. \ref{fig:infl-sp2-geom}) has become irrelevant. This is entirely consistent with our interpretation of the initial state: as we increase our knowledge of the initial momentum (chosen to represent an expanding universe), only the expanding solution survives. Thus we see that the path integral gives sensible results for transitions in which the scale factor expands and the scalar field rolls down the potential. At the same time, we can appreciate the importance of the Robin initial condition in determining the outcome of future evolution.

\section{Jumping Up the Potential} \label{jump}

Inflation may be able to sustain itself indefinitely if the scalar field can jump up the potential, thus inducing a phase of enhanced accelerated expansion. In order to understand the true consequences of eternal inflation, it seems likely that a more fully quantum understanding of such transitions, and the associated issues of measures, must be developed. Here we take a step in that direction, by investigating the semi-classical geometries of such up-jumps. Thus we will now consider boundary conditions of the form
\begin{align}
a_1 > a_0 > \sqrt{\frac{3}{V(\phi_0)}}\,, \qquad  \qquad \phi_1 > \phi_0\,.
\end{align}
Again we must find the relevant saddle points, so that we can look at their geometries. Just as in the previous case in the Dirichlet limit $\sigma_{x,y} = 0$ we have four saddle points out of which two will be relevant for the path integral (the other two being the time reverses of the relevant two). 

\begin{figure}[ht!]
	\centering
	\includegraphics[width=0.9\textwidth]{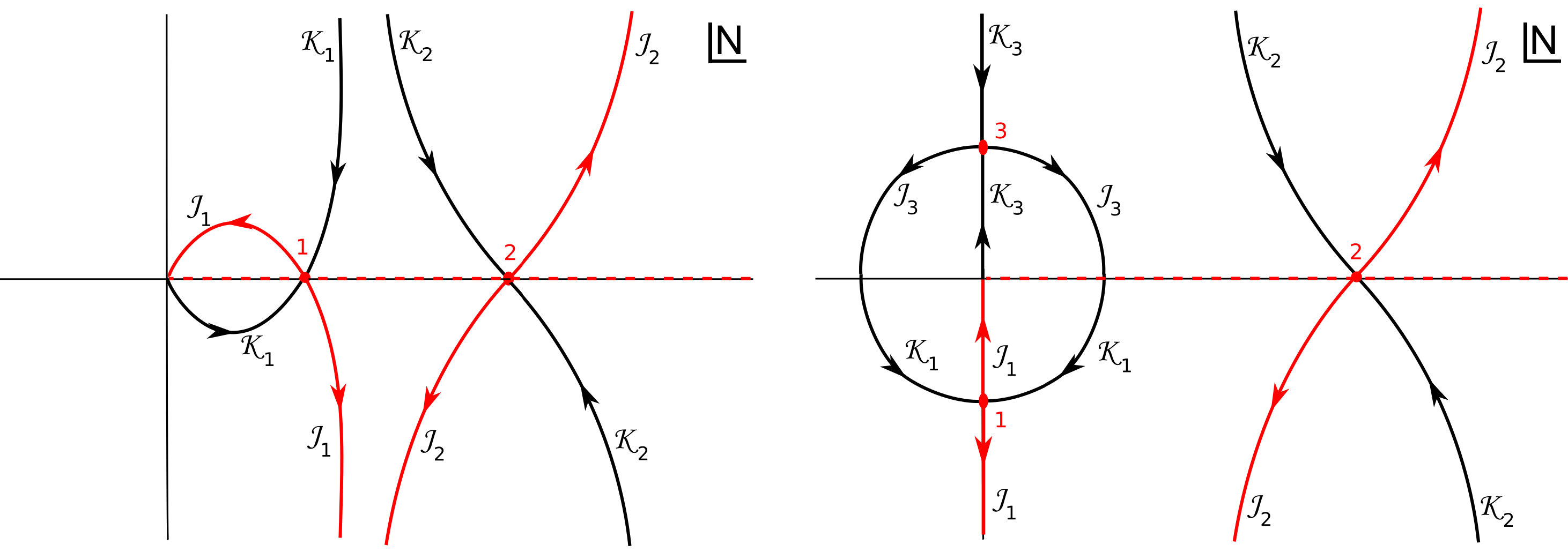}
	\caption{The figures show the structure of the steepest ascent and descent path ($\mathcal{J}_s$ and $\mathcal{K}_s$) for $\sigma_{x,y} = 0 $ and $ \phi_1 > \phi_0$. The left (case A) and the right (case B) panels show the only two inequivalent qualitative structures allowed for by up-jumping boundary conditions. In both cases the action has 4 critical points but only those with $Re(N_s) \geq 0$ are plotted. The original integration contour (dashed red line) can be deformed smoothly to the Lefschetz thimble $\mathcal{J}_1 + \mathcal{J}_2$ so that two saddle points contribute to the path integral. The geometries of these saddle points are plotted in Figs. \ref{fig:conv-sp-no-sigma} and \ref{fig:conv-sp2-no-sigma} for case B.}  
\label{fig:rollingup}
\end{figure} 

\begin{figure}[ht!]
	\centering
	\includegraphics[width=0.4\textwidth]{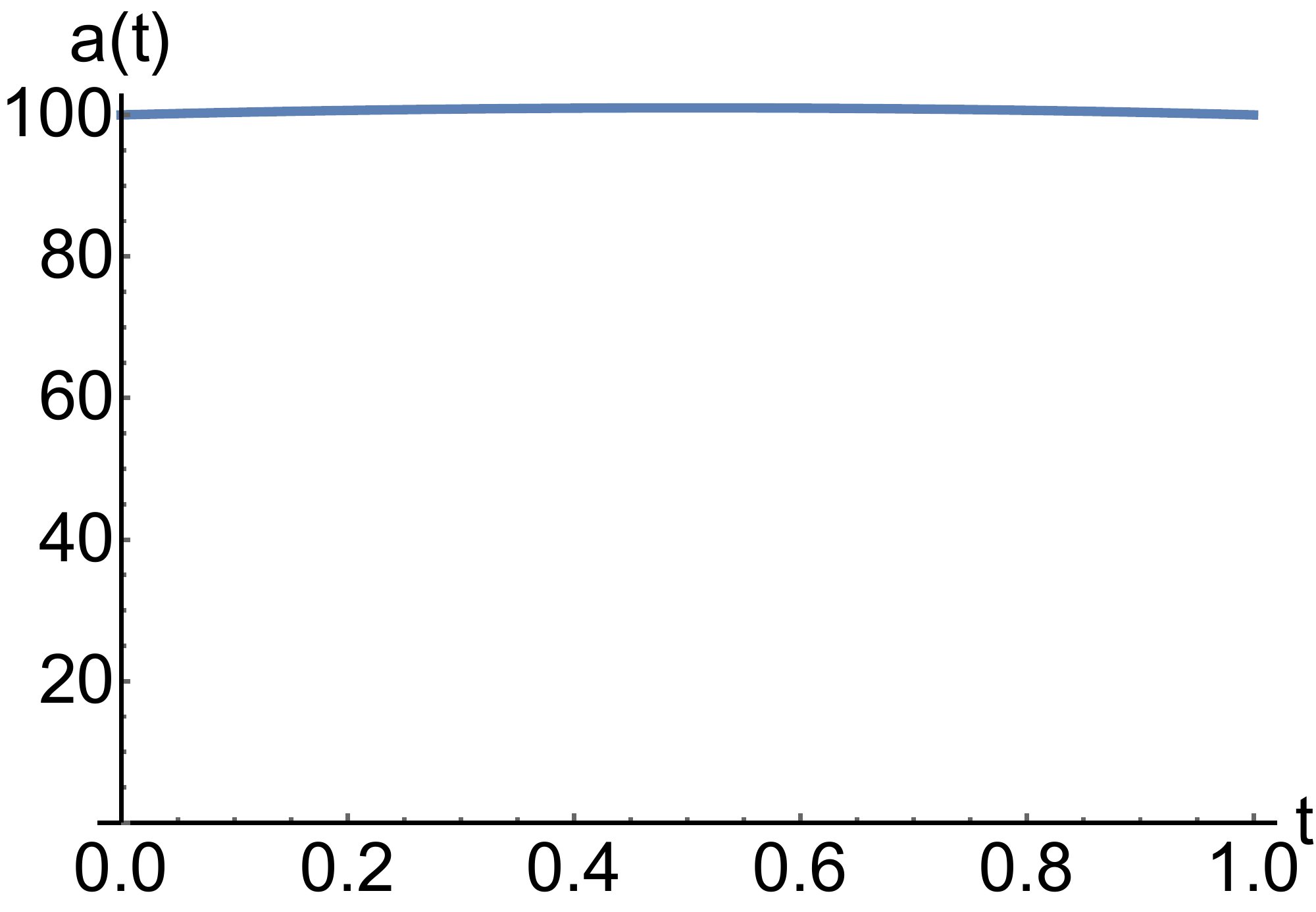}
	\includegraphics[width=0.4\textwidth]{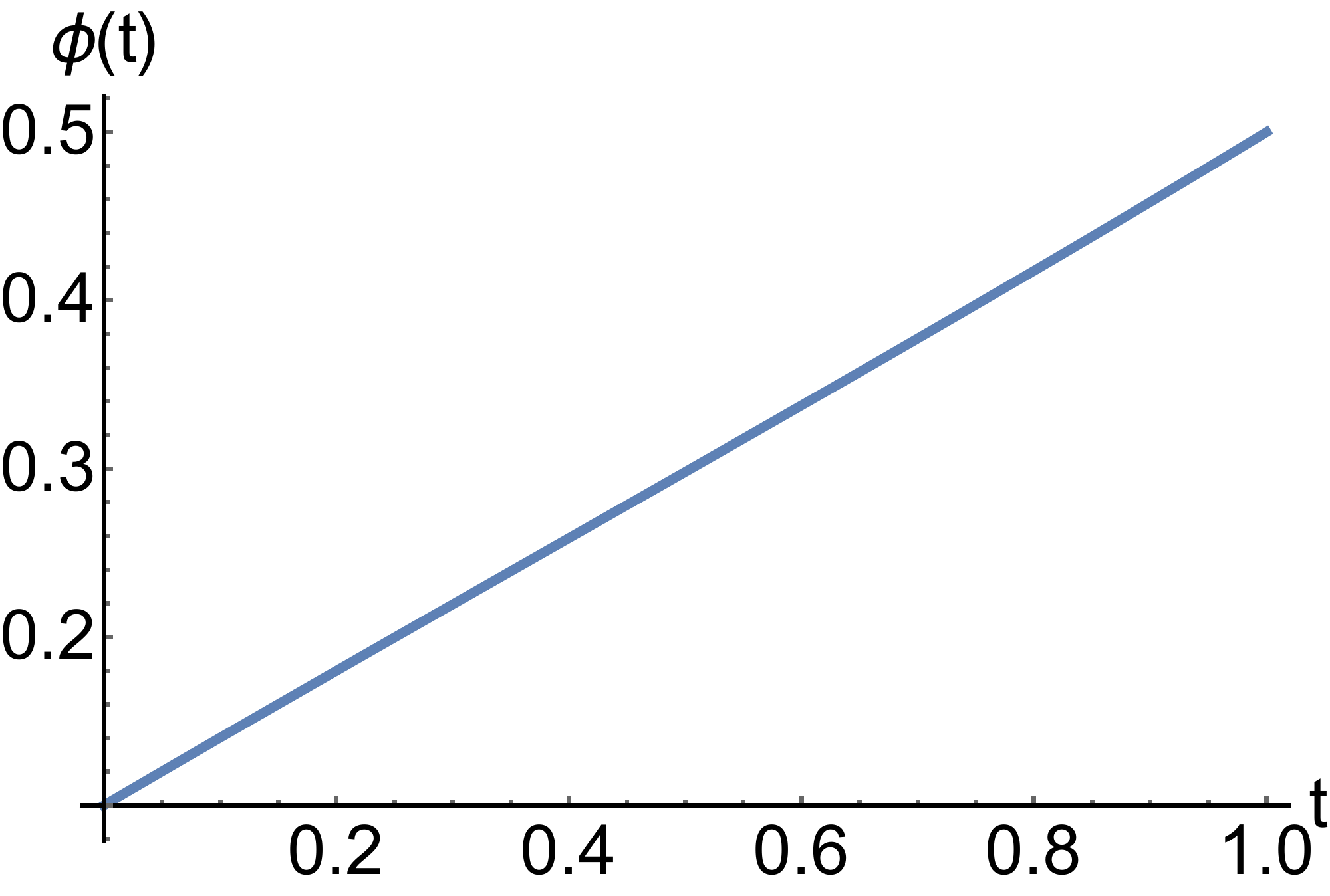}
	\caption{Geometry of saddle point number $1$ in the right panel of Fig. \ref{fig:rollingup}. Plotted here are the scale factor and scalar field with respect to coordinate time where we have chosen $\alpha = 1/10$, $\phi_i = 1/10$, $\phi_1 = 1/2$, $a_0 = 100$, $a_1 = 100$ and $\sigma_{\phi} = 0$. The saddle point is purely imaginary, and consequently the scale factor is Euclidean here.} \label{fig:conv-sp-no-sigma}
\end{figure}

\begin{figure}[ht!]
	\centering
	\includegraphics[width=0.4\textwidth]{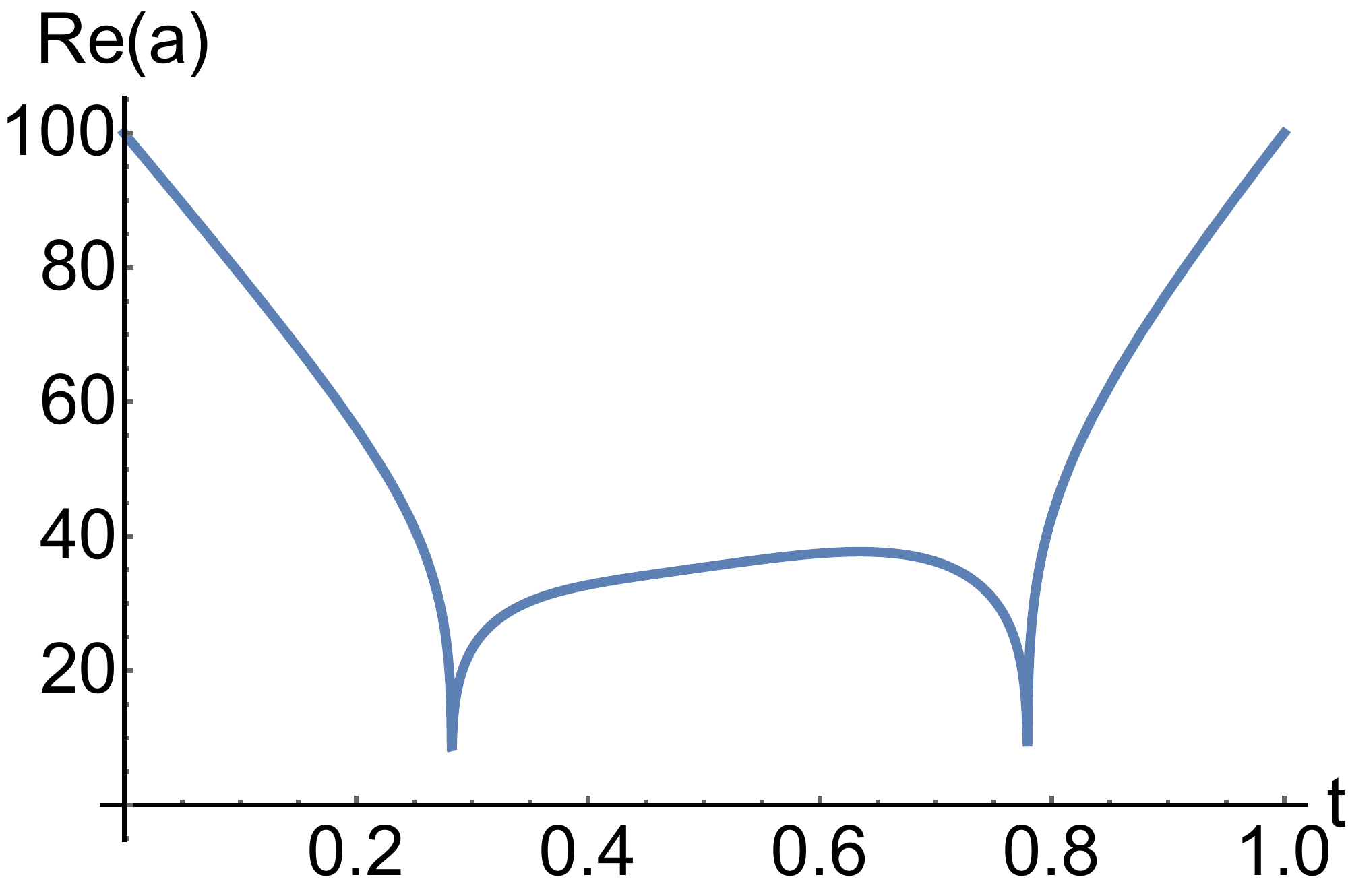}
	\includegraphics[width=0.4\textwidth]{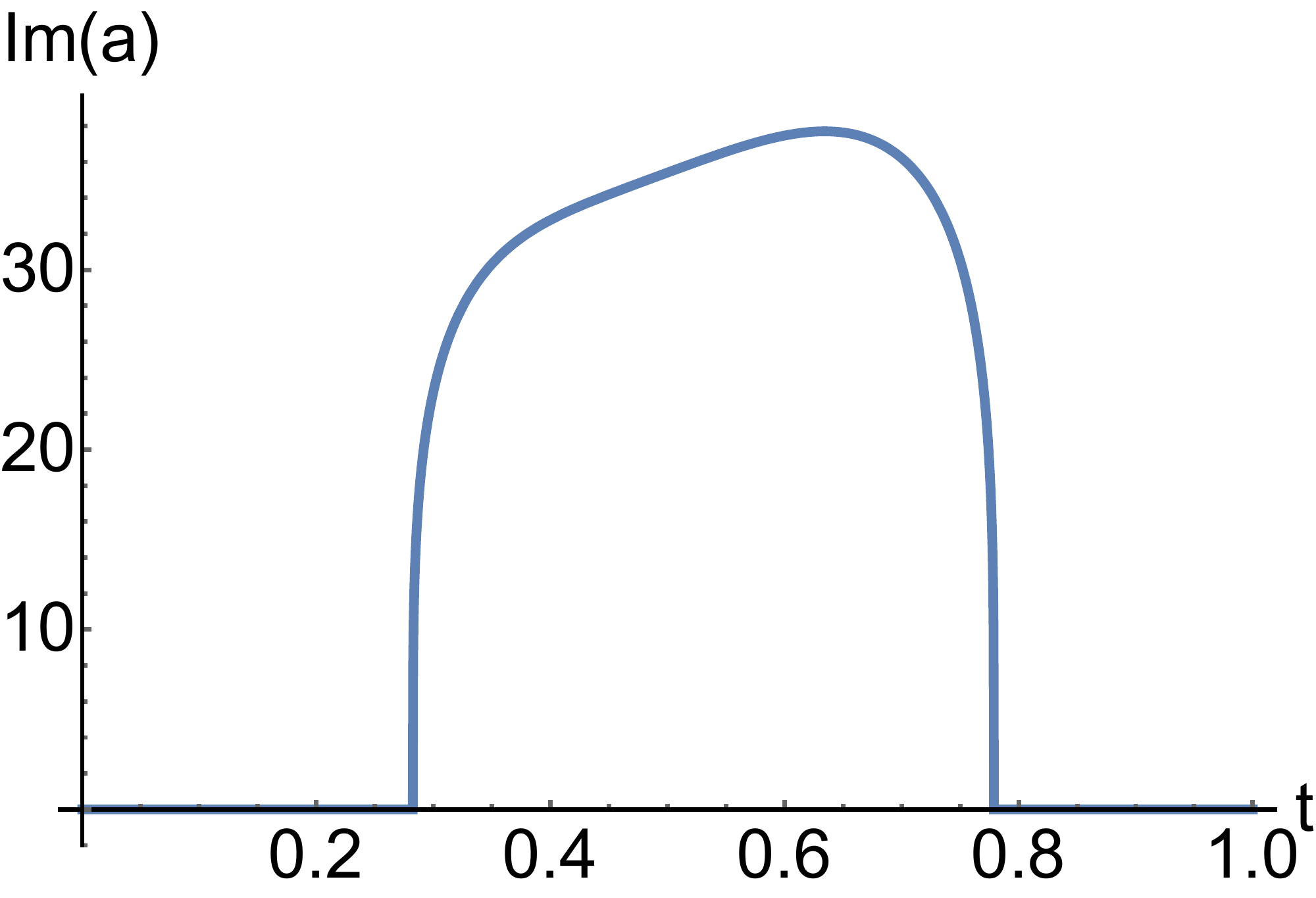}\\
	\includegraphics[width=0.4\textwidth]{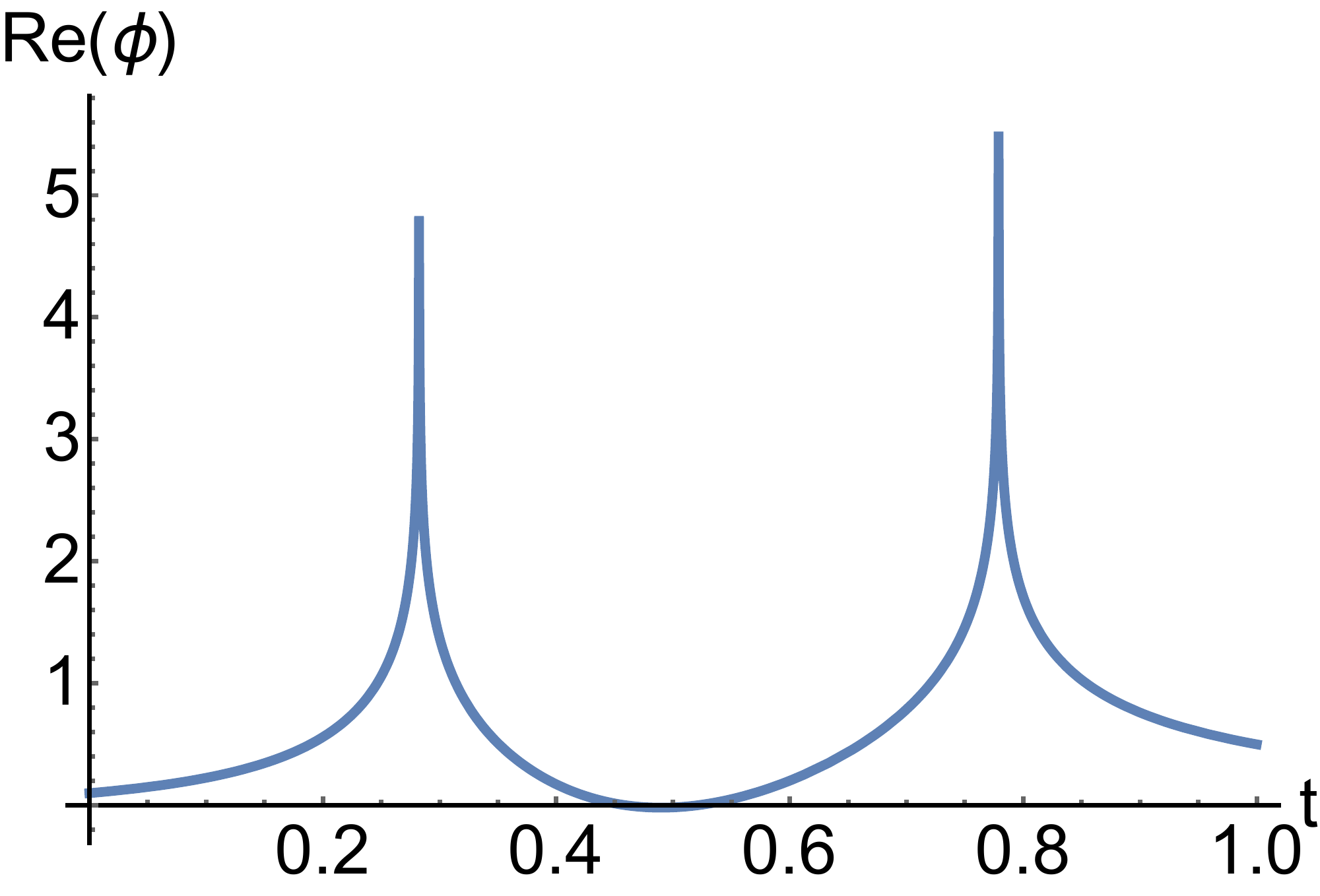}
	\includegraphics[width=0.4\textwidth]{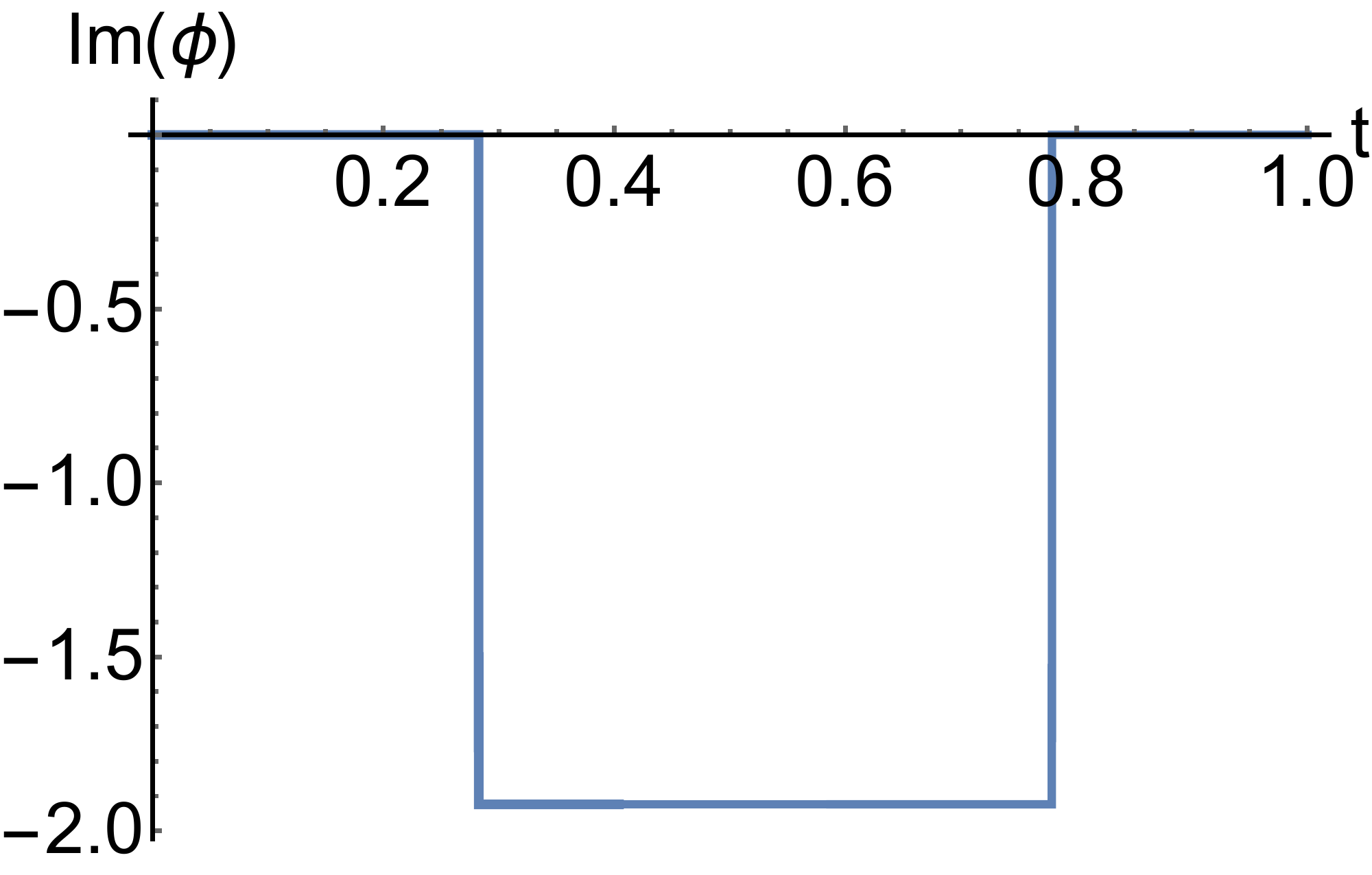}
	\caption{Geometry of saddle point number $2$ in the right panel of Fig. \ref{fig:rollingup}. The scalar field passes through zero twice, and at these singularities the scalar field blows up. Moreover, the scalar field starts out rolling up the hill, so that this geometry could only be relevant for a physical situation in which the scalar field would already have a large initial velocity up the hill, while we are interested in a prior state with the scalar slowly rolling down the potential.} \label{fig:conv-sp2-no-sigma}
\end{figure}

In fact, for different values of the initial conditions the saddle point $N_1$ can be either purely real or purely imaginary: we call these two possibilities case A and case B respectively. Case A is obtained for $a_1 \geq a_0 e^{\sqrt{6}(\phi_1-\phi_0)},$ otherwise we have case B. The second relevant saddle point $(N_2)$ always turns out to be real. Fig. \ref{fig:rollingup} shows the flow lines for the two possible inequivalent cases, while Figs. \ref{fig:conv-sp-no-sigma} and \ref{fig:conv-sp2-no-sigma} show the associated geometries for case B.

Note that the field values for all saddle points are strictly real, although for case B the saddle point $N_1$ is at a purely imaginary value, implying that the geometry is in fact Euclidean. This also means that the action at this saddle point is imaginary, and consequently the contribution to the path integral will be significantly suppressed compared to the saddle point $N_2.$ This second saddle point also has some peculiarities: it contains two singularities where the scale factor $a(t)$ passes through zero and where the scalar field blows up. On top of this difficulty, the scalar field starts out by rolling up the potential. Thus such a geometry may not be smoothly linked to a prior phase of inflation where the field is rolling down the potential. We can in fact show that no saddle point exists for which the scalar field is initially rolling down, but where it ends up higher in the potential. To see this consider the physical time derivatives of the scalar field and scale factor,
\begin{align}
\dot{\phi}(t_p) &= \frac{\phi'(t)a(t)}{N} = \sqrt{\frac{3}{2}}\frac{x(t)y'(t)-y(t)x'(t) }{N \left(x(t)^2 - y(t)^2 \right)^{3/4}} \,,\\
 \dot{a}(t_p) &= \frac{a'(t)a(t)}{N} = \frac{x(t)x'(t)-y(t)y'(t) }{2 N \left(x(t)^2 - y(t)^2 \right)^{1/2}}\,,
\end{align}
which at $t = 0$ reduce to
\begin{equation}
\begin{split}
\dot{\phi}_0 &= \frac{\alpha N^2 y_0 + 3x_0y_1 - 3x_1y_0}{\sqrt{6} N \left(x_0^2 - y_0^2 \right)^{3/4}}  \\
&= \frac{1}{\sqrt{6} N a_0}\left( N^2 \alpha \sinh\left( \sqrt{\frac{2}{3}} \phi_0\right) + 3 a_1^2\sinh\left( \sqrt{\frac{2}{3}} (\phi_1 - \phi_0) \right)  \right) \,,\label{phidot}
\end{split}
\end{equation}
\begin{equation}
\begin{split}
\dot{a}_0 &= \frac{x_0 (3 (x_1 - x_0) - \alpha N^2)- 3 y_0 (y_1 -y_0)}{6 N((x_0 - y_0)(x_0 + y_0))^{1/2}}  \\
&= \frac{1}{6 N } \left( 3 a_1^2 \cosh \left( \sqrt{\frac{2}{3} }(\phi_1 - \phi_0)  \right) - 3 a_0^2 -  \alpha N^2 \cosh \left( \sqrt{\frac{2}{3}} \phi_0   \right)  \right)\,.
\end{split}
\end{equation}
Since we assume a transition up the potential, $\phi_1 - \phi_0 > 0$ and this makes the second term in \eqref{phidot} positive. Thus $\dot{\phi}_0$ can never be real and positive for the considered boundary conditions. The reason for this stumbling block is simply that we are working in the pure position representation here, where we have not included any information about the momenta of the fields. But we are actually interested in the situation in which we have a prior inflationary state, with the scale factor growing and the inflaton rolling down the potential. Once we include this information, we will see that much more sensible results are obtained.

Thus we must repeat the same procedure as in the last section, i.e. we introduce Robin boundary conditions or, equivalently, convolve the propagator with an initial wavefunction as in Eq. (\ref{eq:wavefunction}), yielding the effective action \eqref{actionfull}, where the momenta are chosen to correspond to an inflating universe. Let us be more specific about which form of the spreads $\sigma_{x,y}$ we will consider. From the definitions of the variables $x,y$ we have to leading order
\begin{align}
\sigma_x = 2 a_0 \cosh \left( \sqrt{\frac{2}{3}} \phi_0 \right)  \sigma_a + \sqrt{\frac{2}{3}} a_0^2 \sinh \left( \sqrt{\frac{2}{3}} \phi_0 \right) \sigma_\phi\,, \label{sigtrf1}\\
\sigma_y = 2 a_0 \sinh \left( \sqrt{\frac{2}{3}} \phi_0 \right)  \sigma_a + \sqrt{\frac{2}{3}} a_0^2 \cosh \left( \sqrt{\frac{2}{3}} \phi_0 \right) \sigma_\phi\,. \label{sigtrf2}
\end{align}
While these relations are only accurate for small spreads, we will simply use them as definitions, even when the spread is large. Our discussion in section \ref{review} indicated that we can expect that for flat potentials the metric changes little, and most of the perturbation is expressed as a change in the scalar field value. This would suggest the choice $\sigma_a=0$ with the entire spread relegated to $\phi.$ In this case
\begin{align}
\sigma_x = \sqrt{\frac{2}{3}} a_0^2 \sinh \left( \sqrt{\frac{2}{3}} \phi_0 \right) \sigma_\phi\,, \label{sigxphi}\\
\sigma_y = \sqrt{\frac{2}{3}} a_0^2 \cosh \left( \sqrt{\frac{2}{3}} \phi_0 \right) \sigma_\phi\,.\label{sigyphi}
\end{align}
Note that this case corresponds to a specific choice of initial state, a choice that is motivated by the calculations of section \ref{review}. If not specified otherwise, this will be our default choice of initial state. Thus when we quote results in terms of $\sigma_\phi$ alone, this should be understood as shorthand for $\sigma_{x,y}$ given by Eqs. \eqref{sigxphi} and \eqref{sigyphi}. However, we will also be led to consider other choices, with both $\sigma_a$ and $\sigma_\phi$ turned on.

\begin{figure}
\includegraphics[width=1\textwidth]{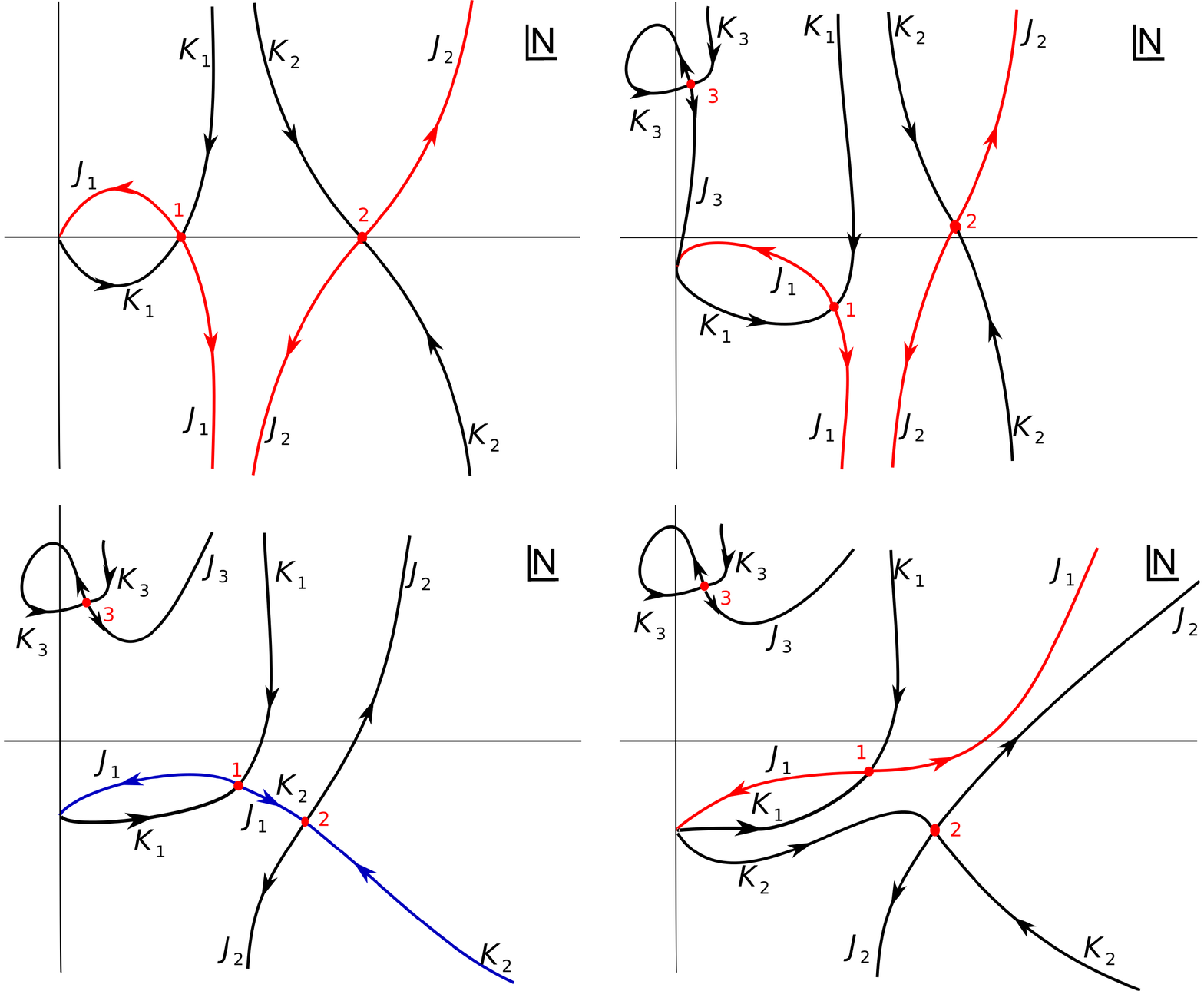}
\caption{Evolution of the saddle points and their associated flow lines in the complex $N$ plane as $\sigma_\phi$ is increased. The original integration contour (positive real line) can be smoothly deformed to the complex Lefschetz thimble (the red line) leaving the value of the path integral unchanged. The Lefschetz thimble runs through one or more saddle points of the action. One of the initially relevant saddle points ($N_1$) is relevant for all values of $\sigma_\phi$ but its position and therefore the geometry associated to it changes. The other saddle point becomes irrelevant after the Stokes phenomenon (the Stokes line is the blue line in the bottom left panel). The third saddle point never contributes to the path integral. In order to draw these graphs, we have used the boundary conditions $a_0=100, \phi_0=1/10, a_1=200, \phi_1=1/2,$ while the values of the spread for these four plots are respectively $\sigma_\phi=0, 0.0100, \sigma_c \approx 0.0154, 0.0700.$}
 \label{fig:conv-sp-evolution-srei}
\end{figure} 

\begin{figure}[ht!]
	\centering
	\includegraphics[width=0.4\textwidth]{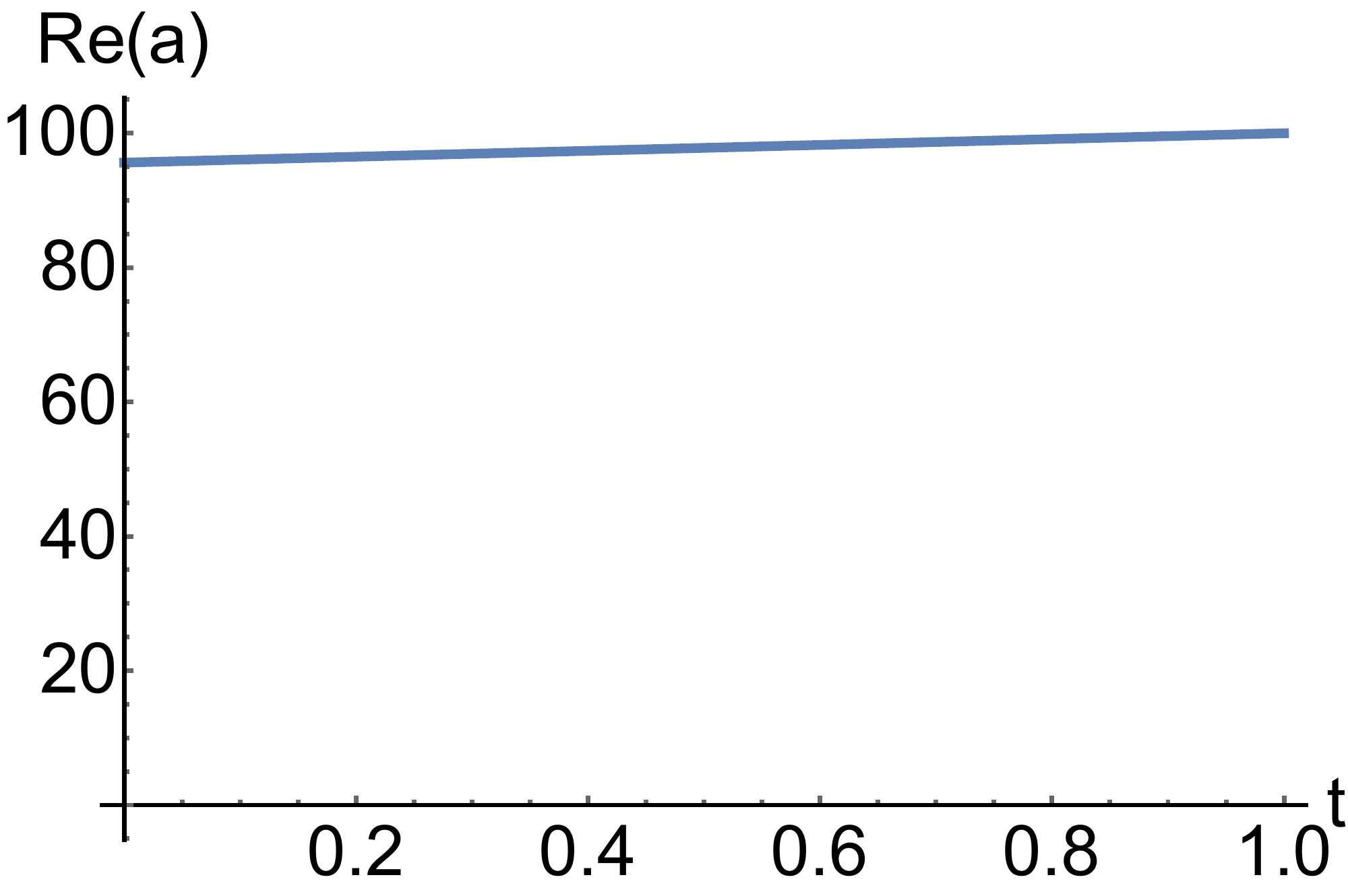}
	\includegraphics[width=0.4\textwidth]{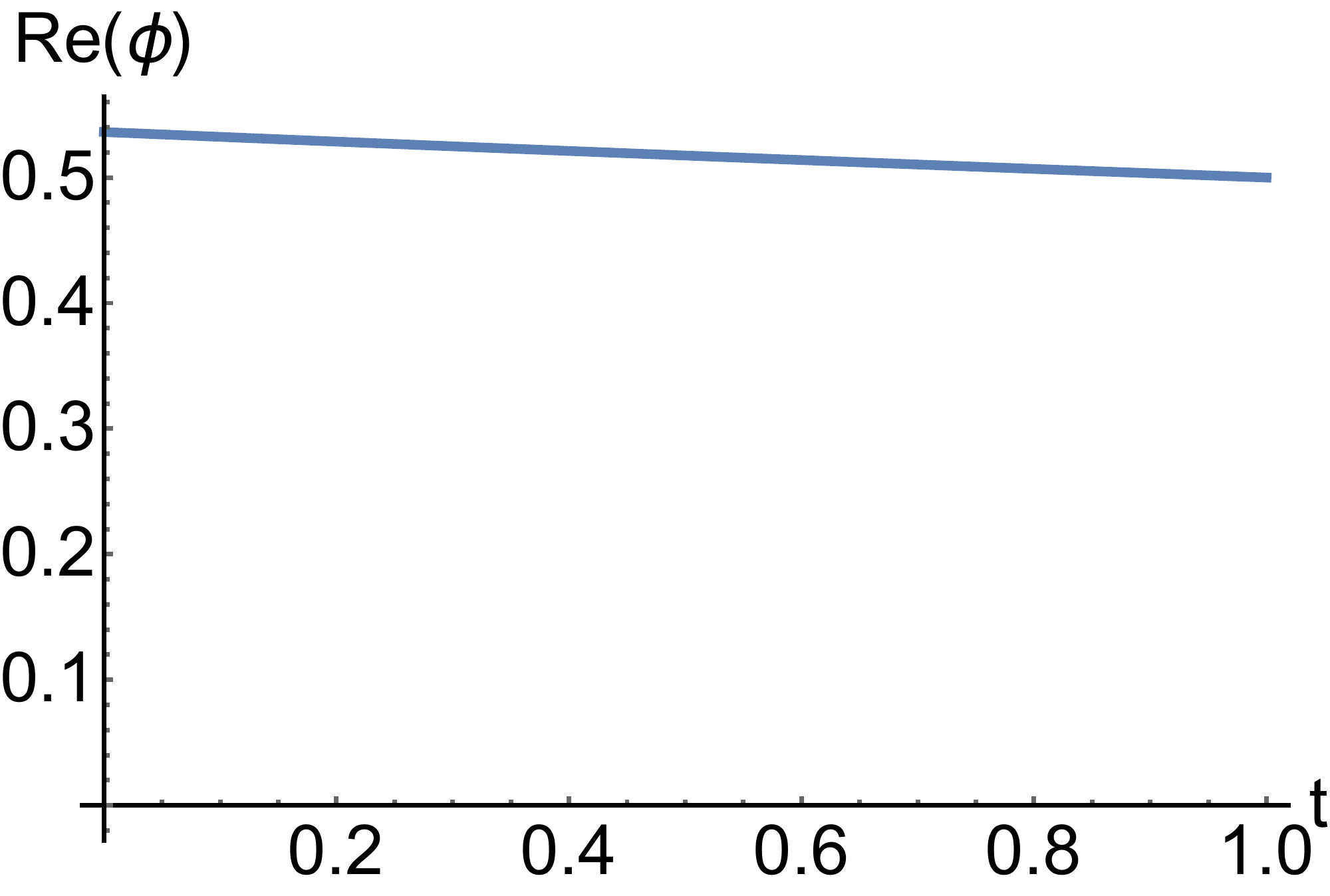}\\
	\includegraphics[width=0.42\textwidth]{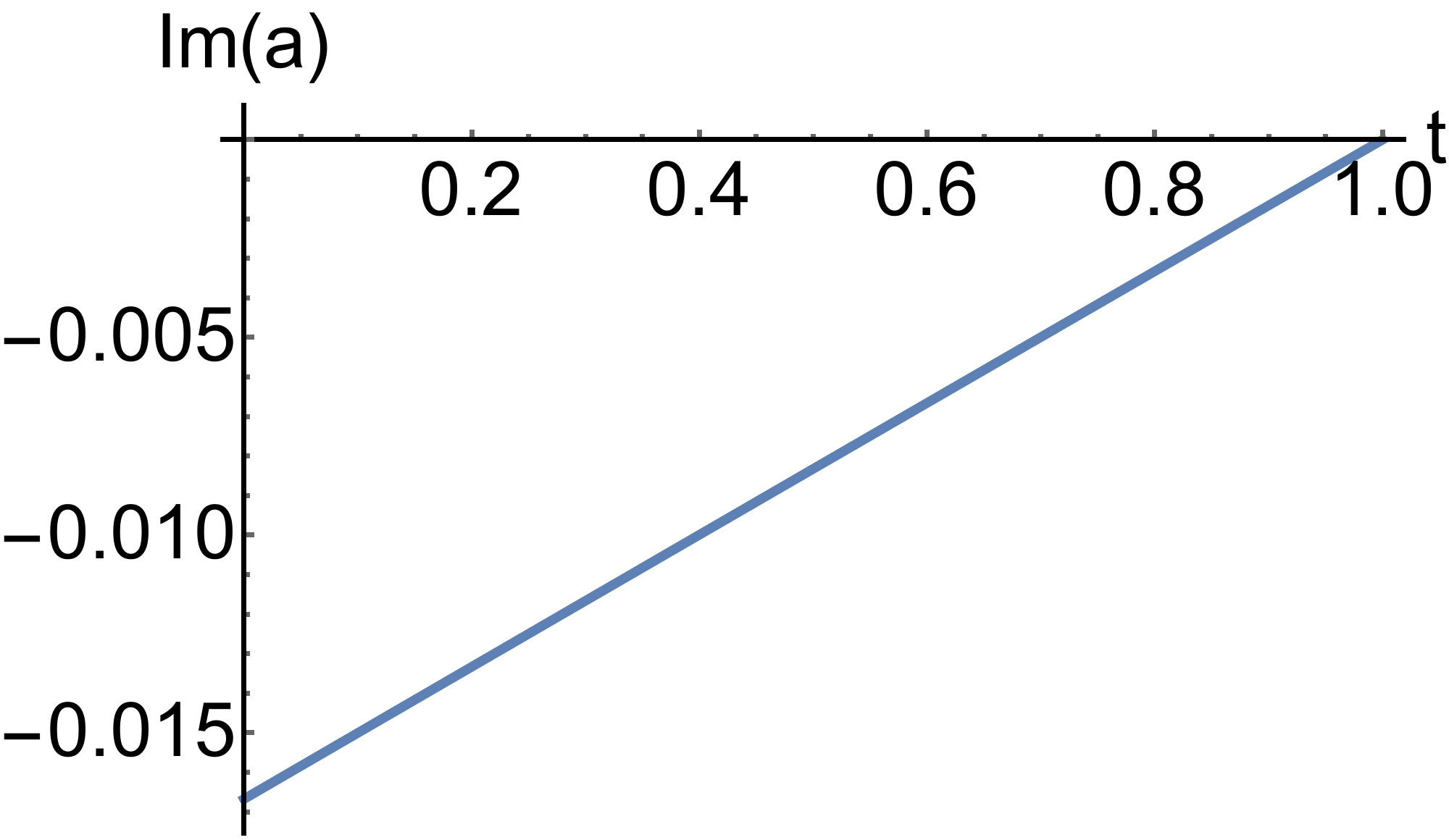}
	\includegraphics[width=0.42\textwidth]{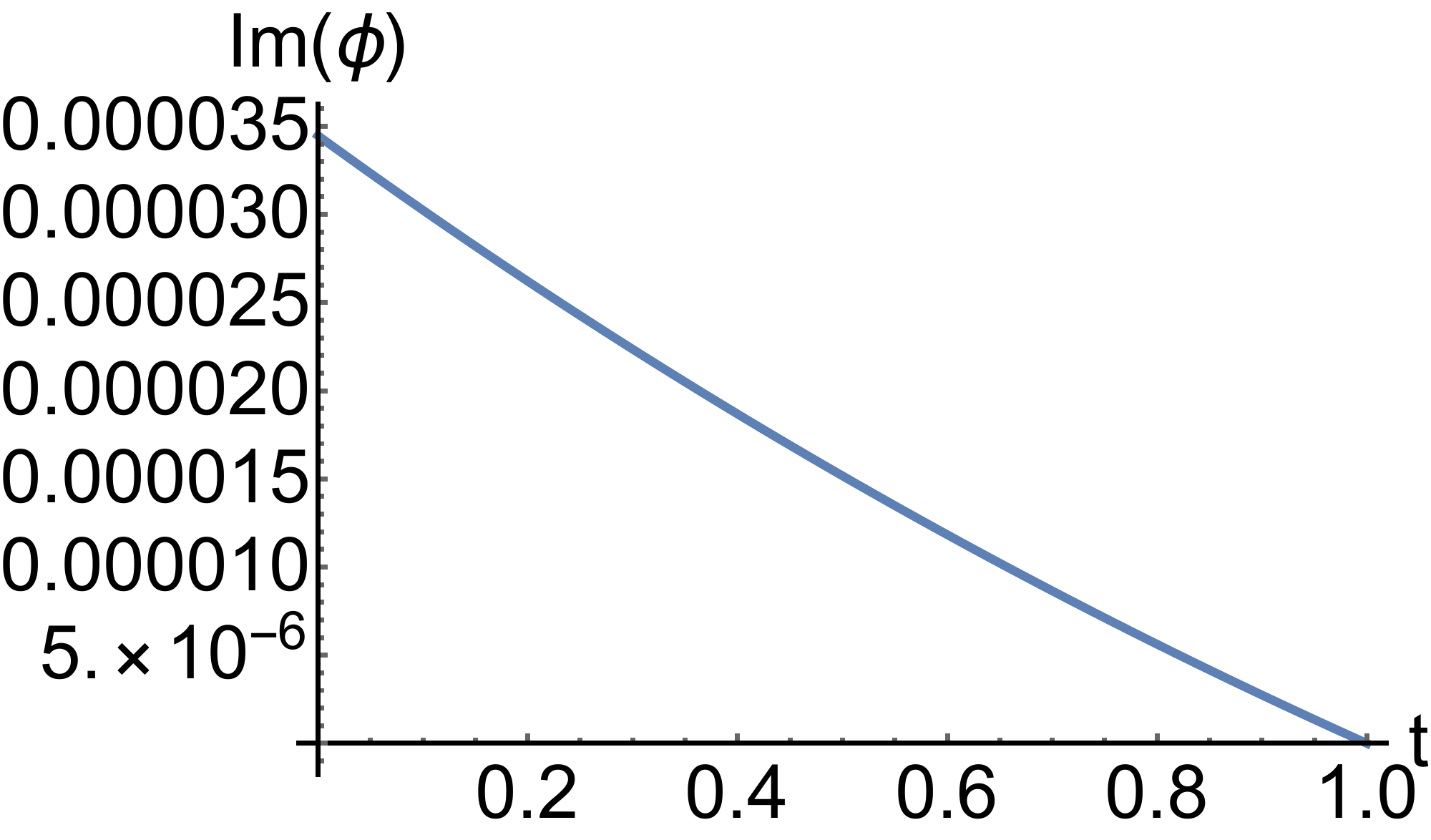}
	\caption{Geometry of the relevant saddle point for $\sigma > \sigma_c$. Plotted here are the real and imaginary parts of the scale factor and scalar field with respect to coordinate time where we have chosen $\alpha = 1/10$, $\phi_0 = 1/10$, $\phi_1 = 1/2$, $a_0 = 100$, $a_1 = 100$ and $\sigma_{\phi} = 2/100$. The final relevant solution is seen to be a slightly complexified version of an ordinary inflationary solution, with the scale factor expanding and the scalar field rolling \emph{down} the potential, even though our boundary conditions are such that we consider an up-jump from the central value of the inflaton.} \label{fig:conv-sp}
\end{figure}

The evolution of the saddle point locations and their associated flow lines as a function of $\sigma_\phi$ is illustrated in Fig. \ref{fig:conv-sp-evolution-srei}. For $\sigma_\phi = 0$, two of the four critical points of the action are relevant to the Lorentzian propagator. As we turn on $\sigma_\phi$, the four saddle points smoothly change their location in the complex $N$ plane and two extra saddle points appear, which however turn out to never give a dominant contribution to the path integral. For a critical value of the uncertainty $\sigma_\phi = \sigma_c,$ a Stokes phenomenon happens which changes the topology of the flow lines. The process is shown in Fig. \ref{fig:conv-sp-evolution-srei} for case A. The final result is entirely analogous for case B, the only difference lying in the fact that in that case the saddle point $N_1$ travels from the imaginary line to the real line as $\sigma_\phi$ increases. After the Stokes phenomenon ($\sigma > \sigma_c$), the only relevant saddle point is $N_1$ and this saddle point becomes more and more real as $\sigma_\phi$ is further increased. The geometry of the relevant saddle point $N_1$ after the Stokes phenomenon has occurred, i.e. for $\sigma_\phi > \sigma_c$, is shown in Fig. \ref{fig:conv-sp}. An interesting aspect is that the initial position of the scalar field $\bar{x}_0$ is no longer close to the original initial position $x_0,$ but is significantly larger -- in fact it has become larger than the final value $x_1$ (and $\phi$ also contains a small imaginary part, which is a reflection of the transition being a quantum transition).

What does this mean? In the Dirichlet formulation of the Feynman propagator we calculate a transition between two fixed geometries and matter content. In that setting it is not possible to continuously link an inflationary evolution with an evolution where the scalar field tunnels up the potential. However, by introducing Robin boundary conditions, thus allowing for a spread in field values and momenta, we do find solutions. Analysing them in more detail, we find that the scalar field \textit{already} starts higher up the potential and then simply rolls down according to an inflationary solution.  Thus, instead of choosing a solution that rolls up the potential, the system has picked out a (comparatively unlikely) configuration contained within the initial state in which the inflaton is already higher up in the potential than required, so as to allow a slow-roll solution to the final configuration. In complete analogy, the scale factor starts out at a smaller value and then grows as the scalar field rolls down. 

\begin{figure}[ht!]
	\centering
	\includegraphics[width=0.48\textwidth]{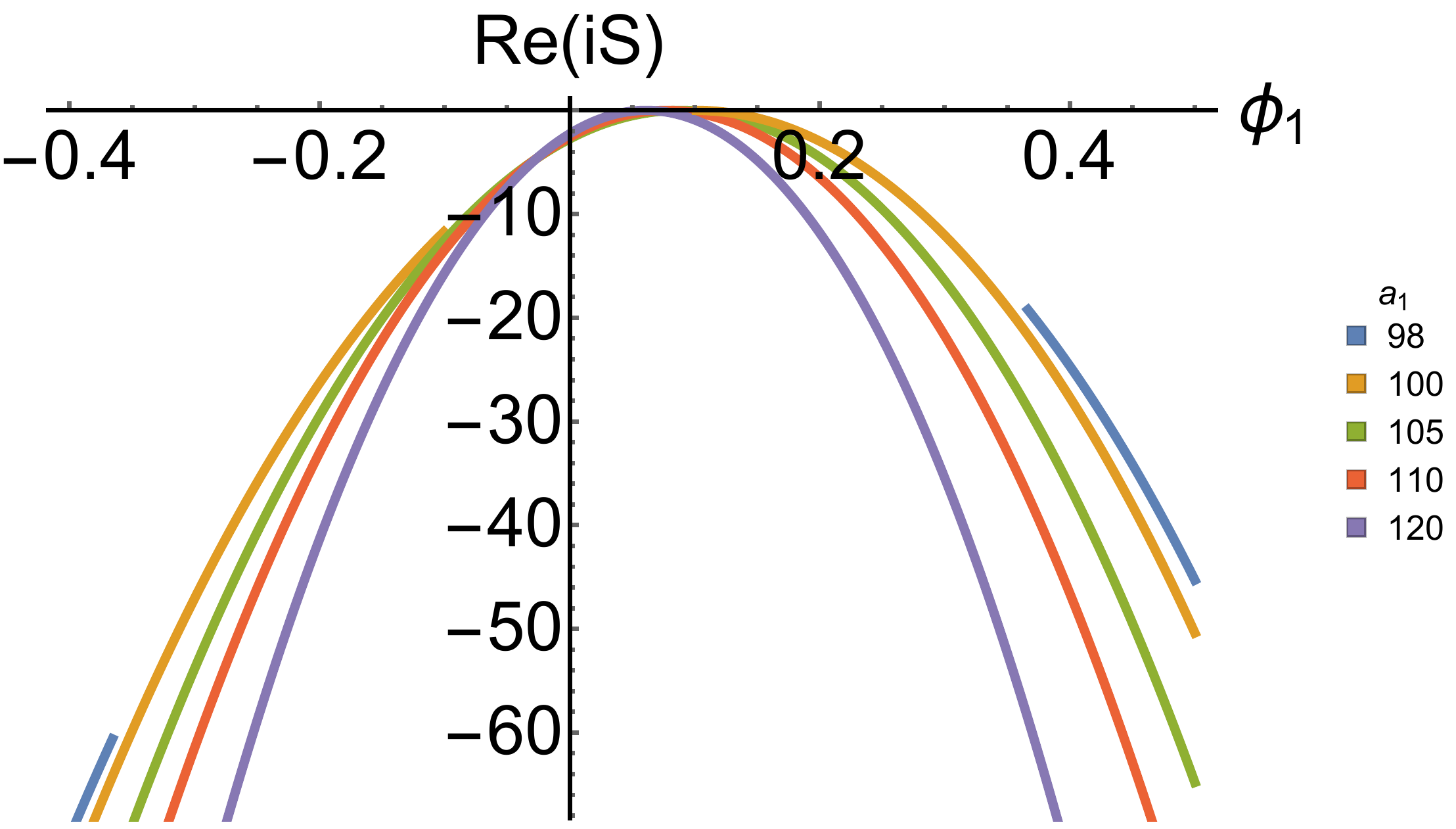} 
	\includegraphics[width=0.44\textwidth]{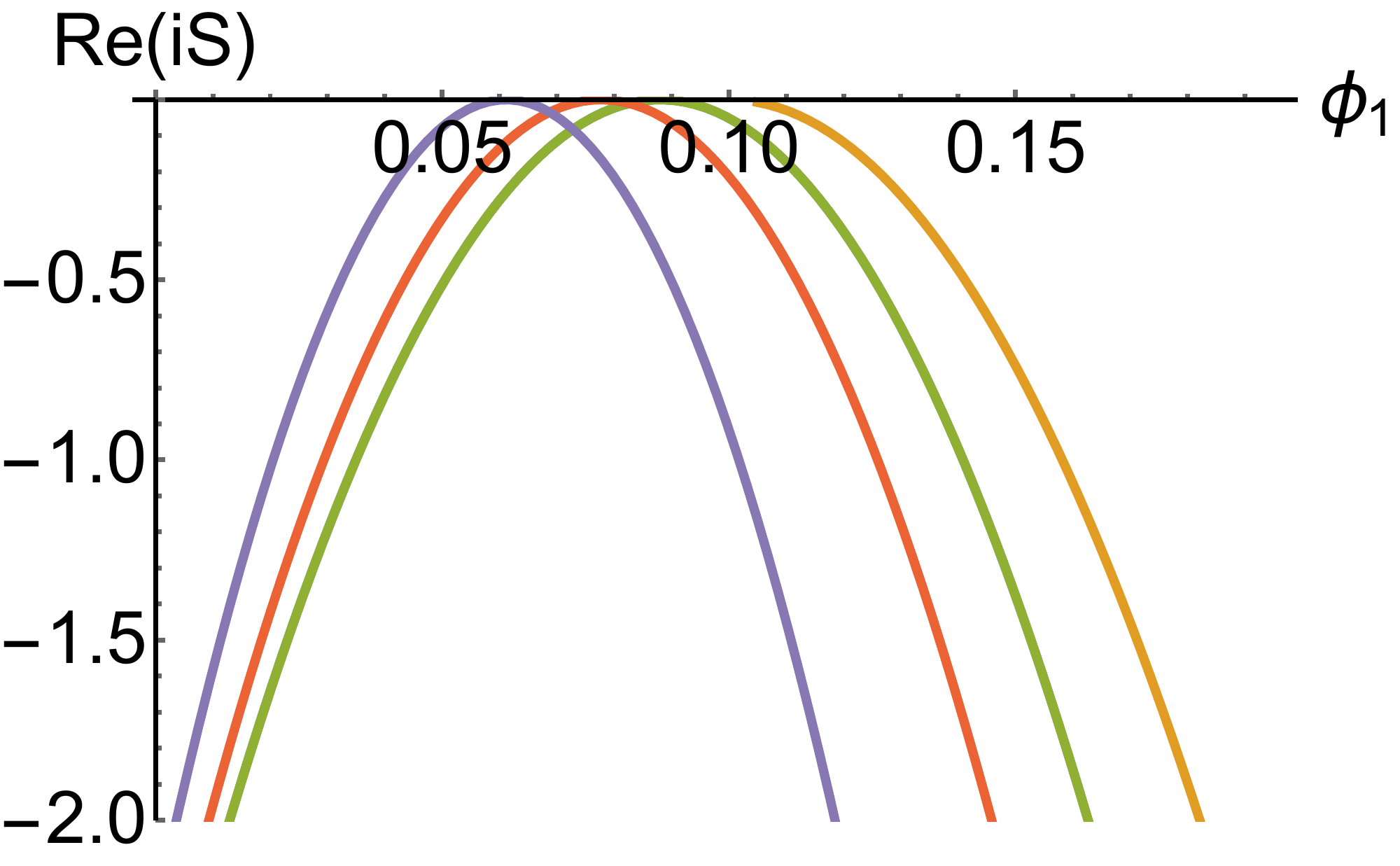}
	\caption{The logarithm of the transition amplitude going from $(a_0, \phi_0) = (100,1/10)$ to various values of $a_1$ and $\phi_1$. Here the x-axis represents $\phi_1$ while the different colours refer to different values of $a_1$, ranging between 98 and 120. The action was evaluated for a spread of $\sigma_\phi = \frac{H}{2\pi}$. Interestingly, for $a_1 \leq 100$ there are areas where no transition is possible and hence there are gaps in the parabola. This is because the usually relevant saddle point has negative real $N$ for these transitions, and no other saddle point is relevant. In such cases, the transition would be more than exponentially suppressed. The picture on the right is the same as on the left except zoomed in onto the top of the curves. As the final scale factor value $a_1$ is increased, the peak of the distribution shifts and the spread in $\phi$ narrows.} \label{fig-action-series-a-phi1}
\end{figure}

\begin{figure}[ht!]
	\centering
	\includegraphics[width=0.32\textwidth]{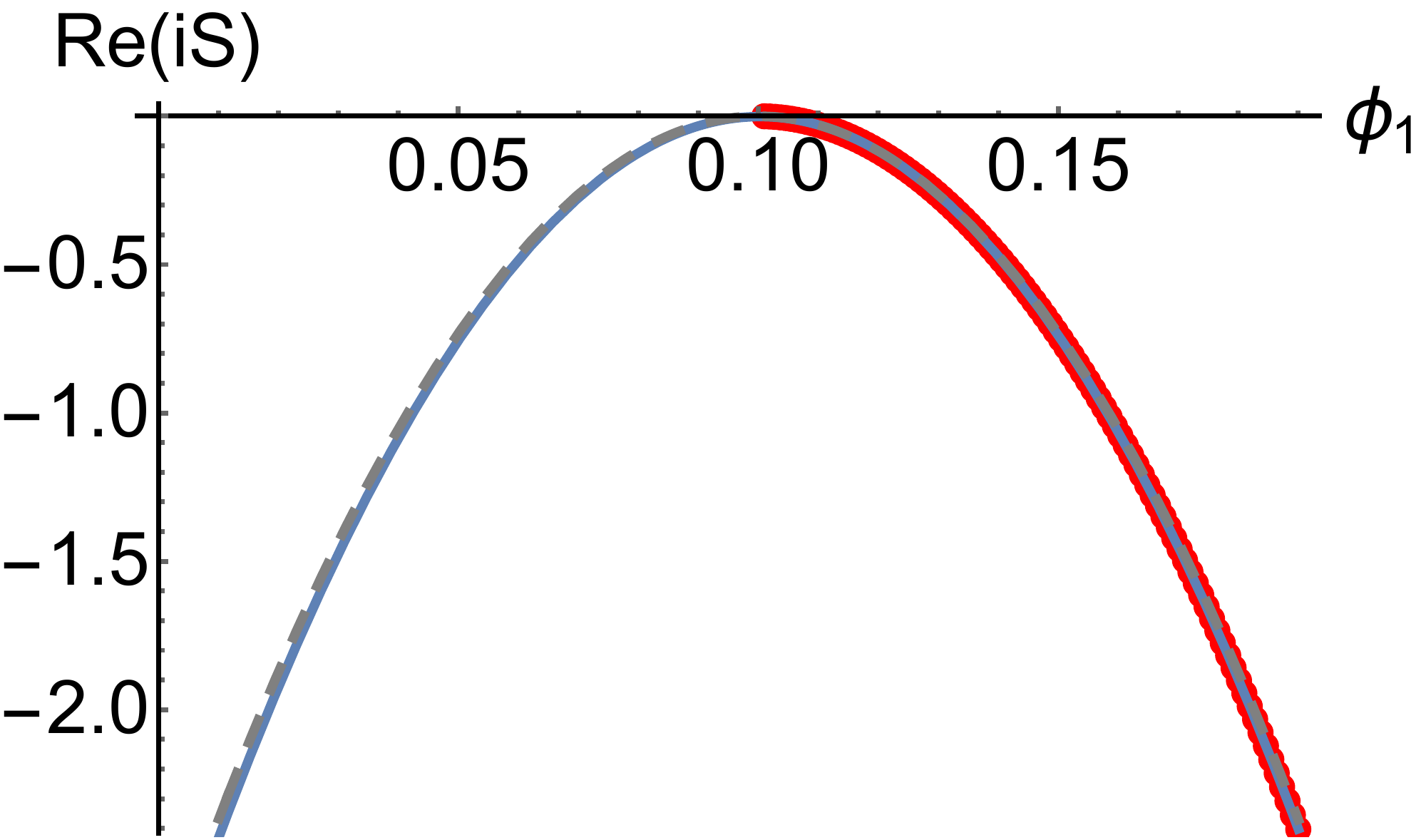}
		\includegraphics[width=0.32\textwidth]{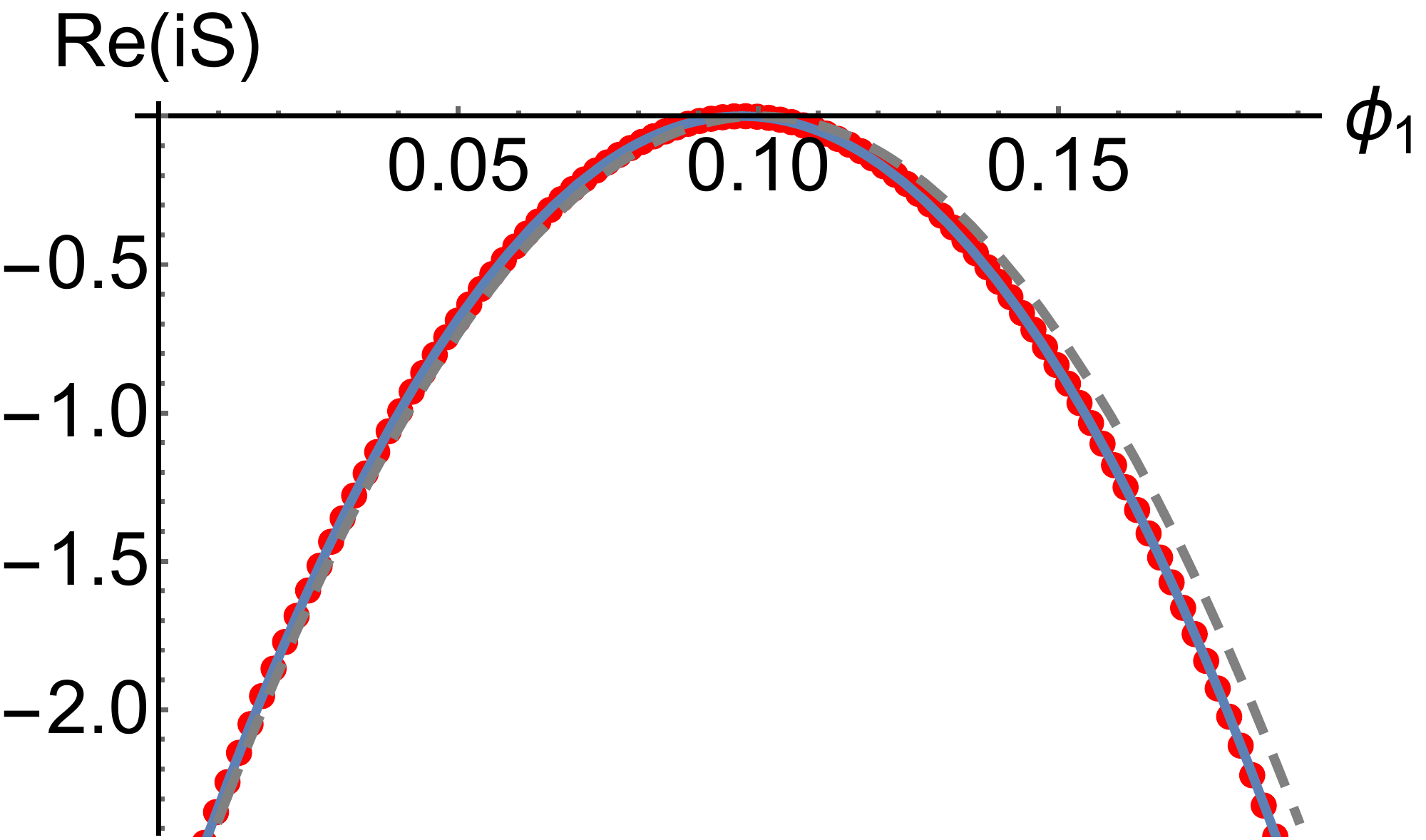}
		\includegraphics[width=0.32\textwidth]{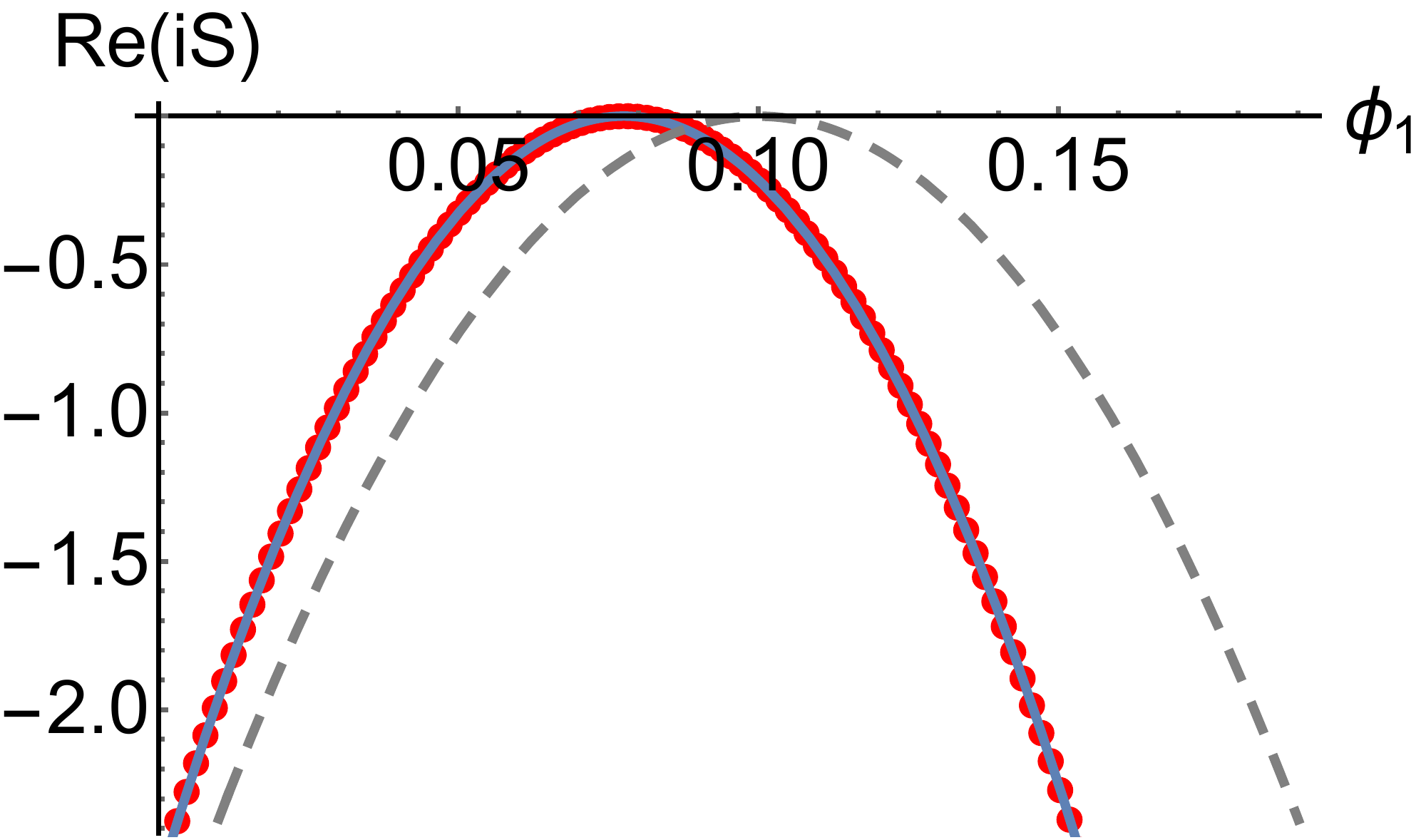}\\
	\caption{An example of the distribution of fluctuations after the quantum transition in red with a fitted parabola in blue. In all three graphs $\alpha = 1/10$, $\phi_0 = 1/10$, $a_0 = 100$. However $a_1 = 100, 101, 110$ in the left, centre and right graphs respectively. In dashed grey we have plotted the initial spread in $\sigma_\phi = H/(2 \pi)$ centred around the classical value $\phi_{top}$. From the picture it is clear that the peak of the distribution shifts and the spread in $\phi$ narrows as we increase the final scale factor -- see also Fig. \ref{fig-spread-af} for more details.} \label{fig-fluctuation-distribution-parabola}
\end{figure}

\begin{figure}[ht!]
	\centering
	\includegraphics[width=0.45\textwidth]{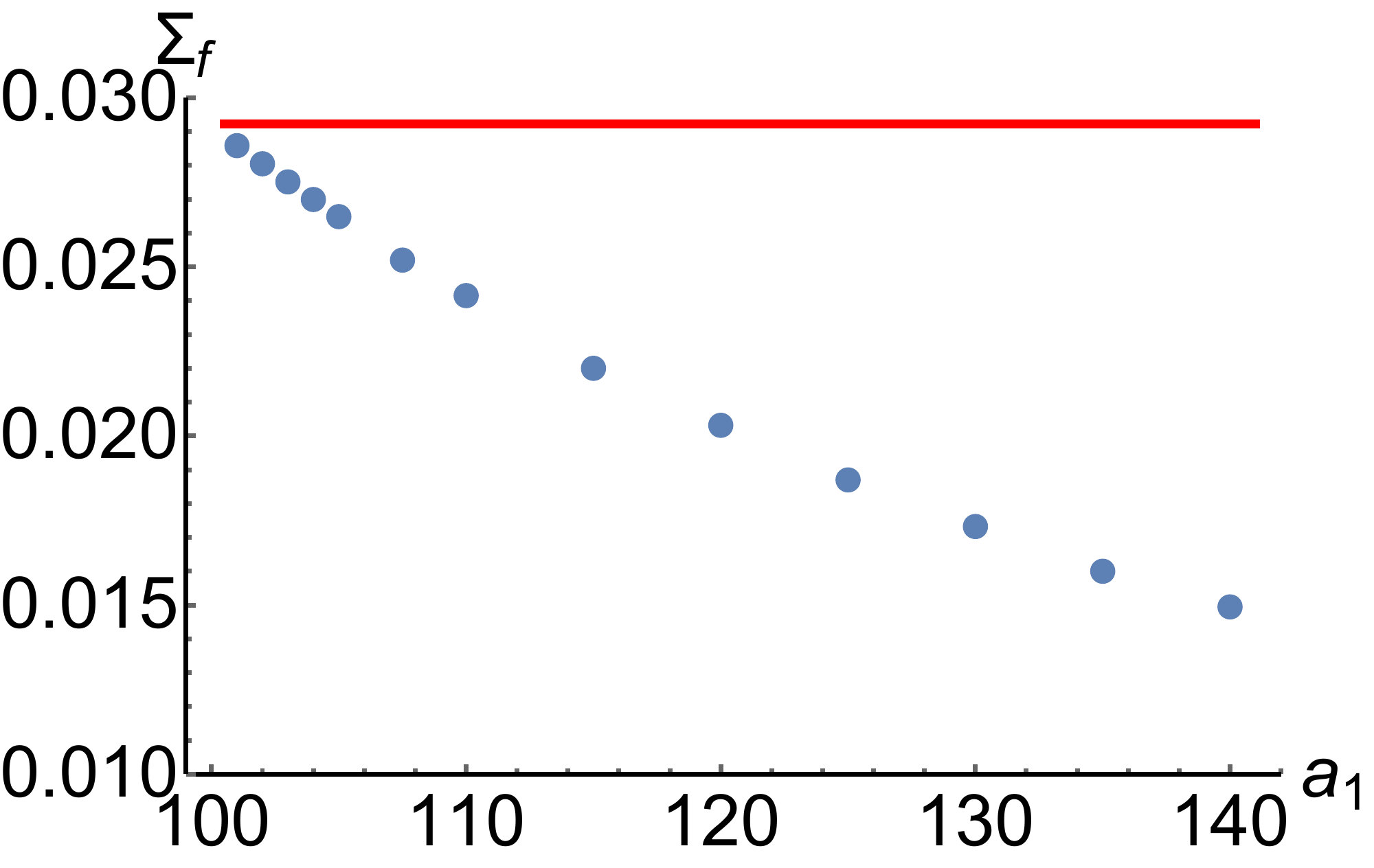}
	\caption{For transitions in which the final scale factor is only slightly larger than the initial one, the weighting for different final configurations is essentially equal to the weighting implied by the initial state. But as the final scale factor value $a_1$ increases the spread (of the weighting) is reduced as a result of the inflationary attractor. The numerical example shown here is the same one as in Fig.\ref{fig-action-series-a-phi1}. The red line is the spread in the inflaton value imposed before the transition occurs.} \label{fig-spread-af}
\end{figure}

This result can be further quantified by analysing probabilities of the geometry and scalar field undergoing transitions to various values of $\phi_1$ and $a_1$ as depicted in Fig. \ref{fig-action-series-a-phi1}. It is obvious that for larger values of $a_1$ and $\phi_1$, transitions become less and less likely. In fact, the most likely transitions occur for a tiny increase in the scale factor, in our example from $a_0=100$ to $a_1 \approx 101.$ This confirms the expectation from QFT in curved spacetime that the geometry ultimately changes very little when the scalar field jumps up the potential. Here we should note that when we impose a final scale factor value that is equal to or smaller than the initial one, then transitions to certain values of the scalar field are impossible (semi-classically). This is reflected in some of the curves in Fig. \ref{fig-action-series-a-phi1} having gaps in them. What happens in these cases is that the relevant saddle point moves to the region where $Re(N)<0,$ i.e. these solutions then actually correspond to time-reversed solutions. This is consistent with the fact that the system prefers to choose inflationary, expanding solutions and requiring the final scale factor to be small then clashes with this preference. In line with this observation is the fact that if we look at increasing values of the final scale factor, then the spread is actually reduced due to the inflationary attractor. In fact, the final weighting remains Gaussian to a good approximation, with only the peak value having shifted and the spread shrinking. We can see this more quantitatively in Fig. \ref{fig-fluctuation-distribution-parabola}, where we plot the final weighting alongside a fitted parabola with final width $\Sigma_f,$ defined via
\begin{align}
Re(i\tilde{S}/\hbar) = h(\phi_1) = h(\phi_{top}) - \frac{(\phi_1-\phi_{top})}{4\Sigma_f^2} + \cdots\,,
\end{align}
where $\phi_{top}$ denotes that value of $\phi$ at which the weighting (Morse function $h$) is maximal for a given final scale factor value $a_1.$ From the figure we can see that the parabola provides an excellent fit. The decrease in the width as the universe expands is plotted in Fig.\ref{fig-spread-af}.

\begin{figure}[ht!]
	\centering
	\includegraphics[width=0.75\textwidth]{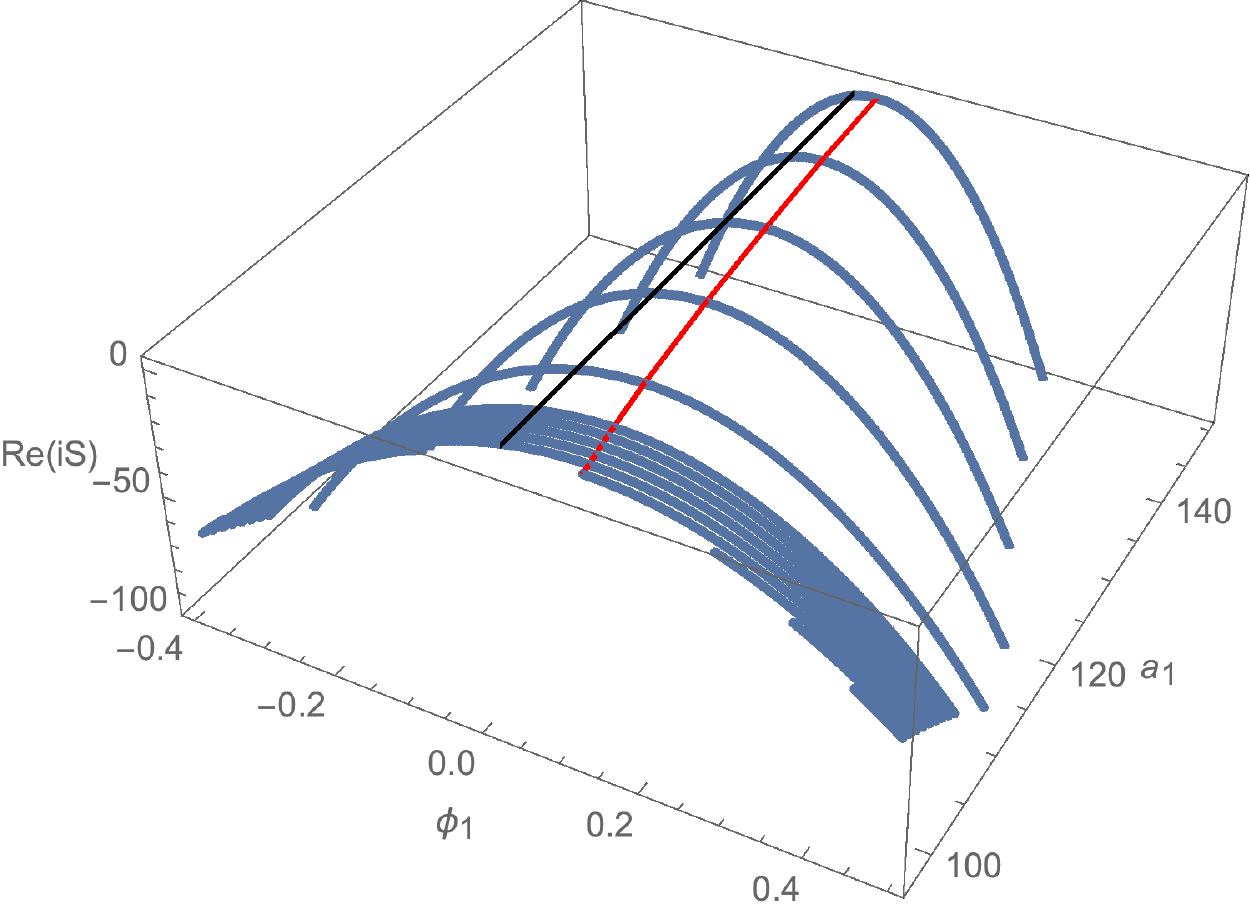}
	\caption{A 3-dimensional version of Fig. \ref{fig-action-series-a-phi1} illustrating how the peak of the weighting (red line) follows a slow-roll solution down towards the minimum of the potential at $\phi=0$ (black line), while large excursions of the inflaton away from the classical solution become less and less likely as the universe expands.} \label{fig-action-series-a-phi1-3d}
\end{figure}

All this is nicely visible in a 3-dimensional version of these plots in Fig. \ref{fig-action-series-a-phi1-3d}. Accompanied with this shrinking of the width is a displacement of the peak of the weighting. The 3-dimensional picture shows that the peak slowly approaches $\phi=0$ as the universe expands -- in other words, the peak of the weighting follows the classical slow-roll trajectory associated with the initial central values of the fields and their momenta that we imposed via the initial state. As the universe expands, the wavefunction narrows around this classical solution, and we attribute this feature to the inflationary attractor. Thus, starting from a fixed initial state and as the universe grows larger, inflaton excursions away from the classical solution become less likely.

\begin{figure}[h]
	\centering
	\includegraphics[width=0.45\textwidth]{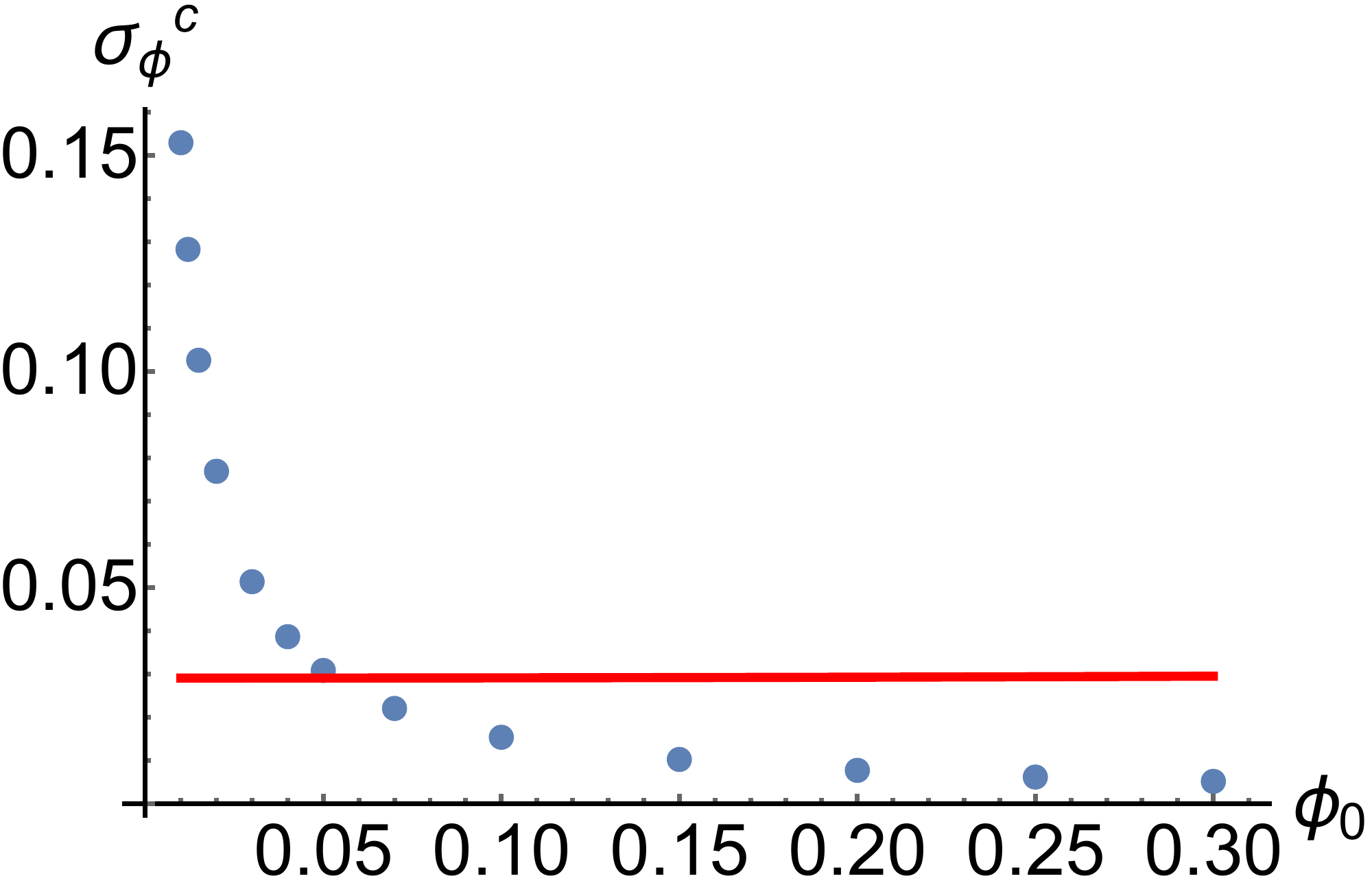}
	\includegraphics[width=0.45\textwidth]{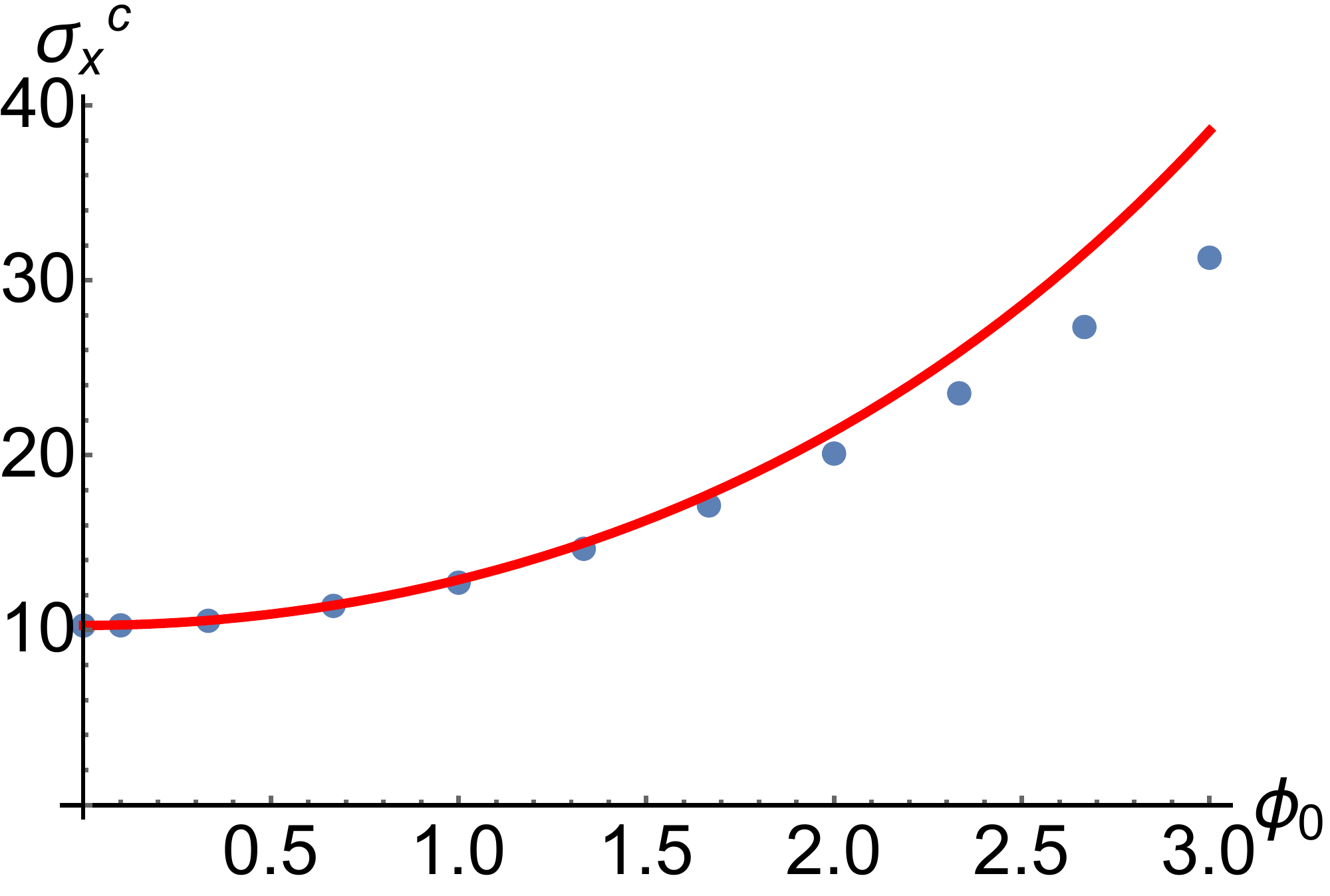}
	\caption{{\it Left panel:} The blue dots indicate the value of $\sigma_\phi$ at which the Stokes phenomenon appears while the value of $H/(2\pi)$ is given by the red line. Here $\alpha = 1/10$ and $a_0 = 100$. The critical value of $\sigma$ depends only on the initial value $\phi_0$ and not on the final values $a_1$ and $\phi_1.$ This graph shows that for a sufficiently large initial scalar field value, only one saddle point is relevant at $\sigma=H/(2\pi).$ {\it Right panel:} The critical spread expressed in terms of the canonical variables. This figure shows the critical value $\sigma_x^c$ as a function of the initial inflaton value $\phi_0.$ Near $\phi_0=0$ we recover the exact value for de Sitter space given in Eq. \eqref{eq:CritdS}. The red curve shows the expected value if one were to assume only an uncertainty in the initial scale factor, and not in the inflaton value.  For larger $\phi_0$ we can see that the critical value lies below this curve, implying that for sufficiently large $\phi_0,$ where the potential is less flat, the uncertainty in the inflaton value can induce an earlier Stokes phenomenon. To calculate the points, we have fixed $a_0 = a_1 = 100$ with varying $\phi_0$ and a $\phi_1$ that corresponds to the field jumping up the potential. The initial state's momenta were calculated by keeping $\phi_0'$ fixed, finding the corresponding $a_0'$ via the Friedmann equation and then converting to the momenta in $x$ and $y$.} \label{fig-spread-af2}
\end{figure}

An important effect that we saw earlier was that beyond some critical value of the spread a Stokes phenomenon happens and only a single saddle point remains relevant. When this occurs, we automatically obtain a situation in which quantum field theory in curved spacetime is a reasonable approximation, as only a single background geometry is relevant to the path integral. We can now make this discussion more quantitative -- see Fig. \ref{fig-spread-af2}. An important aspect of this discussion concerns the relationship between the original variables $a, \phi$ and the canonical variables $x, y,$ expressed via the transformations Eqs. \eqref{redef1} - \eqref{redef2} and the relations between the spreads \eqref{sigtrf1} and \eqref{sigtrf2}. We argued in section \ref{review} that the standard calculation in a fixed background suggests that the inflaton should have a significant spread, of order $H/(2\pi),$ with the scale factor being kept essentially fixed. This would amount to setting $\sigma_a = 0.$ The left panel in Fig. \ref{fig-spread-af2} shows the critical value of $\sigma_\phi$ that is required under those circumstances in order to obtain the Stokes phenomenon, as a function of the initial inflaton value $\phi_0.$ (An important point is that the critical spread does not depend on the final inflaton value $\phi_1$.) What the figure shows is that for large enough $\phi_0$ the Stokes phenomenon always occurs before the spread is increased to $H/(2\pi).$ Thus, in regions where the potential is not too flat (but including regions where the density perturbations that are generated may be large), the standard intuition is vindicated. 

\begin{figure}[ht!]
	\centering
	\includegraphics[width=0.4\textwidth]{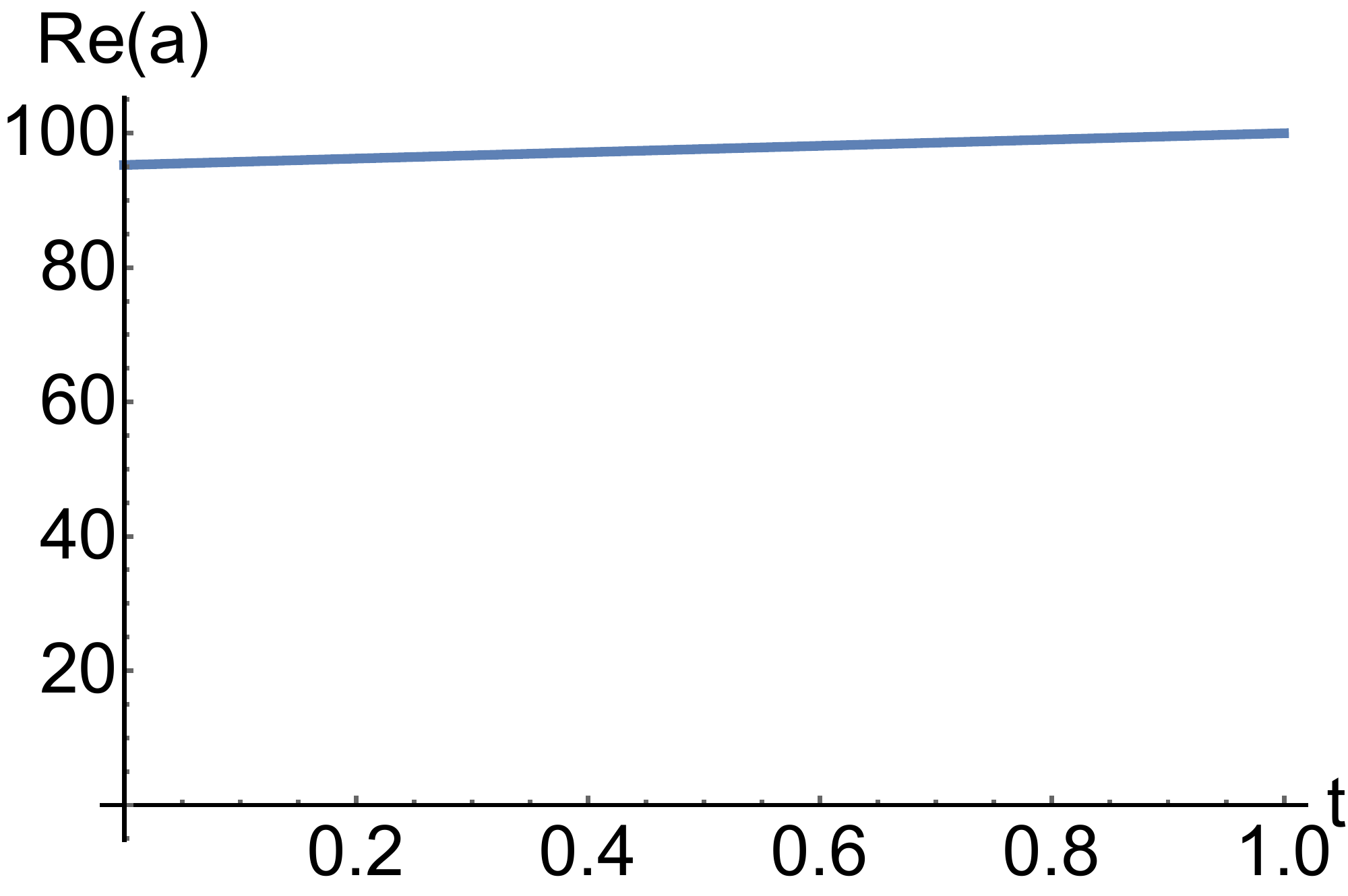} \includegraphics[width=0.4\textwidth]{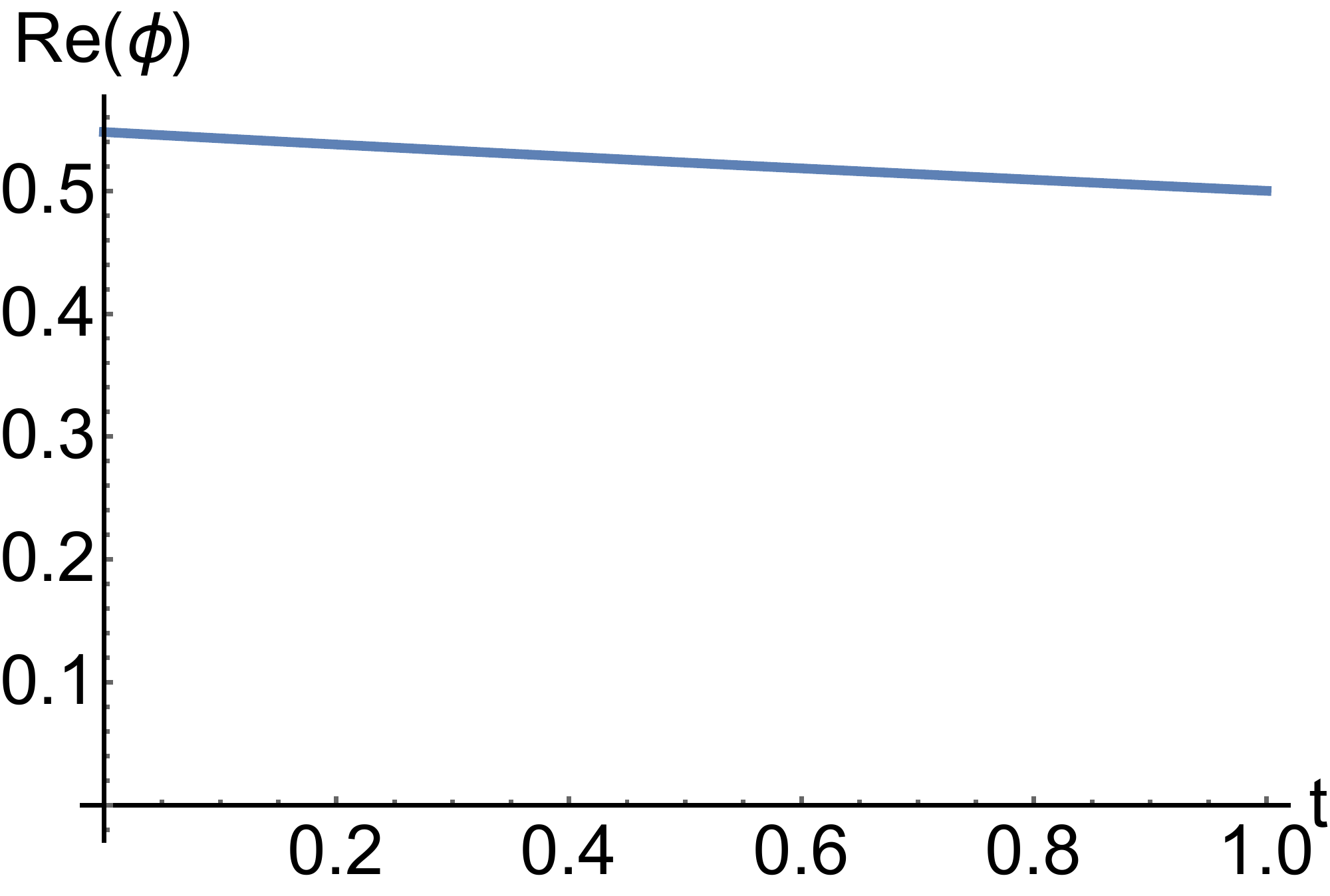}\\
	\includegraphics[width=0.4\textwidth]{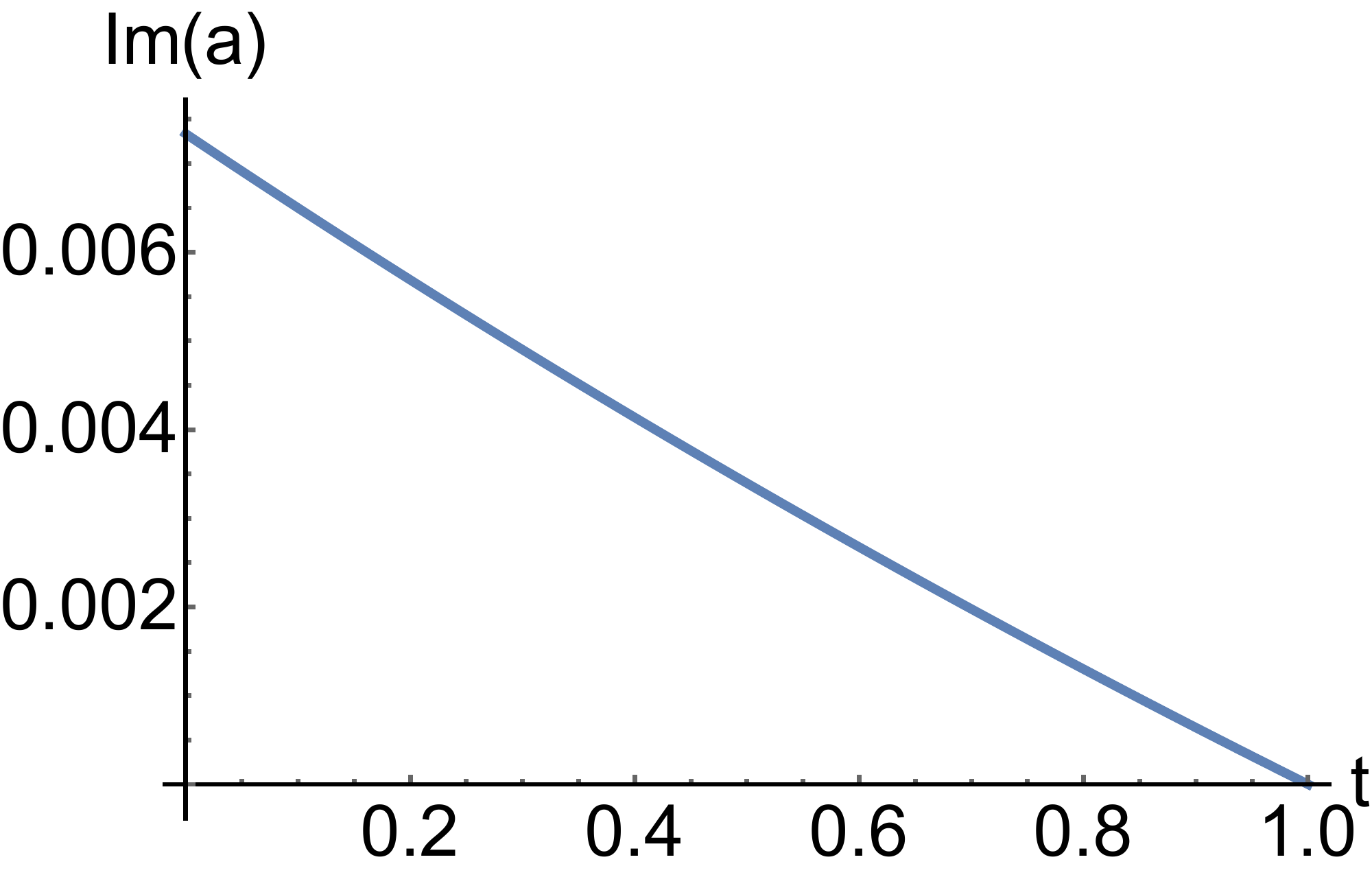} \includegraphics[width=0.4\textwidth]{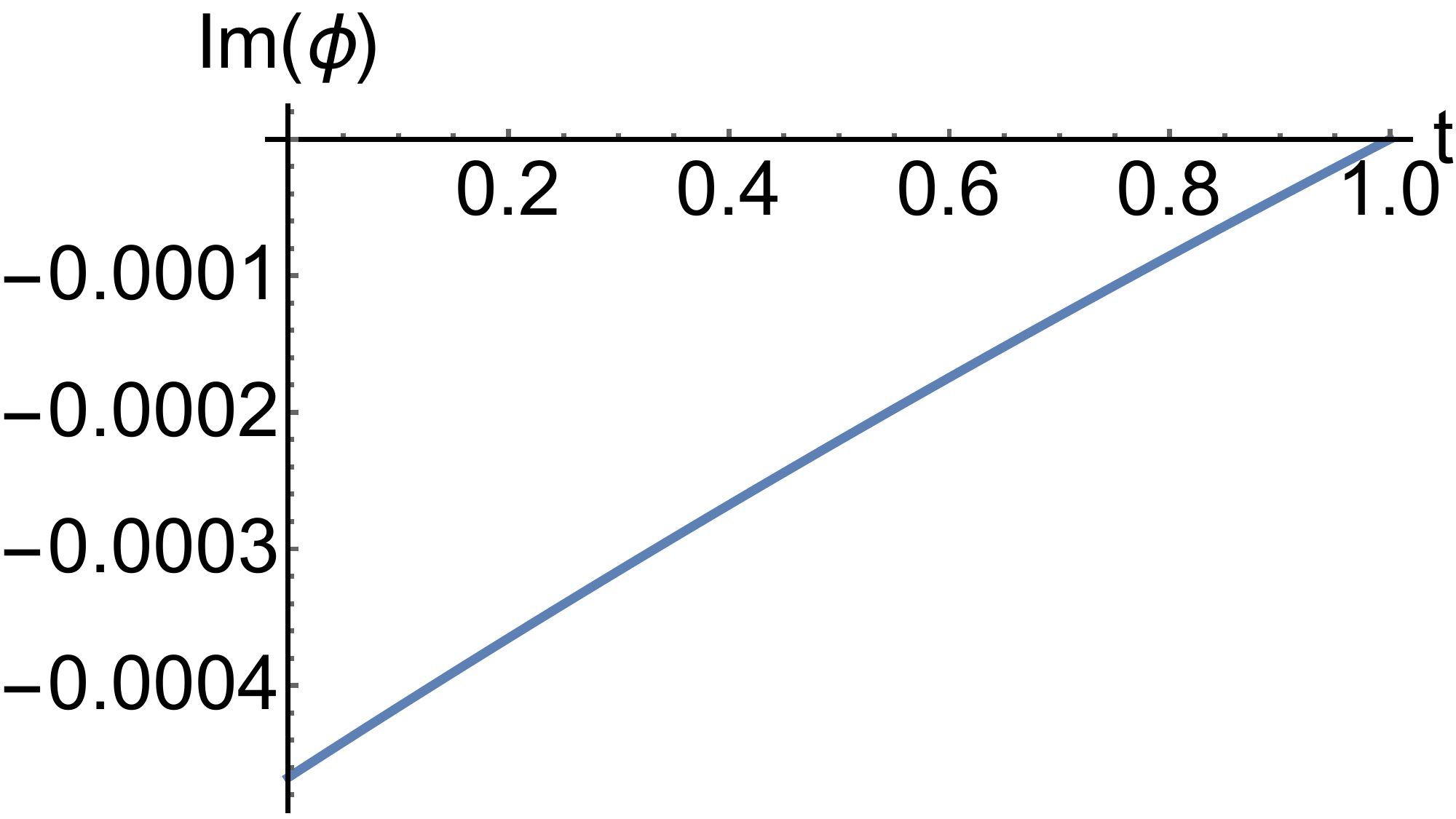}
	\caption{Geometry of the relevant saddle point where the scale factor is kept constant $a_0 = a_1 = 100$ and the scalar field transitions from $\phi_0 = 1/1000$ to $\phi_0 = 1/2$. The initial state's momenta were chosen to be $p_x \approx -54.7$ and $p_y \approx -0.0134$ with uncertainties $\sigma_x = 11$ and $\sigma_y = 100$ which implies that the Stokes phenomenon has already happened. Notice that this geometry closely resembles the one of Fig. \ref{fig:conv-sp}, where $\phi_0$ is larger.} \label{fig:largesigy}
\end{figure}

\begin{figure}[ht!]
	\centering
	\includegraphics[width=0.4\textwidth]{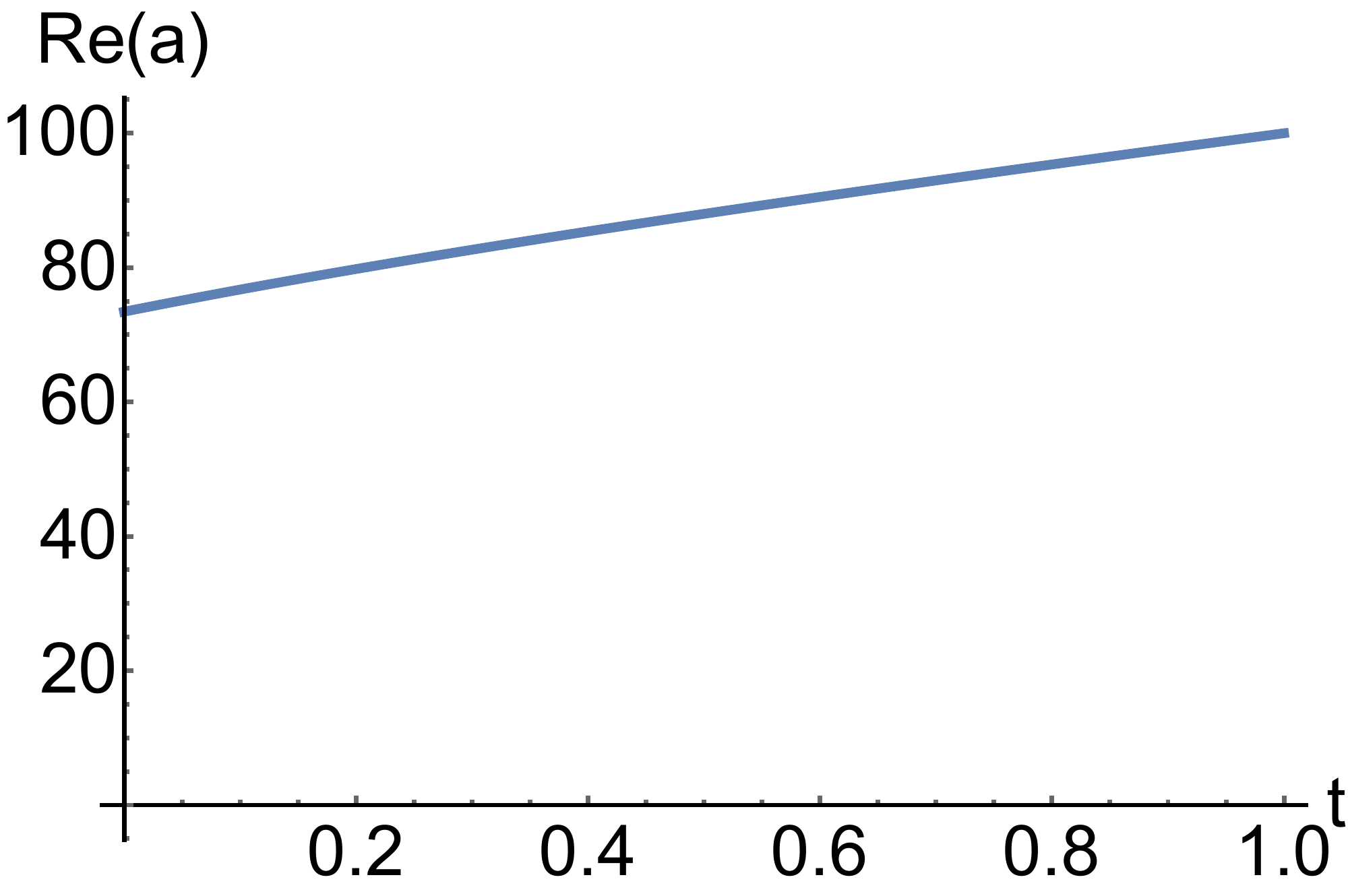} \includegraphics[width=0.4\textwidth]{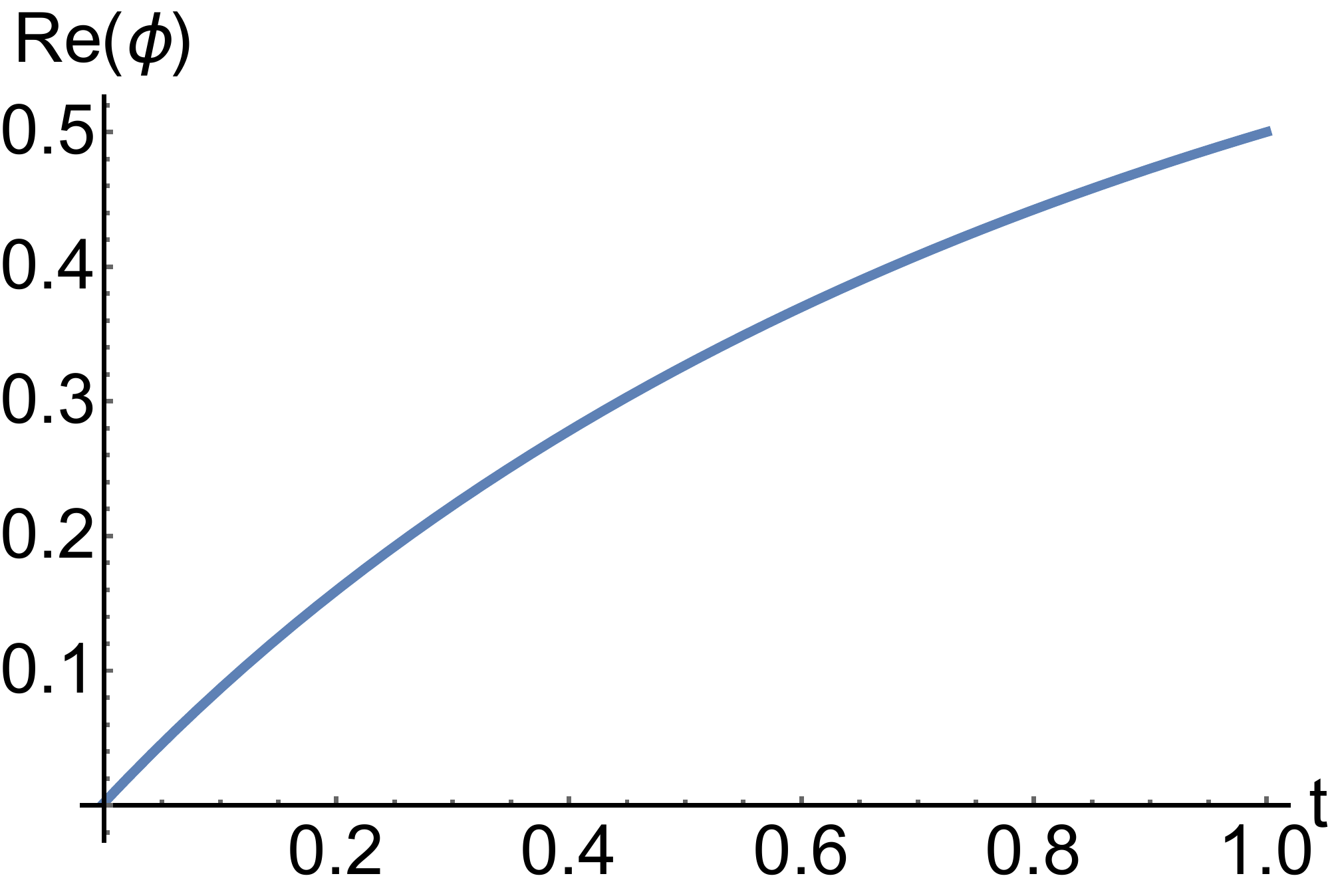}\\
	\includegraphics[width=0.4\textwidth]{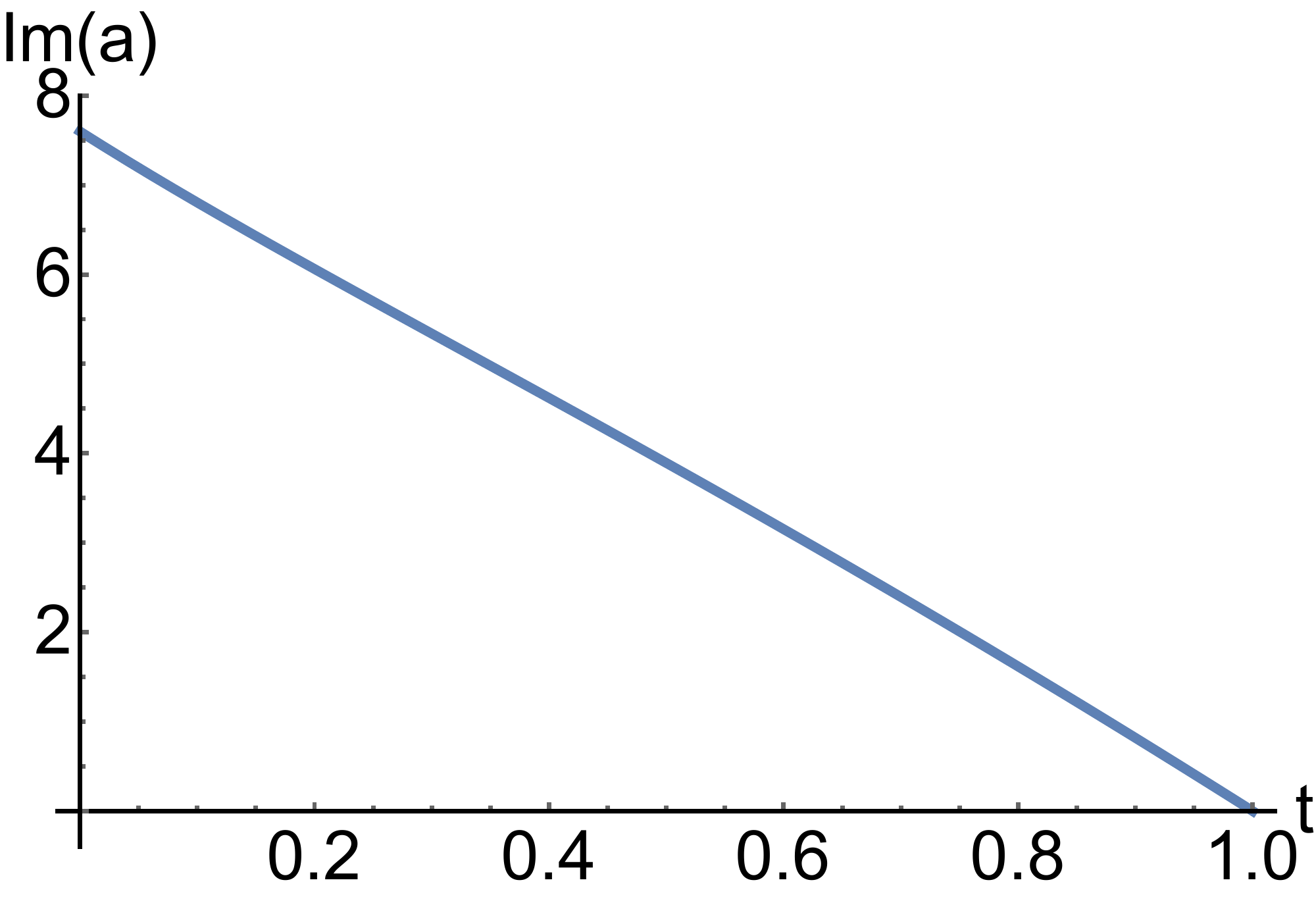} \includegraphics[width=0.4\textwidth]{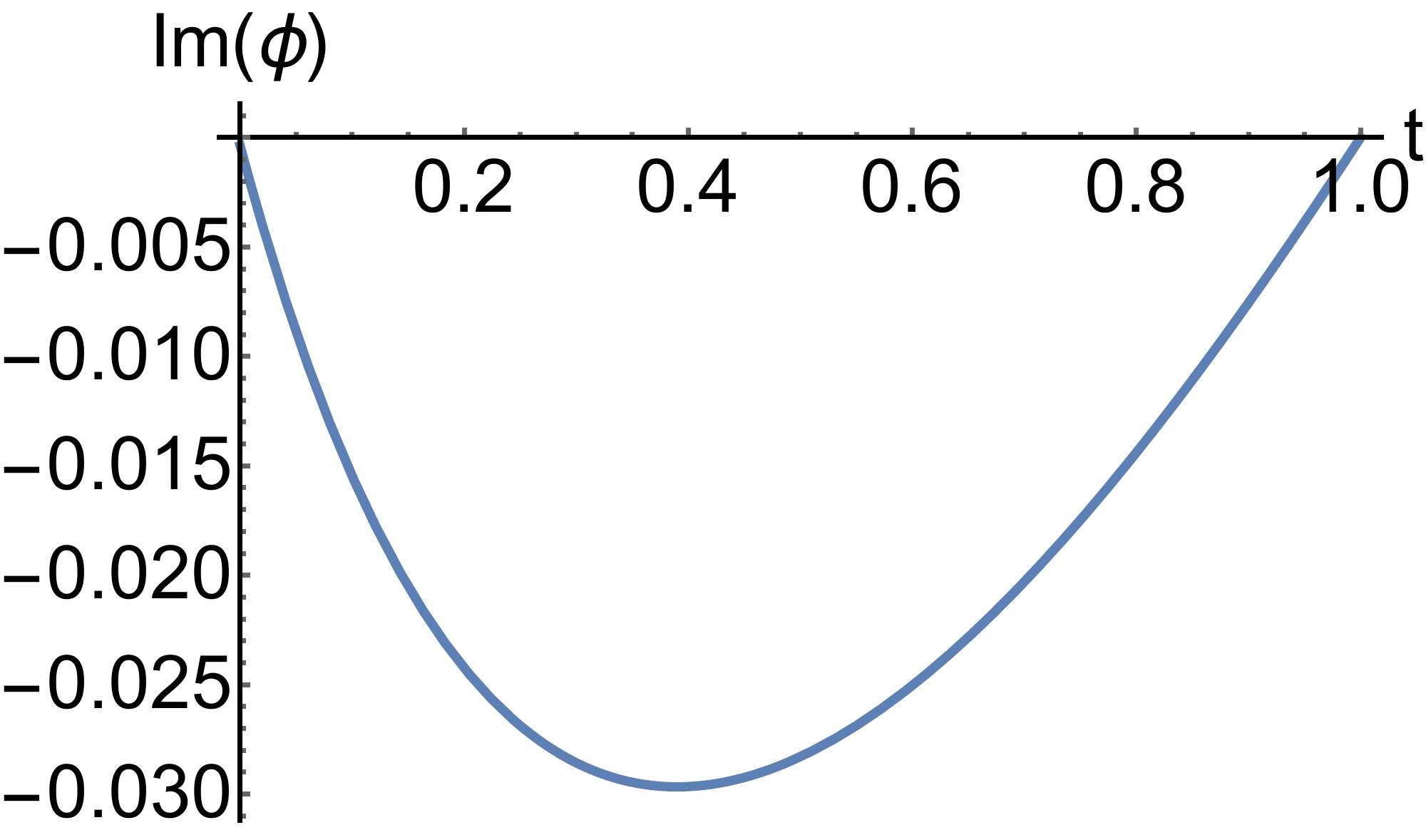}
	\caption{Geometry of the relevant saddle point where the numerical values are identical to the ones of  Fig. \ref{fig:largesigy} except that now $\sigma_y = 0$.} \label{fig:smallsigy}
\end{figure}

For small values of $\phi_0$ however we see a departure from this behaviour, in that the minimum value of $\sigma_\phi$ that would be required to obtain a Stokes phenomenon becomes larger and larger. At this point it is advantageous to switch to a description in terms of the canonical variables $x, y.$ Note that when $\sigma_a=0,$ we have that $\sigma_x \propto \sinh\left(\sqrt{2/3} \phi_0\right) \sigma_\phi$ and thus, for small $\phi_0$ a large inflaton uncertainty $\sigma_\phi$ may still correspond to a much smaller spread $\sigma_x.$ The right panel in Fig. \ref{fig-spread-af2} now shows the critical spread expressed in terms of $\sigma_x$ as a function of $\phi_0.$ Here we are departing from the assumption that $\sigma_a=0,$ and in fact in the plot we have chosen a constant value for $\sigma_y$.\footnote{It turns out that the precise value of $\sigma_y$ is not so important, except when $\sigma_y$ is very small (a case which we will discuss below). We believe that the relative insensitivity to $\sigma_y,$ and the importance of $\sigma_x,$ are simply a reflection of the fact that the potential depends solely on $x.$} What we see is that for small initial scalar field values the critical spread is reduced, rather than enhanced, compared to larger $\phi_0.$ Moreover, the limiting value at $\phi_0=0$ corresponds exactly to the critical value calculated for pure de Sitter space in \cite{DiTucci:2019xcr}, and where the scale factor of the universe was the only degree of freedom, 
\begin{align}
\sigma_x^c(\phi_0=0) = \left( \frac{a_0^2}{9 \alpha}\right)^{1/4}\,. \label{eq:CritdS}
\end{align}
Thus we see that in the region where the potential is flattest, we {\emph{require}} a minimum uncertainty in the size of the universe $\sigma_a \neq 0,$ and it appears not to be sufficient to only have a large enough uncertainty in the inflaton value. Based on the formula \eqref{eq:CritdS}, we might guess that the critical uncertainty should be given, as long as the slow-roll approximation holds, by replacing $\alpha$ by $V(\phi_0),$ and taking into account the transformation formula \eqref{sigtrf1}. Hence, if we assumed that now on the contrary $\sigma_\phi$ was set to zero, and we would consider only an initial spread in the scale factor, then we might expect the critical spread to be given by
\begin{align}
\sigma_x^c(\sigma_a \neq 0, \sigma_\phi=0) = \cosh\left(\sqrt{\frac{2}{3}\phi_0} \right) \, \left( \frac{a_0^2}{9 V(\phi_0)}\right)^{1/4} \,. \label{CritdS-corr}
\end{align}
The corresponding curve is plotted in red in the right panel of Fig. \ref{fig-spread-af2}. We can see that the true critical spread in fact lies somewhat below this curve. This can be understood in terms of the previous discussion where we showed that for large enough $\phi_0$ even a small $\sigma_\phi$ is already enough to cause the Stokes phenomenon. Thus, away from the very flat region of the potential near $\phi_0=0$ we find that an inflaton uncertainty $\sigma_\phi \lessapprox H/(2\pi)$ is sufficient to lead to a consistent description of quantum transitions, both down and up the potential.  However, in the flattest region of the potential  (where the slow-roll parameter is smaller than $\epsilon \lesssim 5 \times 10^{-4}$), which may be the region of most interest in terms of applications to eternal inflation, this is not enough, and the initial quantum state must contain a significant uncertainty in the scale factor too, of a magnitude indicated by the de Sitter result \eqref{eq:CritdS}. 

As discussed above and shown in Fig. \ref{fig:conv-sp}, the relevant saddle point geometry is typically similar to a standard slow-roll inflationary solution, albeit one with slightly complexified field values. This is certainly the case whenever the initial inflaton value $\phi_0$ is large enough, and $\sigma_\phi$ has been chosen to lie above the critical value $\sigma_c.$ However, as we just saw, in the flattest part of the potential a significant uncertainty in the size of the universe is also required in order to achieve a Stokes phenomenon. We may thus expect the relevant saddle point geometry to change character, and in closing this discussion we will briefly illustrate this effect. Near $\phi_0=0$ we still have the possibility of having a large uncertainty in the inflaton value too, i.e. we may still have a large $\sigma_\phi$ and thus, in combination with $\sigma_a,$ we may still have large values of both $\sigma_x$ and $\sigma_y.$ In this case we still have a roughly slow-roll saddle point geometry, where as before the inflaton starts with a comparatively unlikely value high up on the  potential and slowly rolls down - see Fig. \ref{fig:largesigy}. However, the uncertainty in the inflaton value could also be small, with a correspondingly well determined initial expansion rate, so that once again a Stokes phenomenon is achieved. This corresponds to having a very small (or vanishing) value for $\sigma_y.$ In this case the scalar field is forced to roll up the potential, since its initial and final values are specified with great certainty. But we showed in Eq. \eqref{phidot} that it is not possible for the inflaton to roll up, as long as the field values are real. The resolution is that in this case the saddle point becomes highly complex, and the field evolution also correspondingly complex - see Fig. \eqref{fig:smallsigy}. Moreover, at the end of the transition the scalar field is still rolling up the potential. These two cases thus nicely illustrate the importance of the initial Robin conditions, or equivalently the initial state, in determining the most likely subsequent evolutions. The most appropriate form of the initial state will of course depend on the physical situation under consideration, and  determining the appropriate form of the initial state will be the most important ingredient in applying our results to situations of interest, such as eternal inflation.

\section{Avoiding Off-Shell Singularities} \label{zeros}

For every value of $\sigma_\phi$ there are regions in the complex $N$ plane where the scale factor $a(t)$ vanishes for some $t \in [0,1]$ and the scalar field $\phi(t)$ correspondingly diverges \cite{DiTucci:2019xcr}. These configurations are irregular in terms of the physical variables $a$ and $\phi$ and as a consequence the action functional diverges. Note that this irregularity has no counterpart in terms of the canonical variables $x$, $y$ and the corresponding action is analytic. In fact, what becomes singular when the scale factor vanishes is the map which connects the two sets of variables. Thus, these singularities would  appear in the Jacobian factor the we have been ignoring in the saddle point approximation, because it usually plays a sub-leading role. However, in the special case where the map becomes singular, the Jacobian would render the path integral ill-defined. Therefore, in order to deal with a well defined path integral  we will require that such a curve of zeros (of the scale factor) in the complex $N$ plane does not lie on the defining integration contour, the Lefschetz thimble nor region in between the two. 

\begin{figure}
\includegraphics[width=0.32\textwidth]{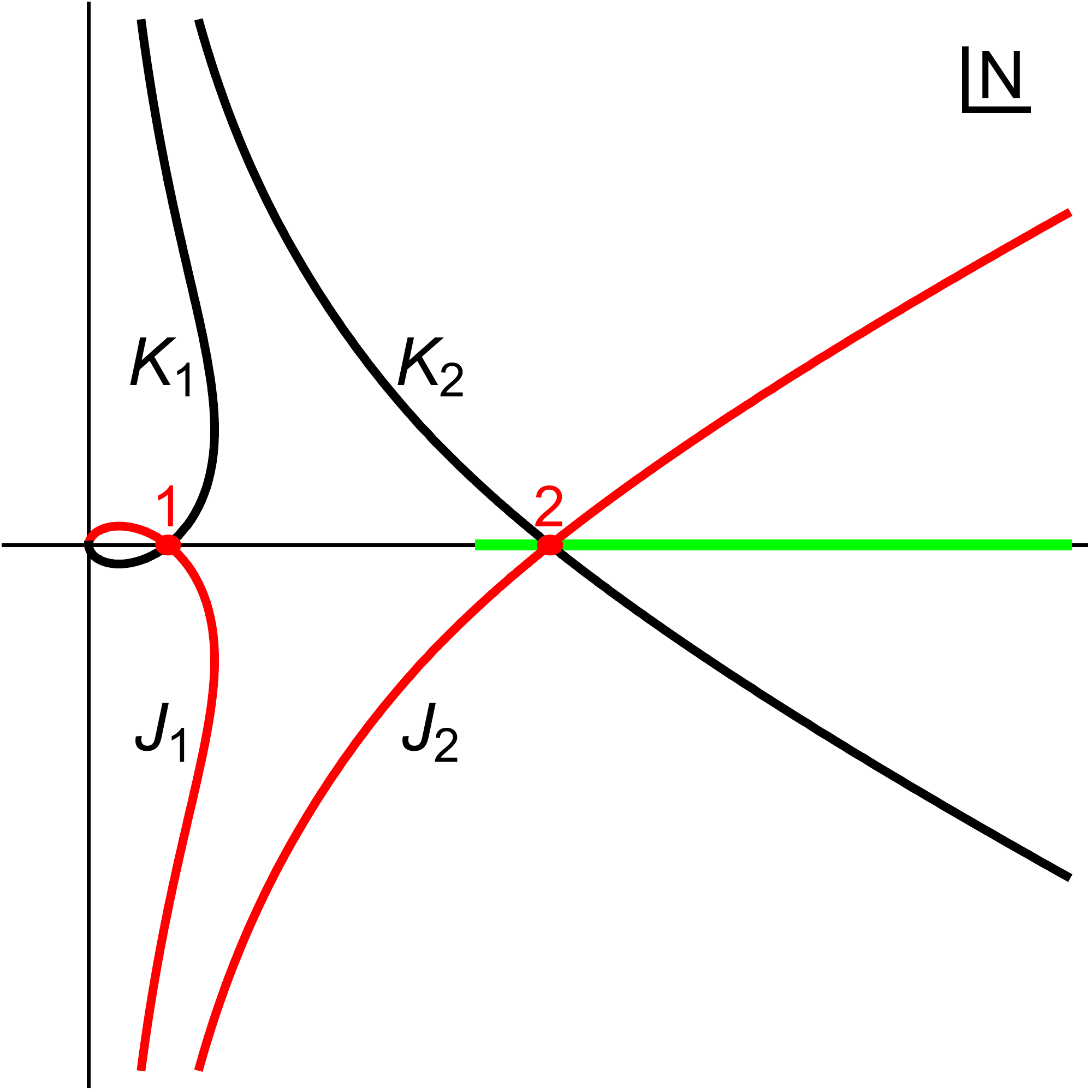}
\includegraphics[width=0.32\textwidth]{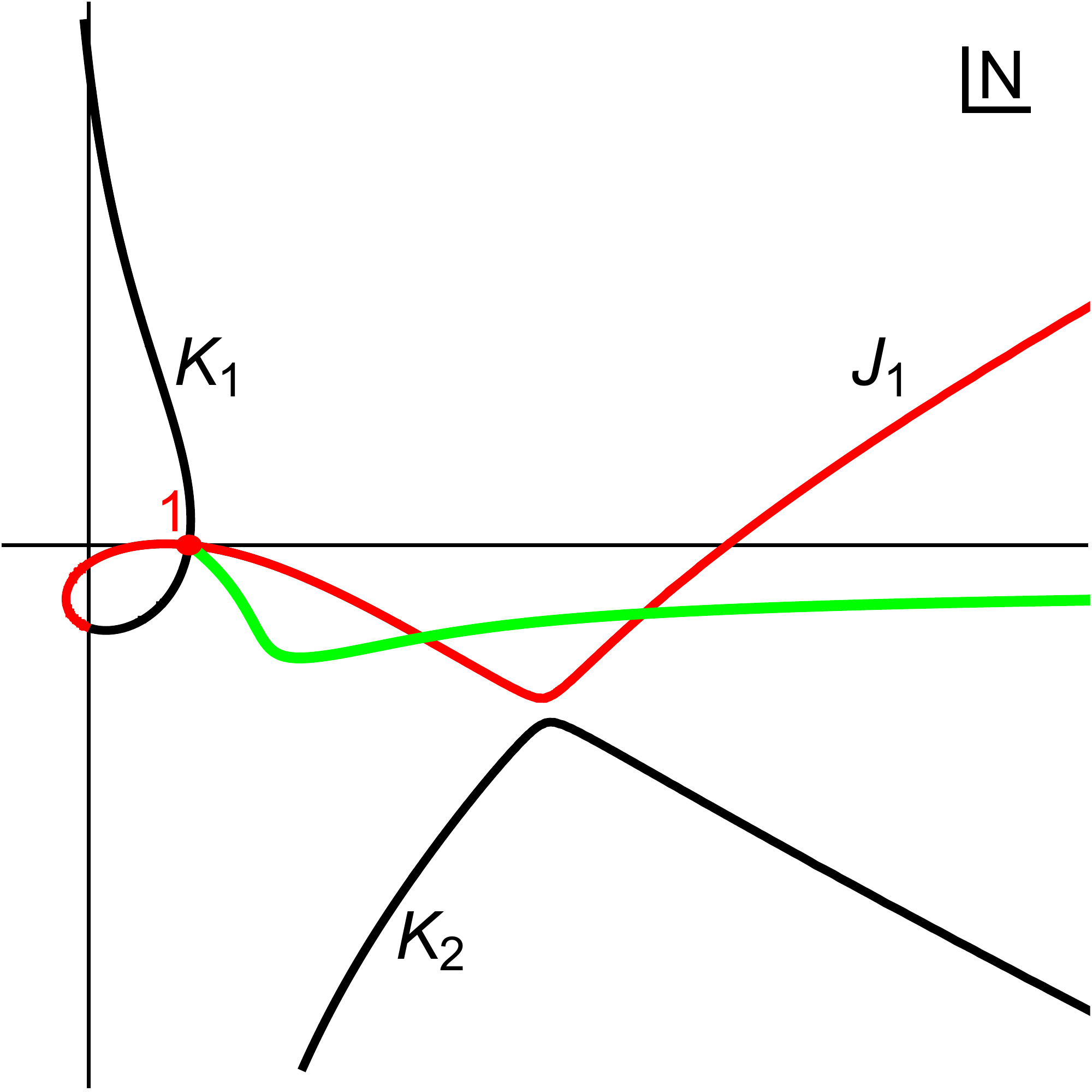}
\includegraphics[width=0.32\textwidth]{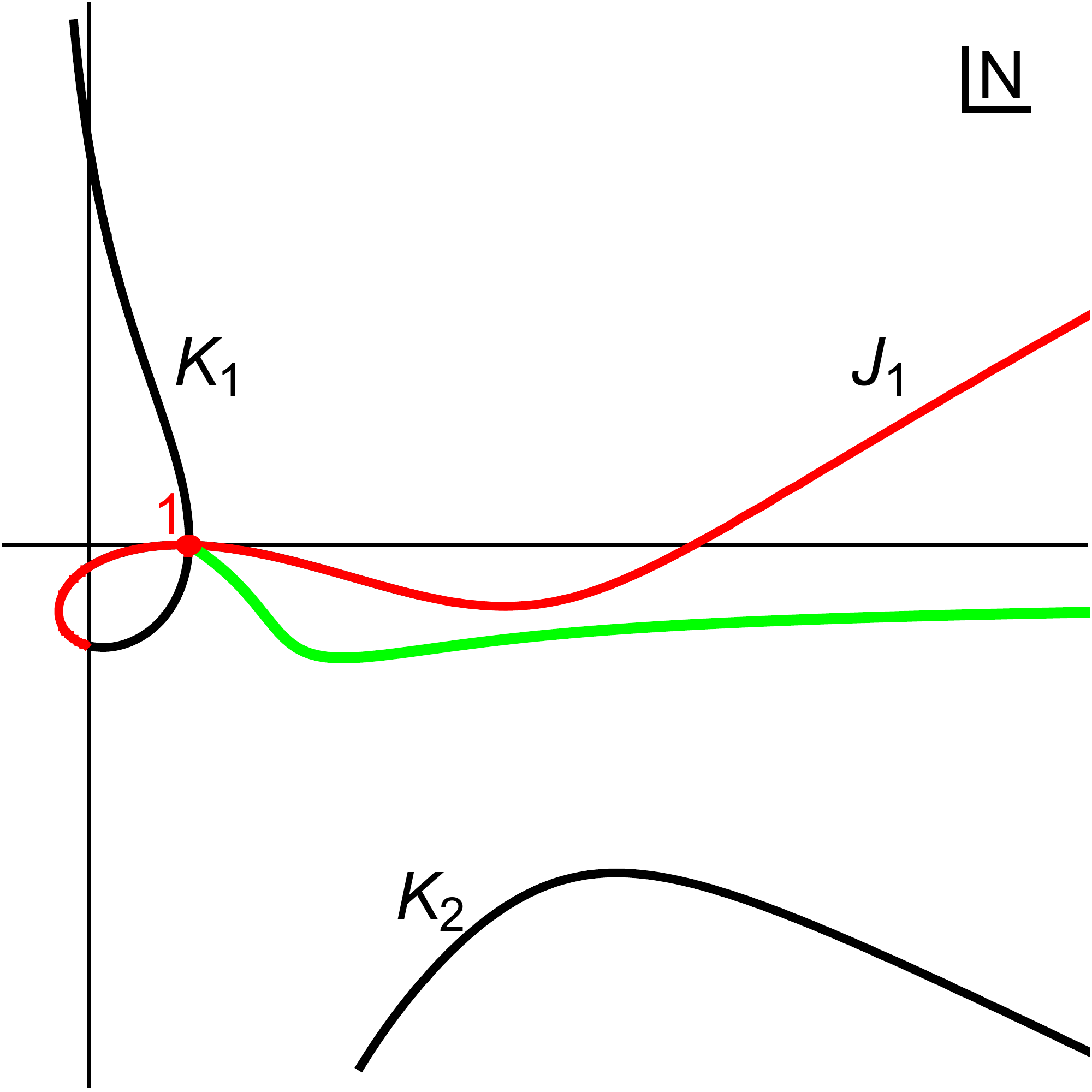}
\caption{The plots show the flow lines (in black) and the scale factor's curve of zeros (in green) for increasing values of $\sigma_{\phi}$. The left, middle and right panels correspond to $\sigma_\phi = 0$, $ \sigma_{\phi} = \sigma_c=0.0154$, and $\sigma_{\phi} = 0.0170 > \sigma_z$ respectively. The other parameters read $a_0 = 100$, $a_1 = 200$, $\phi_0 = 1/10$, $\phi_1=1/2$, $\alpha = 1/10$, $p_x = -54.79$, and $p_y= -1.34$ similar to the previous examples. The curves of zeros still crosses the Lefschetz thimble when the Stokes phenomenon happens, but after a further, modest increase in the spread to $\sigma_{\phi} \approx 0.0164$, the lines do not cross anymore, and the path integral is well defined.} \label{fig-zeros}
\end{figure}

The curve of zeros, just like the flow lines associated with the various saddle points, changes as a function of $\sigma_\phi$. The typical behaviour is shown in Fig. \ref{fig-zeros}. For $0 \leq \sigma_\phi \leq \sigma_c$, the curve of zeros crosses the Lefschetz thimble but there is no more crossing starting from $\sigma > \sigma_z > \sigma_c $. For larger values of the uncertainty, the path integral is well approximated by one saddle point and the variables $a$ and $\phi$ are well defined. From our numerical studies we found that $\sigma_z$ is only modestly larger than $\sigma_c,$ leading to a small increase of the spread required to recover QFT in curved space-time.

\section{Discussion} \label{discussion}

In this work we have taken the first steps in analysing inflationary quantum transitions in semi-classical gravity, more specifically in the path integral formulation of gravity. Such an analysis is of interest since inflationary fluctuations are regularly considered as having momentous implications: they may be the source of the primordial density fluctuations, and they are thought to be able to alter the global structure of spacetime. Since they are typically treated using the framework of QFT in curved spacetime, an important question is whether this approximate treatment is justified. We have analysed this question making use of a specific minisuperspace model containing a scalar field $\phi$ in a potential of the form $V(\phi) = \alpha \cosh\left( \sqrt{\frac{2}{3}} \phi\right),$ where the potential is chosen such that a transformation of variables is possible that enables the action to become quadratic. This potential has the interesting feature of interpolating between a very flat region near $\phi=0,$ where the potential is approximately constant, and a region with a larger slow-roll parameter $\epsilon \approx 1/3.$ 

Our results, which only deal with the simplified case of homogeneous transitions, in fact largely support the results of QFT in curved spacetime, under the assumption that an appropriate initial state of the universe is considered. The way in which the ``standard'' results are recovered is however rather surprising: for instance, we are led to think of a transition up the inflationary potential not so much as involving the inflaton rolling up the potential, but rather as the selection of an unlikely, but otherwise perfectly ordinary, inflationary solution that was already ``hidden'' in the initial state (our results share some conceptual similarities with the framework of Braden et al. in \cite{Braden:2018tky}). In other words, the semi-classical picture that is emerging is that an unlikely large value of the scalar field is picked out (typically containing a small imaginary part as well), such that the desired final value of the scalar field can be reached from it by ordinary slow-roll down the potential. 

In order to obtain consistent results, it is crucial however that an appropriate initial state is imposed. We have done this by using Robin initial conditions, which may equivalently be seen as the imposition of an initial coherent state for the canonical variables of the model. We find that in potential regions that are not too flat, the initial state must contain a sufficient uncertainty in the inflaton value in order for a single saddle point to be relevant to the transition amplitude, implying that an approximate description in terms of QFT in curved spacetime is justified. The critical minimal uncertainty in such potential regions is moreover below the expected scale $H/(2\pi),$ where $H$ denotes the Hubble rate at the start of the transition. More surprising is perhaps our finding that in very flat potential regions (in our model where the slow-roll parameter $\epsilon$ is smaller than about $5 \times 10^{-4}$), considering only an uncertainty in the inflaton is not sufficient: one must also allow for a sufficiently large uncertainty in the size of the universe. This may have consequences for models of eternal inflation, since it remains to be demonstrated that an appropriate state is generated prior to the up-jumping transitions that are usually considered in this framework. The generation of an appropriate initial state remains an interesting topic for future work.

There are in fact many other avenues for future work. An important extension of the present work will be to add inhomogeneous perturbations. Another aspect that will be worth studying will be the difference between transitions that occur while inflation is already underway, compared to transitions right at the beginning of inflation. This latter study will of course require the additional input from a theory of initial conditions, such as the no-boundary proposal \cite{Halliwell:2018ejl,DiTucci:2019dji}. In addition, it may be of interest to clarify what goes wrong when two saddle points remain relevant to a particular transition. Based on the earlier study in pure de Sitter space \cite{DiTucci:2019xcr} we expect fluctuations around the two saddle point geometries to be incompatible with each other and to lead to problematic interference effects or instabilities. Understanding such interference may help in clarifying what happens for transitions in very flat potential regions when the uncertainty in the size of the universe is insufficient. Finally, one can use the semi-classical techniques employed here to investigate other  physical setups, such as quantum transitions across the big bang \cite{Gielen:2015uaa,Bramberger:2017cgf}. We hope to report on progress along those lines in the future.

\acknowledgments

We would like to thank Job Feldbrugge, Matt Johnson, Laura Sberna, Neil Turok and Alex Vilenkin for informative and stimulating discussions. ADT and JLL gratefully acknowledge the support of the European Research Council in the form of the ERC Consolidator Grant CoG 772295 ``Qosmology''. The work of SFB is supported in part by a fellowship from the Studienstiftung des Deutschen Volkes.

\bibliography{Literature}

\begin{thebibliography}{38}
\expandafter\ifx\csname natexlab\endcsname\relax\def\natexlab#1{#1}\fi
\expandafter\ifx\csname bibnamefont\endcsname\relax
  \def\bibnamefont#1{#1}\fi
\expandafter\ifx\csname bibfnamefont\endcsname\relax
  \def\bibfnamefont#1{#1}\fi
\expandafter\ifx\csname citenamefont\endcsname\relax
  \def\citenamefont#1{#1}\fi
\expandafter\ifx\csname url\endcsname\relax
  \def\url#1{\texttt{#1}}\fi
\expandafter\ifx\csname urlprefix\endcsname\relax\def\urlprefix{URL }\fi
\providecommand{\bibinfo}[2]{#2}
\providecommand{\eprint}[2][]{\url{#2}}

\bibitem[{\citenamefont{Sakharov}(1966)}]{Sakharov:1966aja}
\bibinfo{author}{\bibfnamefont{A.~D.} \bibnamefont{Sakharov}},
  \bibinfo{journal}{Zh. Eksp. Teor. Fiz.} \textbf{\bibinfo{volume}{49}},
  \bibinfo{pages}{345} (\bibinfo{year}{1966}), \bibinfo{note}{[Sov. Phys.
  JETP22,241(1966)]}.

\bibitem[{\citenamefont{Mukhanov and Chibisov}(1981)}]{Mukhanov:1981xt}
\bibinfo{author}{\bibfnamefont{V.~F.} \bibnamefont{Mukhanov}} \bibnamefont{and}
  \bibinfo{author}{\bibfnamefont{G.~V.} \bibnamefont{Chibisov}},
  \bibinfo{journal}{JETP Lett.} \textbf{\bibinfo{volume}{33}},
  \bibinfo{pages}{532} (\bibinfo{year}{1981}), \bibinfo{note}{[Pisma Zh. Eksp.
  Teor. Fiz.33,549(1981)]}.

\bibitem[{\citenamefont{Starobinsky}(1979)}]{Starobinsky:1979ty}
\bibinfo{author}{\bibfnamefont{A.~A.} \bibnamefont{Starobinsky}},
  \bibinfo{journal}{JETP Lett.} \textbf{\bibinfo{volume}{30}},
  \bibinfo{pages}{682} (\bibinfo{year}{1979}), \bibinfo{note}{[,767(1979)]}.

\bibitem[{\citenamefont{Guth and Pi}(1982)}]{Guth:1982ec}
\bibinfo{author}{\bibfnamefont{A.~H.} \bibnamefont{Guth}} \bibnamefont{and}
  \bibinfo{author}{\bibfnamefont{S.~Y.} \bibnamefont{Pi}},
  \bibinfo{journal}{Phys. Rev. Lett.} \textbf{\bibinfo{volume}{49}},
  \bibinfo{pages}{1110} (\bibinfo{year}{1982}).

\bibitem[{\citenamefont{Hawking}(1982)}]{Hawking:1982cz}
\bibinfo{author}{\bibfnamefont{S.~W.} \bibnamefont{Hawking}},
  \bibinfo{journal}{Phys. Lett.} \textbf{\bibinfo{volume}{115B}},
  \bibinfo{pages}{295} (\bibinfo{year}{1982}).

\bibitem[{\citenamefont{Bardeen et~al.}(1983)\citenamefont{Bardeen, Steinhardt,
  and Turner}}]{Bardeen:1983qw}
\bibinfo{author}{\bibfnamefont{J.~M.} \bibnamefont{Bardeen}},
  \bibinfo{author}{\bibfnamefont{P.~J.} \bibnamefont{Steinhardt}},
  \bibnamefont{and} \bibinfo{author}{\bibfnamefont{M.~S.}
  \bibnamefont{Turner}}, \bibinfo{journal}{Phys. Rev.}
  \textbf{\bibinfo{volume}{D28}}, \bibinfo{pages}{679} (\bibinfo{year}{1983}).

\bibitem[{\citenamefont{Finelli and Brandenberger}(2002)}]{Finelli:2001sr}
\bibinfo{author}{\bibfnamefont{F.}~\bibnamefont{Finelli}} \bibnamefont{and}
  \bibinfo{author}{\bibfnamefont{R.}~\bibnamefont{Brandenberger}},
  \bibinfo{journal}{Phys. Rev.} \textbf{\bibinfo{volume}{D65}},
  \bibinfo{pages}{103522} (\bibinfo{year}{2002}), \eprint{hep-th/0112249}.

\bibitem[{\citenamefont{Finelli}(2002)}]{Finelli:2002we}
\bibinfo{author}{\bibfnamefont{F.}~\bibnamefont{Finelli}},
  \bibinfo{journal}{Phys. Lett.} \textbf{\bibinfo{volume}{B545}},
  \bibinfo{pages}{1} (\bibinfo{year}{2002}), \eprint{hep-th/0206112}.

\bibitem[{\citenamefont{Lehners et~al.}(2007)\citenamefont{Lehners, McFadden,
  Turok, and Steinhardt}}]{Lehners:2007ac}
\bibinfo{author}{\bibfnamefont{J.-L.} \bibnamefont{Lehners}},
  \bibinfo{author}{\bibfnamefont{P.}~\bibnamefont{McFadden}},
  \bibinfo{author}{\bibfnamefont{N.}~\bibnamefont{Turok}}, \bibnamefont{and}
  \bibinfo{author}{\bibfnamefont{P.~J.} \bibnamefont{Steinhardt}},
  \bibinfo{journal}{Phys. Rev.} \textbf{\bibinfo{volume}{D76}},
  \bibinfo{pages}{103501} (\bibinfo{year}{2007}), \eprint{hep-th/0702153}.

\bibitem[{\citenamefont{Qiu et~al.}(2013)\citenamefont{Qiu, Gao, and
  Saridakis}}]{Qiu:2013eoa}
\bibinfo{author}{\bibfnamefont{T.}~\bibnamefont{Qiu}},
  \bibinfo{author}{\bibfnamefont{X.}~\bibnamefont{Gao}}, \bibnamefont{and}
  \bibinfo{author}{\bibfnamefont{E.~N.} \bibnamefont{Saridakis}},
  \bibinfo{journal}{Phys. Rev.} \textbf{\bibinfo{volume}{D88}},
  \bibinfo{pages}{043525} (\bibinfo{year}{2013}), \eprint{1303.2372}.

\bibitem[{\citenamefont{Li}(2013)}]{Li:2013hga}
\bibinfo{author}{\bibfnamefont{M.}~\bibnamefont{Li}}, \bibinfo{journal}{Phys.
  Lett.} \textbf{\bibinfo{volume}{B724}}, \bibinfo{pages}{192}
  (\bibinfo{year}{2013}), \eprint{1306.0191}.

\bibitem[{\citenamefont{Ijjas et~al.}(2014{\natexlab{a}})\citenamefont{Ijjas,
  Lehners, and Steinhardt}}]{Ijjas:2014fja}
\bibinfo{author}{\bibfnamefont{A.}~\bibnamefont{Ijjas}},
  \bibinfo{author}{\bibfnamefont{J.-L.} \bibnamefont{Lehners}},
  \bibnamefont{and} \bibinfo{author}{\bibfnamefont{P.~J.}
  \bibnamefont{Steinhardt}}, \bibinfo{journal}{Phys. Rev.}
  \textbf{\bibinfo{volume}{D89}}, \bibinfo{pages}{123520}
  (\bibinfo{year}{2014}{\natexlab{a}}), \eprint{1404.1265}.

\bibitem[{\citenamefont{Kiefer et~al.}(1998)\citenamefont{Kiefer, Polarski, and
  Starobinsky}}]{Kiefer:1998qe}
\bibinfo{author}{\bibfnamefont{C.}~\bibnamefont{Kiefer}},
  \bibinfo{author}{\bibfnamefont{D.}~\bibnamefont{Polarski}}, \bibnamefont{and}
  \bibinfo{author}{\bibfnamefont{A.~A.} \bibnamefont{Starobinsky}},
  \bibinfo{journal}{Int. J. Mod. Phys.} \textbf{\bibinfo{volume}{D7}},
  \bibinfo{pages}{455} (\bibinfo{year}{1998}), \eprint{gr-qc/9802003}.

\bibitem[{\citenamefont{Battarra and Lehners}(2014)}]{Battarra:2013cha}
\bibinfo{author}{\bibfnamefont{L.}~\bibnamefont{Battarra}} \bibnamefont{and}
  \bibinfo{author}{\bibfnamefont{J.-L.} \bibnamefont{Lehners}},
  \bibinfo{journal}{Phys. Rev.} \textbf{\bibinfo{volume}{D89}},
  \bibinfo{pages}{063516} (\bibinfo{year}{2014}), \eprint{1309.2281}.

\bibitem[{\citenamefont{Birrell and Davies}(1984)}]{Birrell:1982ix}
\bibinfo{author}{\bibfnamefont{N.~D.} \bibnamefont{Birrell}} \bibnamefont{and}
  \bibinfo{author}{\bibfnamefont{P.~C.~W.} \bibnamefont{Davies}},
  \emph{\bibinfo{title}{{Quantum Fields in Curved Space}}}, Cambridge
  Monographs on Mathematical Physics (\bibinfo{publisher}{Cambridge Univ.
  Press}, \bibinfo{address}{Cambridge, UK}, \bibinfo{year}{1984}), ISBN
  \bibinfo{isbn}{0521278589, 9780521278584, 9780521278584}.

\bibitem[{\citenamefont{Steinhardt}(1982)}]{Steinhardt:1982kg}
\bibinfo{author}{\bibfnamefont{P.~J.} \bibnamefont{Steinhardt}}, in
  \emph{\bibinfo{booktitle}{{Nuffield Workshop on the Very Early Universe
  Cambridge, England, June 21-July 9, 1982}}} (\bibinfo{year}{1982}), pp.
  \bibinfo{pages}{251--266}.

\bibitem[{\citenamefont{Vilenkin}(1983)}]{Vilenkin:1983xq}
\bibinfo{author}{\bibfnamefont{A.}~\bibnamefont{Vilenkin}},
  \bibinfo{journal}{Phys. Rev.} \textbf{\bibinfo{volume}{D27}},
  \bibinfo{pages}{2848} (\bibinfo{year}{1983}).

\bibitem[{\citenamefont{Aguirre}(2007)}]{Aguirre:2007gy}
\bibinfo{author}{\bibfnamefont{A.}~\bibnamefont{Aguirre}}
  (\bibinfo{year}{2007}), \eprint{0712.0571}.

\bibitem[{\citenamefont{Johnson and Lehners}(2012)}]{Johnson:2011aa}
\bibinfo{author}{\bibfnamefont{M.~C.} \bibnamefont{Johnson}} \bibnamefont{and}
  \bibinfo{author}{\bibfnamefont{J.-L.} \bibnamefont{Lehners}},
  \bibinfo{journal}{Phys. Rev.} \textbf{\bibinfo{volume}{D85}},
  \bibinfo{pages}{103509} (\bibinfo{year}{2012}), \eprint{1112.3360}.

\bibitem[{\citenamefont{Teitelboim}(1982)}]{Teitelboim:1981ua}
\bibinfo{author}{\bibfnamefont{C.}~\bibnamefont{Teitelboim}},
  \bibinfo{journal}{Phys. Rev.} \textbf{\bibinfo{volume}{D25}},
  \bibinfo{pages}{3159} (\bibinfo{year}{1982}).

\bibitem[{\citenamefont{Teitelboim}(1983)}]{Teitelboim:1983fk}
\bibinfo{author}{\bibfnamefont{C.}~\bibnamefont{Teitelboim}},
  \bibinfo{journal}{Phys. Rev.} \textbf{\bibinfo{volume}{D28}},
  \bibinfo{pages}{297} (\bibinfo{year}{1983}).

\bibitem[{\citenamefont{Garay et~al.}(1991)\citenamefont{Garay, Halliwell, and
  Mena~Marugan}}]{Garay:1990re}
\bibinfo{author}{\bibfnamefont{L.~J.} \bibnamefont{Garay}},
  \bibinfo{author}{\bibfnamefont{J.~J.} \bibnamefont{Halliwell}},
  \bibnamefont{and} \bibinfo{author}{\bibfnamefont{G.~A.}
  \bibnamefont{Mena~Marugan}}, \bibinfo{journal}{Phys. Rev.}
  \textbf{\bibinfo{volume}{D43}}, \bibinfo{pages}{2572} (\bibinfo{year}{1991}).

\bibitem[{\citenamefont{Starobinsky}(1986)}]{Starobinsky:1986fx}
\bibinfo{author}{\bibfnamefont{A.~A.} \bibnamefont{Starobinsky}},
  \bibinfo{journal}{Lect. Notes Phys.} \textbf{\bibinfo{volume}{246}},
  \bibinfo{pages}{107} (\bibinfo{year}{1986}).

\bibitem[{\citenamefont{Koehn et~al.}(2016)\citenamefont{Koehn, Lehners, and
  Ovrut}}]{Koehn:2015vvy}
\bibinfo{author}{\bibfnamefont{M.}~\bibnamefont{Koehn}},
  \bibinfo{author}{\bibfnamefont{J.-L.} \bibnamefont{Lehners}},
  \bibnamefont{and} \bibinfo{author}{\bibfnamefont{B.}~\bibnamefont{Ovrut}},
  \bibinfo{journal}{Phys. Rev.} \textbf{\bibinfo{volume}{D93}},
  \bibinfo{pages}{103501} (\bibinfo{year}{2016}), \eprint{1512.03807}.

\bibitem[{\citenamefont{Ijjas et~al.}(2014{\natexlab{b}})\citenamefont{Ijjas,
  Steinhardt, and Loeb}}]{Ijjas:2014nta}
\bibinfo{author}{\bibfnamefont{A.}~\bibnamefont{Ijjas}},
  \bibinfo{author}{\bibfnamefont{P.~J.} \bibnamefont{Steinhardt}},
  \bibnamefont{and} \bibinfo{author}{\bibfnamefont{A.}~\bibnamefont{Loeb}},
  \bibinfo{journal}{Phys. Lett.} \textbf{\bibinfo{volume}{B736}},
  \bibinfo{pages}{142} (\bibinfo{year}{2014}{\natexlab{b}}),
  \eprint{1402.6980}.

\bibitem[{\citenamefont{Feldbrugge
  et~al.}(2017{\natexlab{a}})\citenamefont{Feldbrugge, Lehners, and
  Turok}}]{Feldbrugge:2017kzv}
\bibinfo{author}{\bibfnamefont{J.}~\bibnamefont{Feldbrugge}},
  \bibinfo{author}{\bibfnamefont{J.-L.} \bibnamefont{Lehners}},
  \bibnamefont{and} \bibinfo{author}{\bibfnamefont{N.}~\bibnamefont{Turok}},
  \bibinfo{journal}{Phys. Rev.} \textbf{\bibinfo{volume}{D95}},
  \bibinfo{pages}{103508} (\bibinfo{year}{2017}{\natexlab{a}}),
  \eprint{1703.02076}.

\bibitem[{\citenamefont{Feldbrugge
  et~al.}(2018{\natexlab{a}})\citenamefont{Feldbrugge, Lehners, and
  Turok}}]{Feldbrugge:2017mbc}
\bibinfo{author}{\bibfnamefont{J.}~\bibnamefont{Feldbrugge}},
  \bibinfo{author}{\bibfnamefont{J.-L.} \bibnamefont{Lehners}},
  \bibnamefont{and} \bibinfo{author}{\bibfnamefont{N.}~\bibnamefont{Turok}},
  \bibinfo{journal}{Phys. Rev.} \textbf{\bibinfo{volume}{D97}},
  \bibinfo{pages}{023509} (\bibinfo{year}{2018}{\natexlab{a}}),
  \eprint{1708.05104}.

\bibitem[{\citenamefont{Witten}(2011)}]{Witten:2010cx}
\bibinfo{author}{\bibfnamefont{E.}~\bibnamefont{Witten}},
  \bibinfo{journal}{AMS/IP Stud. Adv. Math.} \textbf{\bibinfo{volume}{50}},
  \bibinfo{pages}{347} (\bibinfo{year}{2011}), \eprint{1001.2933}.

\bibitem[{\citenamefont{Feldbrugge
  et~al.}(2017{\natexlab{b}})\citenamefont{Feldbrugge, Lehners, and
  Turok}}]{Feldbrugge:2017fcc}
\bibinfo{author}{\bibfnamefont{J.}~\bibnamefont{Feldbrugge}},
  \bibinfo{author}{\bibfnamefont{J.-L.} \bibnamefont{Lehners}},
  \bibnamefont{and} \bibinfo{author}{\bibfnamefont{N.}~\bibnamefont{Turok}},
  \bibinfo{journal}{Phys. Rev. Lett.} \textbf{\bibinfo{volume}{119}},
  \bibinfo{pages}{171301} (\bibinfo{year}{2017}{\natexlab{b}}),
  \eprint{1705.00192}.

\bibitem[{\citenamefont{Feldbrugge
  et~al.}(2018{\natexlab{b}})\citenamefont{Feldbrugge, Lehners, and
  Turok}}]{FLT4}
\bibinfo{author}{\bibfnamefont{J.}~\bibnamefont{Feldbrugge}},
  \bibinfo{author}{\bibfnamefont{J.-L.} \bibnamefont{Lehners}},
  \bibnamefont{and} \bibinfo{author}{\bibfnamefont{N.}~\bibnamefont{Turok}},
  \bibinfo{journal}{Universe} \textbf{\bibinfo{volume}{4}},
  \bibinfo{pages}{100} (\bibinfo{year}{2018}{\natexlab{b}}),
  \eprint{1805.01609}.

\bibitem[{\citenamefont{Di~Tucci and Lehners}(2018)}]{DiTucci:2018fdg}
\bibinfo{author}{\bibfnamefont{A.}~\bibnamefont{Di~Tucci}} \bibnamefont{and}
  \bibinfo{author}{\bibfnamefont{J.-L.} \bibnamefont{Lehners}},
  \bibinfo{journal}{Phys. Rev.} \textbf{\bibinfo{volume}{D98}},
  \bibinfo{pages}{103506} (\bibinfo{year}{2018}), \eprint{1806.07134}.

\bibitem[{\citenamefont{Di~Tucci et~al.}(2019)\citenamefont{Di~Tucci,
  Feldbrugge, Lehners, and Turok}}]{DiTucci:2019xcr}
\bibinfo{author}{\bibfnamefont{A.}~\bibnamefont{Di~Tucci}},
  \bibinfo{author}{\bibfnamefont{J.}~\bibnamefont{Feldbrugge}},
  \bibinfo{author}{\bibfnamefont{J.-L.} \bibnamefont{Lehners}},
  \bibnamefont{and} \bibinfo{author}{\bibfnamefont{N.}~\bibnamefont{Turok}}
  (\bibinfo{year}{2019}), \eprint{1906.09007}.

\bibitem[{\citenamefont{Di~Tucci and Lehners}(2019)}]{DiTucci:2019dji}
\bibinfo{author}{\bibfnamefont{A.}~\bibnamefont{Di~Tucci}} \bibnamefont{and}
  \bibinfo{author}{\bibfnamefont{J.-L.} \bibnamefont{Lehners}},
  \bibinfo{journal}{Phys. Rev. Lett.} \textbf{\bibinfo{volume}{122}},
  \bibinfo{pages}{201302} (\bibinfo{year}{2019}), \eprint{1903.06757}.

\bibitem[{\citenamefont{Feldbrugge et~al.}(2019)\citenamefont{Feldbrugge,
  Fertig, Sberna, and Turok}}]{FFST}
\bibinfo{author}{\bibfnamefont{J.}~\bibnamefont{Feldbrugge}},
  \bibinfo{author}{\bibfnamefont{A.}~\bibnamefont{Fertig}},
  \bibinfo{author}{\bibfnamefont{L.}~\bibnamefont{Sberna}}, \bibnamefont{and}
  \bibinfo{author}{\bibfnamefont{N.}~\bibnamefont{Turok}}, \bibinfo{journal}{in
  preparation}  (\bibinfo{year}{2019}).

\bibitem[{\citenamefont{Braden et~al.}(2018)\citenamefont{Braden, Johnson,
  Peiris, Pontzen, and Weinfurtner}}]{Braden:2018tky}
\bibinfo{author}{\bibfnamefont{J.}~\bibnamefont{Braden}},
  \bibinfo{author}{\bibfnamefont{M.~C.} \bibnamefont{Johnson}},
  \bibinfo{author}{\bibfnamefont{H.~V.} \bibnamefont{Peiris}},
  \bibinfo{author}{\bibfnamefont{A.}~\bibnamefont{Pontzen}}, \bibnamefont{and}
  \bibinfo{author}{\bibfnamefont{S.}~\bibnamefont{Weinfurtner}}
  (\bibinfo{year}{2018}), \eprint{1806.06069}.

\bibitem[{\citenamefont{Halliwell et~al.}(2019)\citenamefont{Halliwell, Hartle,
  and Hertog}}]{Halliwell:2018ejl}
\bibinfo{author}{\bibfnamefont{J.~J.} \bibnamefont{Halliwell}},
  \bibinfo{author}{\bibfnamefont{J.~B.} \bibnamefont{Hartle}},
  \bibnamefont{and} \bibinfo{author}{\bibfnamefont{T.}~\bibnamefont{Hertog}},
  \bibinfo{journal}{Phys. Rev.} \textbf{\bibinfo{volume}{D99}},
  \bibinfo{pages}{043526} (\bibinfo{year}{2019}), \eprint{1812.01760}.

\bibitem[{\citenamefont{Gielen and Turok}(2016)}]{Gielen:2015uaa}
\bibinfo{author}{\bibfnamefont{S.}~\bibnamefont{Gielen}} \bibnamefont{and}
  \bibinfo{author}{\bibfnamefont{N.}~\bibnamefont{Turok}},
  \bibinfo{journal}{Phys. Rev. Lett.} \textbf{\bibinfo{volume}{117}},
  \bibinfo{pages}{021301} (\bibinfo{year}{2016}), \eprint{1510.00699}.

\bibitem[{\citenamefont{Bramberger et~al.}(2017)\citenamefont{Bramberger,
  Hertog, Lehners, and Vreys}}]{Bramberger:2017cgf}
\bibinfo{author}{\bibfnamefont{S.~F.} \bibnamefont{Bramberger}},
  \bibinfo{author}{\bibfnamefont{T.}~\bibnamefont{Hertog}},
  \bibinfo{author}{\bibfnamefont{J.-L.} \bibnamefont{Lehners}},
  \bibnamefont{and} \bibinfo{author}{\bibfnamefont{Y.}~\bibnamefont{Vreys}},
  \bibinfo{journal}{JCAP} \textbf{\bibinfo{volume}{1707}}, \bibinfo{pages}{007}
  (\bibinfo{year}{2017}), \eprint{1701.05399}.

\end{thebibliography}
\end{document}